\documentclass[]{article}
\usepackage[a4paper,top=2cm,bottom=2cm,left=1.5cm,right=1.5cm,marginparwidth=1.75cm]{geometry}
\usepackage{authblk}
\usepackage[numbers, sort&compress]{natbib} 
\usepackage[english]{babel}
\usepackage{url,microtype}
\usepackage[hidelinks]{hyperref} 
\usepackage{amsmath}
\usepackage{accents}
\usepackage{amssymb}
\usepackage{mathtools}
\usepackage{graphicx}
\usepackage[dvipsnames]{xcolor}
\usepackage{arydshln}
\usepackage{pdfpages}
\usepackage{soul}
\usepackage{units}
\usepackage[shortlabels]{enumitem}
\usepackage{bbm}
\usepackage{subcaption}
\usepackage[percent]{overpic}
\usepackage{algorithm}
\usepackage[noEnd=true,endLComment=]{algpseudocodex}
\usepackage{tikz}
\usetikzlibrary{shapes.geometric, arrows}
\usepackage{multirow}
\usepackage{multicol}
\usepackage{rotating}
\usepackage{pgfplots}

\newcommand{\ve}[1]{\text{\boldmath${#1}$}} 
\newcommand{\te}[1]{\text{\boldmath$\mathrm{#1}$}} 

\newcommand{\grad}{\nabla\,}
\renewcommand{\div}{{\grad\cdot\,}}
\newcommand{\grads}{\nabla^\mathrm{S}}
\renewcommand{\top}{\mathrm{T}} 

\newcommand{\id}{\te{I}}

\newcommand{\n}{\ve{n}} 
\newcommand{\GD}{{\Gamma^\mathrm{D}}}
\newcommand{\GN}{{\Gamma^\mathrm{N}}}

\newcommand{\gu}{{\ve{g}_u}}
\newcommand{\gp}{{g_p}}
\newcommand{\hu}{{\ve{h}_u}}
\newcommand{\hp}{{h_p}}

\newcommand{\dt}{{\Delta t}}

\newlength{\dhatheight}
\newcommand{\hhat}[1]{%
	\settoheight{\dhatheight}{\ensuremath{\hat{#1}}}%
	\addtolength{\dhatheight}{-0.35ex}%
	\hat{\vphantom{\rule{1pt}{\dhatheight}}%
		\smash{\hat{#1}}}}

\newcommand{\avg}[1]{{\left\{\!\!\left\{#1\right\}\!\!\right\}}}
\newcommand{\jump}[1]{{\left[\!\left[#1\right]\!\right]}}
\newcommand{\ojump}[1]{{\left[#1\right]}}

\newcommand{\convectiveterm}[2]{\left(#2\cdot \nabla\right)#1} 

\newcommand{\draftAll}{false}
\newcommand{\draftSec}{\draftAll}

\AfterEndEnvironment{prob}{\noindent\ignorespaces}

\newcommand{\logLogSlopeTriangle}[6]
{

	\pgfplotsextra
	{
		\pgfkeysgetvalue{/pgfplots/xmin}{\xmin}
		\pgfkeysgetvalue{/pgfplots/xmax}{\xmax}
		\pgfkeysgetvalue{/pgfplots/ymin}{\ymin}
		\pgfkeysgetvalue{/pgfplots/ymax}{\ymax}

		\pgfmathsetmacro{\xArel}{#1}
		\pgfmathsetmacro{\yArel}{#3}
		\pgfmathsetmacro{\xBrel}{#1-#2}
		\pgfmathsetmacro{\yBrel}{\yArel}
		\pgfmathsetmacro{\xCrel}{\xArel}

		\pgfmathsetmacro{\lnxB}{\xmin*(1-(#1-#2))+\xmax*(#1-#2)} 
		\pgfmathsetmacro{\lnxA}{\xmin*(1-#1)+\xmax*#1} 
		\pgfmathsetmacro{\lnyA}{\ymin*(1-#3)+\ymax*#3} 
		\pgfmathsetmacro{\lnyC}{\lnyA+#4*(\lnxA-\lnxB)}
		\pgfmathsetmacro{\yCrel}{\lnyC-\ymin)/(\ymax-\ymin)} 

		\coordinate (A) at (rel axis cs:\xArel,\yArel);
		\coordinate (B) at (rel axis cs:\xBrel,\yBrel);
		\coordinate (C) at (rel axis cs:\xCrel,\yCrel);

		\draw[#5]   (A)-- node[pos=0.5,anchor=north] {}
		(B)--
		(C)-- node[pos=0.5,anchor=west] {#6}
		cycle;
	}
}

\newcommand{\s}{\phantom{0}}
\newcommand{\previouslyrevised}[1]{{#1}} %
\newcommand{\previouslychanged}[1]{{#1}} 
\newcommand{\revised}[1]{{#1}}

\title{Higher-Order \previouslyrevised{Discontinuous Galerkin} Splitting Schemes for Fluids with Variable Viscosity}
\author[1]{R. Schussnig$^{\star,}$}
\author[2]{N. Fehn}
\author[3]{D.R.Q. Pacheco}
\author[1]{M. Kronbichler}
\affil[$^\star$]{richard.schussnig@rub.de}
\affil[1]{Faculty of Mathematics, Ruhr University Bochum}
\affil[2]{Institute of Mathematics, University of Augsburg}
\affil[3]{Chair for Computational Analysis of Technical Systems, RWTH Aachen University}

\begin{document}

\maketitle

\begin{abstract}
	This article investigates matrix-free higher-order discontinuous
	Galerkin discretizations of the Navier--Stokes equations for
	incompressible flows with variable viscosity.
	The viscosity field may be prescribed analytically or governed by
	a rheological law, as often found in biomedical or
	industrial applications. The DG discretization of the adapted
	second-order viscous terms is carried out via the
	symmetric interior penalty Galerkin method, obviating
	auxiliary variables. Based on this spatial discretization,
	we compare several linearized variants of saddle point block systems and
	projection-based splitting time integration schemes in terms of
	their computational performance. Compared to the
	velocity-pressure block-system for the former, the splitting scheme allows solving a
	sequence of simple problems such as mass, convection-diffusion and Poisson
	equations. We investigate under which conditions
	the improved temporal stability of fully implicit schemes and
	resulting expensive nonlinear solves outperform the splitting
	schemes and linearized variants that are stable under hyperbolic
	time step restrictions.

	\previouslychanged{
	The key aspects of this work are
	i) a higher-order DG discretization for incompressible flows with variable viscosity,
	ii) accelerated nonlinear solver variants and suitable linearizations
	adopting a matrix-free $hp$-multigrid solver,
	and
	iii) a detailed comparison of the monolithic and projection-based solvers
	in terms of their (non-)linear solver performance.}

	The presented schemes are evaluated in a series of numerical examples
	verifying their spatial and temporal accuracy, and the preconditioner
	performance under increasing viscosity contrasts, while their efficiency
	is showcased in the backward-facing step benchmark.
\end{abstract}

\section{Introduction}

Non-constant viscosities occur in various flow scenarios, such as when non-Newtonian
behavior---e.g., shear thinning, yield stress---or temperature gradients are present.
However, numerical methods tailored specifically to variable-viscosity flows have been available,
until recently, only based on mixed velocity-pressure formulations, while schemes
built upon consistent splitting or projection steps are still
rather scarce in the literature. Some of the latest developments
include mixed formulations~\cite{Anaya2021, Anaya2023}, stabilization
methods~\cite{Schussnig2021_e, Barrenechea2023, Galarce2025}, and efficient
timestepping schemes~\cite{Plasman2020, Stiller2020, Pacheco2021_c, Guesmi2023,
Barrenechea2024, ElAmrani2024}. Notably, some of the most popular and accurate
fractional-step methods were originally conceived for constant viscosity.
An example is the incremental pressure-correction method in its rotational
variant~\cite{Timmermans1996, Guermond2003rotational}, which can reach second-order
accuracy for homogeneous flows but does not even converge for variable
viscosity~\cite{Deteix2018}. To remedy that, \citet{Deteix2019} proposed a so-called
shear-rate projection method, which extends the rotational scheme to variable viscosity
at the price of (many) additional substeps. Moreover, the dual splitting scheme
by~\citet{Karniadakis1991}---one of the few higher-order fractional-step methods,
along with consistent splitting schemes~\cite{Pacheco2021_c, Liu2009}---was also originally
designed for constant viscosity. The extension of the scheme towards generalized Newtonian fluids
\previouslychanged{was carried out by \citet{Karamanos2000} in the general case and
by \citet{Blackburn2025} for turbulent flows.
In this context, the present article focuses on higher-order accurate discontinuous Galerkin (DG)
discretizations of monolithic and projection-based schemes of variable-viscosity
flows and thorough testing of the resulting framework.}

One of the main motivations for the present work and splitting/projection schemes in general
is that the construction of efficient preconditioners within a coupled formulation of the
incompressible Navier--Stokes equations is a
delicate matter even for constant viscosity. Depending on the flow regime, the pressure Schur
complement can be suitably approximated by mass and Poisson operators defined in the
pressure space~\cite{Cahouet1988}, while problems dominated by convective effects
can be captured, e.g., by augmented-Lagrangian-based approaches~\cite{Benzi2006, Benzi2011}
or their continuous equivalent adding a
grad-div term to the velocity block~\cite{Heister2013}. However, these approaches
shift the challenges from approximating the pressure Schur complement to
approximating an ill-conditioned velocity-velocity block, requiring tailored multigrid strategies~\cite{Shih2023}.
More standard alternatives are available with the
pressure convection-diffusion (PCD)~\cite{Elman2014, Kay2003}
and least-squares commutator~\cite{Elman2007, Elman2009} preconditioners,
which are not as robust with respect to the Reynolds number as augmented-Lagrangian approaches,
but tend to perform well given reasonable time step sizes and grids.
Coupled or also called monolithic solvers typically achieve Courant numbers of 1 to 5 or more
on practically relevant grids in a standard hyperbolic Courant--Friedrichs--Lewy (CFL) condition,
meaning the time step size scales with the grid size rather than its square.

Aiming for variable viscosity (Navier--)Stokes problems, ideas from~\citet{Cahouet1988} have
been successfully adopted, e.g., in earth mantle convection~\cite{Clevenger2021}
and multiphase flows~\cite{Kronbichler2018}.
Spectral equivalence of the pressure Schur complement and its continuous Galerkin
approximation via scaled Poisson and mass operators in the pressure space
can be shown for sufficiently smooth viscosity $\nu$~\cite{Grinevich2009}.
However, this approximation typically breaks down for sharp contrasts in the viscosity field
and dominant convection. In the former case, the so-called wBFBT
preconditioner~\cite{Rudi2017} might be considered as a remedy with higher
cost per iteration but significantly enhanced robustness with respect to the parameter field.
Alternatively, augmented-Lagrangian methods are also formidable candidates in this regard.
As shown in~\cite{Shih2023}, tailored multigrid ingredients enable tackling large and abrupt variations in
the viscosity field as well. These ideas can be extended to global-in-time approaches
including variable viscosity and convective terms as recently shown in~\cite{Lohmann2024augmented}.

Given the difficulty of a single solver setup for all flow regimes,
there is room for specialized methods concentrating either on
highly convective transport or purely viscous processes.
Demanding robustness and efficiency in both highly convective regimes and
under highly heterogeneous viscosity is
challenging on the one side, but also in parts contradictory: Where convective effects are dominant,
typically viscous effects are less prevalent, and vice versa. However, prime examples of such
scenarios can be found in technical sciences or biology, where certain pathologies of the vascular
system such as aortic dissection~\cite{RolfPissarczyk2025} show distinct low-Reynolds zones
combined with regions of dominant convection.
Fast physics-based solvers designed for these complex flow scenarios are hence desirable to
capture the practically relevant effects, e.g., of generalized Newtonian fluids~\cite{Schussnig2021_e},
multiphase~\cite{Pacheco2022_a}, or thixoviscoplastic flows~\cite{Begum2025}.
Suitable numerical techniques can further enhance or even enable new applications in
research and industry applications in the medical and technical sciences related to problems such as
fluid--structure interaction~\cite{Schussnig2021_d}, sensitivity analysis and uncertainty
quantification~\cite{Bosnjak2024geometricUncertainty}, or
engineering (bio-)mechanics~\cite{Schussnig2022_a, Schussnig2024}.
Hence, the methods presented herein can directly improve solution times for practically relevant scenarios
under realistic parameter combinations, and allow for larger and more detailed problems to be tackled
potentially leading to new insights in the downstream applications.

\previouslyrevised{
	The paper is organized as follows: Sec.~\ref{sec:problem_formulation} introduces the problem
	of incompressible viscous flow with variable viscosity,
	whose time integration in a standard monolithic setting and via a splitting method are
	discussed in Secs.~\ref{sec:monolithic_scheme} and \ref{sec:splitting_scheme}, respectively.
	Spatial discretization adopting an $L^2$-conforming DG formulation of the involved operators
	including divergence and continuity penalty terms for consistent stabilization
	is discussed in Sec.~\ref{sec:spatial_discretization}.
	Thereafter, Sec.~\ref{sec:solution_algorithms} summarizes the solution strategies based on various
	(semi-)implicit and explicit variants of the nonlinear convective terms and viscosity parameter
	together with nonlinear solvers, matrix-free linear solvers and suitable preconditioners.
	Sec.~\ref{sec:numerical_results} critically examines the capabilites of the devised solver family through:
	i) a manufactured solution, showcasing convergence rates in space and time,
	ii) the lid-driven cavity flow benchmark, highlighting robustness with respect to varying viscosity parameters,
	and
	iii) the backward-facing step benchmark, comparing the schemes in terms of
	throughput achieved in a practically relevant example.
}

\section{Problem formulation}
\label{sec:problem_formulation}
 The incompressible Navier--Stokes system in a domain
$\Omega \subset \mathbb{R}^d$, $d=2, 3$ can be written as
\begin{align}
	\partial_t \ve{u}
	+
	\convectiveterm{\ve{u}}{\ve{u}}
	-
	\div \te{S}
	+
	\grad p
	&
	=
	\ve{b}
	&
	\mathrm{in} \quad \Omega \times (0,T]
	,
	\label{eqn:momentum_balance_strong_form}
	\\
	\div \ve{u}
	&
	= 0
	&
	\mathrm{in} \quad \Omega \times [0,T]
	,
	\label{eqn:continuity_equation}
\end{align}
with flow velocity $\ve{u}$, kinematic pressure $p$, body force $\ve{b}$
and shear stress tensor $\te{S}$ defined as
\begin{align*}
\te{S}(\nu, \ve{u}) \coloneqq 2 \nu \grads\ve{u}
\quad
\mathrm{with}
\quad
\grads(\cdot) = \left[\grad(\cdot)\right]^\mathrm{S}
\quad
\text{and}
\quad
(\cdot)^\mathrm{S} \coloneqq \nicefrac 1 2 \, (\cdot) + \nicefrac 1 2 \, (\cdot)^\top
\end{align*}
Here, the viscosity of the fluid is considered variable, with the source of this variability being, e.g.,
a constitutive model or an analytical function in space and time. Some of the many practically relevant
modeling approaches introducing a variable viscosity are generalized Newtonian laws, which typically
model non-Newtonian effects via a non-linear relation between the fluids' shear rate
$
\dot\gamma \coloneqq  \sqrt{
	2 \grads\ve{u} : \grads\ve{u}
}
$
and the apparent viscosity
$\nu = \eta(\dot\gamma): \mathbb{R}^+ \mapsto \mathbb{R}^+\setminus\{0\}$,
for example considering~\cite{Galdi2008}
\begin{align}
	\eta(\dot\gamma) \coloneqq
	\eta_\infty
	+
	(\eta_0 - \eta_\infty)
	\left[
		\kappa + (\lambda \dot\gamma)^a
	\right]^{\frac{b-1}{a}}
	.
	\label{eqn:viscosity_general}
\end{align}
In Eqn.~\eqref{eqn:viscosity_general}, the asymptotic viscosity limits
are denoted by $\eta_0$ and $\eta_\infty$, and fitting parameters
$\kappa$, $\lambda$, $a$ and $b$ allow recovering the Newtonian
($\eta_0=\eta_\infty$), the Power law ($\kappa=\eta_\infty=0$),
Carreau ($2\,\kappa=2=a$), or Carreau-Yasuda ($\kappa=1$) models.
The Lipschitz-continuous boundary $\Gamma \coloneqq \partial \Omega$
of the domain is decomposed into non-overlapping and non-empty Dirichlet and Neumann parts,
$\GD$ and $\GN$, such that $\Gamma = \GD \cup \GN$ and
$\GD \cap \GN = \emptyset$. The respective boundary conditions
\begin{align}
	\ve{u}
	&= \gu
	&\text{on } \GD \times (0, T],
	\label{eqn:boundary_condition_gu}
	\\
	\left( \te{S} - p \, \id \right) \cdot \ve{n}
	&= \ve{h}
	&\text{on } \GN \times (0,T]
	\label{eqn:boundary_condition_h},
\end{align}
and the initial condition $\ve{u} = \ve{u}_0$, with unit outward normal $\n$, close the problem.
To employ projection methods, we further introduce a splitting of~\eqref{eqn:boundary_condition_h}
into $\ve{h} = \hu - \gp \, \n$, such that
\begin{align}
	\te{S} \cdot \n
	&= \hu
	&\text{on } \GN \times (0,T]
	,
	\label{eqn:boundary_condition_hu}
	\\
	p
	&= \gp
	&\text{on } \GN \times (0,T].
	\label{eqn:boundary_condition_gp}
\end{align}
Note that this assumption of a splitting of the form $\ve{h} = \hu - \gp \n$
being available is in practice a rather minor one, since practical flow
problems are seldom formulated based on the full traction vector $\ve{h}$.
The practically relevant case consists of prescribing either the velocity
$\ve{u}$ or the pressure $p$ on a certain boundary segment.
If the full traction vector shall be imposed on some boundary segment,
reformulations using the decomposition $\ve{h} = \hu - \gp \n$ and suitable
extrapolations in time may be adopted.

\section{Time Integration Methods}
\label{sec:time_integration}
Discretization in time is carried out subdividing the time interval $[0, T]$
into $N_t$ (possibly non-uniform) time steps adopting backward differentiation
formulae of order $m$ (BDF-$m$). The Navier--Stokes equations are treated via a
coupled formulation, which serves as reference and shall be repeated here for the
convenience of the reader. Sec.~\ref{sec:splitting_scheme} presents an alternative
approach based on the velocity-correction scheme by~\cite{Karniadakis1991},
which was further analyzed and characterized in \cite{Guermond2003a, Guermond2006}.

\subsection{Coupled Formulation}
\label{sec:monolithic_scheme}
Employing BDF-$m$ schemes of order $m$ to
system~\eqref{eqn:momentum_balance_strong_form}--\eqref{eqn:continuity_equation}
yields
\begin{align}
	\frac{\gamma_0}{\dt}
	\ve{u}^{n+1}
	-
	\sum_{i=0}^{m-1}
	\frac{\alpha_i}{\dt}
	\ve{u}^{n-i}
	+
	\convectiveterm{\ve{u}^{n+1}}{\ve{u}^\star}
	-
	\div \te{S}\left(\nu^{n+1}\left(\ve{u}^\star\right), \ve{u}^{n+1}\right)
	+
	\grad p^{n+1}
	&
	=
	\ve{b}\left(t_{n+1}\right)
	&
	\mathrm{in} \quad \Omega
	,
	\label{eqn:momentum_balance_time_discrete}
	\\
	\div \ve{u}^{n+1}
	&
	= 0
	&
	\mathrm{in} \quad \Omega
	,
	\label{eqn:continuity_equation_time_discrete}
\end{align}
for time step $n=m,\dots,N_t$ from time $t^n$ to $t^{n+1} \coloneqq t^n + \dt$,
with time step size $\dt$ and suitable coefficients
$\gamma_0$, $\alpha_i$, $i=0, \dots, m-1$ (see,
e.g.,~\cite{Hairer1993}). In~\eqref{eqn:momentum_balance_time_discrete},
the velocity $\ve{u}^\star$ in the convective term and in the rheological
law to recover the viscosity $\nu = \nu(\ve{u})$ may be implicit,
i.e., considered at $t_{n+1}$, leading to a nonlinear
coupled $2\times 2$ block system to be solved at each time step,
or alternatively approximated via a suitable
extrapolation of order $m$ given by
\begin{align}
	\ve{u}^\star
	=
 	\ve{u}^\#
	\coloneqq
	\sum_{i=0}^{m-1}
	\beta_i
	\ve{u}^{n-i}
	=
	\ve{u}\left(t_{n+1}\right) + \mathcal{O}\left(\dt^m\right)
	,
	\label{eqn:extrapolation}
\end{align}
with coefficients $\beta_i$, $i=0,m-1$ (see, e.g.,~\cite{Hairer1993}).
Within this work, $(\cdot)^\#$ denotes an extrapolation in time,
while $(\cdot)^\star$ can denote an implicit quantity or can be defined
depending on the linearization variant chosen.
Non-uniform time steps are easily considered
in~\eqref{eqn:momentum_balance_time_discrete}--\eqref{eqn:extrapolation},
and only omitted here for the sake of presentation. Note also that the employed
BDF-$m$ scheme requires $m$ old time step values, which are not
available in the first $m-1$ steps. This startup dilemma is usually solved
by either sequentially increasing the order starting from $1$
in the first steps, considering the time step values of a precursor simulation,
or adopting single-step schemes of appropriate order.

\subsection{Splitting Scheme}
\label{sec:splitting_scheme}
Two challenges involving system~\eqref{eqn:momentum_balance_time_discrete}--\eqref{eqn:continuity_equation_time_discrete}
are the nonlinearities and the saddle-point structure. To circumvent those features, projection schemes
build on a Helmholtz--Leray decomposition of the velocity into an
irrotational and a solenoidal part to recover the velocity and
pressure in sequential steps. Numerous schemes are available (see, e.g.,
the review paper by~\citet{Guermond2006}), most of which assume (and require) constant viscosity.
\previouslychanged{The present work adopts the scheme by~\citet{Karamanos2000}, which extends the work by~\citet{Karniadakis1991}
to variable viscosity $\nu=\nu(\ve{x},t)$.}

\citet{Blackburn2025} developed a related method also based on
the operator splitting by~\citet{Karniadakis1991}. \previouslychanged{The scheme by~\citet{Karamanos2000}
does not consider for a constant reference viscosity.}
Introducing the constant reference viscosity allows splitting off the spatially variable part
to be treated fully explicitly in time.
Contrarily, the entire viscous term is treated implicitly in \cite{Karamanos2000},
avoiding i) an additional (potentially parabolic) CFL restriction,
and ii) the problem related to choosing reference viscosity appropriately,
while still being iteration free depending on the linearization variant chosen.
\previouslychanged{Hence, in the present work, the viscous contributions in the
steps of the original splitting scheme~\cite{Karniadakis1991}
are considered similar to~\cite{Karamanos2000}.}
Depending on the flow regime, both approaches can be valuable.

The balance of linear
momentum~\eqref{eqn:momentum_balance_time_discrete} is split into a substep advancing
the convective and body force terms
\begin{align}
	\frac{\gamma_0}{\dt}
	\hat{\ve{u}}
	&
	=
	\sum_{i=0}^{m-1}
	\frac{\alpha_i}{\dt}
	\ve{u}^{n-i}
	+
	\ve{b}\left(t_{n+1}\right)
	-
	\left[\convectiveterm{\ve{u}}{\ve{u}}\right]^\#
	&
	\mathrm{in} \quad \Omega
	\label{eqn:convective_step}
	,
	\quad
	\text{with}
	\quad
	\left[\convectiveterm{\ve{u}}{\ve{u}}\right]^\#
	\coloneqq
	\sum_{i=0}^{m-1}
	\beta_i
	\,
	\convectiveterm{\ve{u}^{n-i}}{\ve{u}^{n-i}}
\end{align}
where the convective term is made fully explicit considering the extrapolation of the convective term.
Note that the convective step~\eqref{eqn:convective_step}
is a simple mass matrix solve for $\hat{\ve{u}}$
with the right-hand side containing the known extrapolation $\ve{u}^\#$ and the body force $\ve{b}$.
The intermediate velocity $\hat{\ve{u}}$ is then projected onto a
divergence-free space, yielding $\hhat{\ve{u}}$,
\begin{align}
	\frac{\gamma_0}{\dt}
	\left(
		\hhat{\ve{u}}
		-
		\hat{\ve{u}}
	\right)
	&
	=
	-
	\grad p^{n+1}
	&
	\mathrm{in} \quad \Omega
	\label{eqn:projection_step}
	.
\end{align}
Eqn.~\eqref{eqn:projection_step} can be rewritten exploiting
$\div\hhat{\ve{u}} = 0$ to obtain
\begin{align}
		- \Delta p^{n+1}
		&=
		-
		\frac{\gamma_0}{\dt}
		\div
		\hat{\ve{u}}
		&
		\mathrm{in} \quad \Omega
		\label{eqn:pressure_step}
		,
\end{align}
which enables recovering the velocity and pressure independently,
but is the most troublesome step with respect to boundary conditions.
A suitable pressure boundary condition is derived by forming
the dot product of the momentum balance equation
with the unit outward normal
$\n$ and rewriting the viscous term as
\begin{align*}
    \div \left(
    	2 \, \nu \grads \ve{u}
    \right)
    &=
      \nu \Delta \ve{u}
    + \nu \grad\left(\div \ve{u}\right)
    + 2 \, \grads\ve{u} \cdot \nabla \nu
    \nonumber
    \\
    &\equiv
	- \nu \nabla \times \left(\nabla \times \ve{u}\right)
	+ 2 \, \nu \grad \left(\div \ve{u}\right)
	+ 2 \, \grads\ve{u} \cdot \nabla \nu
	.
\end{align*}
Dropping the term $2\,\nu\grad(\div \ve{u})$, since $\div \ve{u} = 0$,
and extrapolating the remainder of the viscous term and
the convective term with a rule of order $m_p$,
one obtains on the Dirichlet boundary $\GD$, see~\cite{Karamanos2000}
\begin{align}
	\left(\grad p^{n+1}\right) \cdot \n
	=\hp
	\coloneqq
	&
	\,\,
	\n \cdot
	\ve{b}\left(t_{n+1}\right)
	-
	\n \cdot \partial_t \gu\left(t_{n+1}\right)
	-
	\sum_{i=0}^{m_p-1}
	\beta_i
	\n \cdot
	\left[
		\convectiveterm{\ve{u}^{n-i}}{\ve{u}^{n-i}}
	\right]
	\nonumber
	\\
	&
	-
	\nu^\ddag
	\ve{n} \cdot
	\left[
		\nabla \times
		\left(
		\nabla \times \ve{u}^\ddag
		\right)
	\right]
	+
	{
		2
		\n \cdot
		\left(
			\grads\ve{u}^\ddag \cdot \nabla \nu^\ddag
		\right)
	}
	\label{eqn:pressure_Neumann_BC}
	.
\end{align}
Note here that the extrapolation order
$m_p \coloneqq \min \left\{m-1, 2\right\}$ of the terms
within the Neumann boundary condition for the pressure Poisson equation (PPE) may differ from the
overall extrapolation order, following~\cite{Fehn2017},
leading to the definitions
\begin{align*}
	\ve{u}^\ddag
	\coloneqq
	\sum_{i=0}^{m_p-1} \bar{\beta}_i \ve{u}^{n-i}
	=
	\ve{u}\left(t_{n+1}\right) + \mathcal{O}\left(\dt^{m_p}\right)
	\qquad
	\mathrm{and}
	\qquad
	\nu^\ddag
	\coloneqq
	\eta
	\left(
		\dot\gamma\left(
			\grads \ve{u}^\ddag
		\right)
	\right)
	,
\end{align*}
with coefficients $\bar{\beta}_i$, $i=0,\dots,m_p-1$ (see, e.g.,~\cite{Hairer1993}).
The acceleration term on the boundary given by
$\partial_t\gu(t)$ may be derived
analytically, but herein we approximate $\gu(t)$ via a
BDF-$m$ scheme for the sake of generality.
For ease of presentation we assume that the Neumann
boundary is non-empty, such that the pressure is uniquely determined.
\previouslyrevised{For the pure Dirichlet case, the pressure level is fixed by setting the mean value of the
right-hand side to zero prior to solving the linear system.}
Due to the splitting $\ve{h} = \hu - \gp \n$ and
Eqns.~\eqref{eqn:boundary_condition_hu}--\eqref{eqn:boundary_condition_gp},
an essential boundary condition for the pressure Poisson problem is
directly available as
\begin{align}
	p^{n+1} &= \gp(t_{n+1}) & \mathrm{on} \quad \GN
	.
	\label{eqn:discrete_PPE_essential_pressure_BC}
\end{align}

Having recovered $p^{n+1}$, the
velocity projection step~\eqref{eqn:projection_step} then
yields $\hhat{\ve{u}}$.
The last remaining step then involves
advancing the viscous term implicitly in time via
\previouslyrevised{
\begin{align}
	\frac{\gamma_0}{\dt}
	\ve{u}^{n+1}
	-
	\div \te{S}\left(\nu^\star, \ve{u}^{n+1}\right)
	&=
	\frac{\gamma_0}{\dt}
	\hhat{\ve{u}}
	&
	\mathrm{in} \quad \Omega
	\label{eqn:viscous_step}
	,
\end{align}
where $\nu^\star$ can be considered nonlinear implicit or linear implicit.
Alternatively, one can add the convective term in Eqn.~\eqref{eqn:viscous_step}
to avoid the often restrictive time step constraint observed when considering
the convective term fully explicit in Eqn.~\eqref{eqn:convective_step} only.}
This leads to the variant presented in~\cite{Liosi2025}
(see also the related work in~\cite{Sherwin2003, Guermond2003a, Guermond2006})
\begin{align}
	\frac{\gamma_0}{\dt}
	\ve{u}^{n+1}
	+
	\alpha_\mathrm{c}
	\convectiveterm{\ve{u}^{n+1}}{\ve{u}^\star}
	-
	\div \te{S}\left(\nu^\star, \ve{u}^{n+1}\right)
	&=
	\frac{\gamma_0}{\dt} \hhat{\ve{u}}
	+
	\alpha_\mathrm{c}
	\,
	\left[\convectiveterm{\ve{u}}{\ve{u}}\right]^\#
	&
	\mathrm{in} \quad \Omega
	\label{eqn:viscous_step_with_convection}
	,
\end{align}
where $\alpha_\mathrm{c} \in \{0,1\}$ controls whether or not the convective term is
taken into account and corrects the term added in the convective
step~\eqref{eqn:convective_step} accordingly.

In the viscous substep of the splitting scheme, the three possible forms of the convective term are
\begin{enumerate}[label=\roman*)]
	\item \emph{implicit}, based on $\ve{u}^\star = \ve{u}^{n+1}$ and $\alpha_\mathrm{c} = 1$,
	\item \emph{linear implicit} within the viscous step~\eqref{eqn:viscous_step_with_convection},
	based on an extrapolation $\ve{u}^\star = \ve{u}^\#$~\eqref{eqn:extrapolation}
	and $\alpha_\mathrm{c} = 1$, or
	\item \emph{explicit}, by setting $\alpha_\mathrm{c}=0$ and
	taking it into account only in the convective
	step~\eqref{eqn:convective_step}.
\end{enumerate}
Similarly, the viscous term may be
\begin{enumerate}[label=\roman*)]
	\item \emph{implicit}, using
	$\nu^\star = \nu^{n+1}
	\coloneqq
	\eta\left(\dot{\gamma}\left(\grads\ve{u}^{n+1}\right)\right)$, or
	\item \emph{linear implicit}, employing
	$\nu^\star = \nu^\# \coloneqq \eta\left(\dot{\gamma}\left(\grads\ve{u}^\#\right)\right)$.
\end{enumerate}
Note that depending on the choice of $\alpha_\mathrm{c}$, $\ve{u}^\star$
and $\nu^\star$, the viscous step~\eqref{eqn:viscous_step_with_convection}
as a generalization of~\eqref{eqn:convective_step} may be nonlinear,
requiring additional schemes such as Newton's method or a fixed point
iteration to fully resolve the nonlinear effects.
Similar linearization variants can be constructed for the coupled solution scheme.
These aspects will be further discussed in Sec.~\ref{sec:solution_algorithms}.

\section{Spatial Discretization via a Discontinuous Galerkin Method}
\label{sec:spatial_discretization}

We approximate the domain $\Omega$ by a decomposition into $N_e$ non-overlapping hexahedral
elements $\Omega_e$.
Adopting the DG method,
the velocity, pressure and viscosity approximations,
$\ve{u}_h$, $p_h$, and $\nu_h$, are element-wise polynomial,
but globally discontinuous functions. We denote by
$\mathcal{Q}_k(\Omega_\mathrm{ref})$ the polynomial function
space of degree $k$ on the reference finite element $\Omega_\mathrm{ref}$ constructed
by tensor-product shape functions $q_\mathrm{ref}(\ve{\xi})$. Within this work,
only inf-sup stable finite element pairs are considered, setting the
polynomial degree for the pressure one order lower than the velocity
degree, $k_p=k_u-1$, such that $k_u\geq 2$.
\previouslyrevised{The viscosity is precomputed and stored pointwise,
posing less strict regularity
requirements on the viscosity field $\nu(\ve{x}, t)$ and yielding
positive parameters $\nu(\ve{x}, t) > 0$
directly from Eqn.~\eqref{eqn:viscosity_general}}.
Additionally, the element-wise finite element mapping
$\ve{\chi}_e : \ve{\xi} \mapsto \ve{x}$ is of degree $k_u$ to accurately
resolve curved boundaries. This leads to the standard function spaces
\begin{align*}
	\mathcal{V}^u_h
	&=
	\left\{
		\ve{v}_h \in [L^2(\Omega_h)]^d
		:
		\ve{v}_h \circ \ve{\chi}_e(\ve{\xi})\rvert_{\Omega_e}
		=
		\ve{q}_\mathrm{ref}(\ve{\xi})
		\in
		[\mathcal{Q}_{k_u}(\Omega_\mathrm{ref})]^d
		\quad \forall \Omega_e \in \Omega_h
	\right\}
	\\
	\mathcal{V}^p_h
	&=
	\left\{
	q_h \in L^2(\Omega_h)
	:
	q_h \circ \ve{\chi}_e(\ve{\xi})\rvert_{\Omega_e}
	=
	q_\mathrm{ref}(\ve{\xi})
	\in
	\mathcal{Q}_{k_p}(\Omega_\mathrm{ref})
	\quad \forall \Omega_e \in \Omega_h
	\right\}
	,
\end{align*}
We denote by $(\cdot)^-$ quantities on the
current element, while quantities exterior to the element, e.g.~on the neighbor, are indicated by
$\left(\cdot\right)^+$, leading to convenient notations for the average
$\avg{f}\coloneqq\nicefrac{1}{2}\left(f^+ + f^-\right)$,
jump $\jump{f} \coloneqq f^- \otimes \ve{n} - f^+ \otimes \ve{n}$ and
oriented jump operators $\ojump{f} \coloneqq f^- - f^+$. The treatment of
the viscous term and the respective boundary condition to achieve
consistency and higher-order accuracy differ from the Newtonian
case~\cite{Fehn2017}. For the sake of completeness, we present
the derivation of an $L^2$-conforming DG
formulation for the splitting scheme and a coupled solution approach.

\subsection{Splitting Scheme}
The boundary conditions are enforced
weakly by inserting boundary data into the respective numerical flux functions
and adopting the mirror principle as usual for DG methods, see, e.g., \citet{Hesthaven2008,Fehn2017}.

\paragraph{Convective Step}
The weak counterpart of the fully explicit convective
step~\eqref{eqn:convective_step} reads:
Find $\hat{\ve{u}}_h \in \mathcal{V}_h^u$ such that
\begin{align}
	\frac{\gamma_0}{\dt}
	\left(
		\ve{v}_h,
		\hat{\ve{u}}_h
	\right)_{\Omega_e}
	&=
	\left(
		\ve{v}_h,
		\ve{b}\left(t_{n+1}\right)
	\right)_{\Omega_e}
	+
	\sum_{i=0}^{m-1}
	\frac{\alpha_i}{\dt}
	\left(
		\ve{v}_h,
		\ve{u}_h^{n-i}
	\right)_{\Omega_e}
	-
	\sum_{i=0}^{m-1}
	\beta_i
	\,
	c_h^e\left(\ve{v}_h,\ve{u}_h^{n-i};\ve{u}_h^{n-i}\right)
	&
	\forall \ve{v}_h\in\mathcal{V}_h^u,~\Omega_e \in \Omega_h
	,
	\label{eqn:convective_step_weak}
\end{align}
with $\left(\cdot,\cdot\right)_{\Omega_e}$ denoting the
standard inner product of the two arguments integrated over the indicated domain.
The convective term is considered explicitly, where
$c_h^e\left(\ve{v}_h, \ve{u}_h; \ve{w}_h\right)$ is given as
\begin{align}
		c_h^e\left(\ve{v}_h,\ve{u}_h;\ve{w}_h\right)
		\coloneqq
		\left(
		\ve{v}_h,
		\convectiveterm{\ve{u}_h}{\ve{w}_h}
		\right)_{\Omega_e}
		-
		\left(
		\ve{v}_h, \left(\avg{\ve{w_h}} \cdot \n\right)\ve{u}_h
		\right)_{\partial\Omega_e}
		+
		\left(
		\ve{v}_h, \left(\avg{\ve{w}_h} \cdot \n\right)\ve{u}_h^\dag
		\right)_{\partial\Omega_e}
		\label{eqn:convective_term_weak}
\end{align}
with the upwind numerical flux $\ve{u}^\dag \coloneqq \avg{\ve{u}} +
\nicefrac 1 2 \, \mathrm{sign}\left(\avg{\ve{w}}\cdot \n\right) \ojump{\ve{u}}$.
This formulation is often referred to as the strong form
and shows optimal convergence rates in numerical studies.
Due to integrating the convective term by parts in~\eqref{eqn:convective_term_weak},
solving \eqref{eqn:convective_step_weak} requires boundary data for the old time step
values $\ve{u}_h^{n-i} = \gu\left(t_{n-i}\right)$, $i=0,\dots,m-1$ on $\GD$.

\paragraph{Pressure Step}

The Poisson equation in the pressure~\eqref{eqn:pressure_step}
is discretized adopting the symmetric interior penalty Galerkin
(SIPG) method~\cite{Arnold1982, Arnold2000, Arnold2002}.
Therefore, the primal formulation consists of
finding $p_h^{n+1} \in \mathcal{V}_h^p$ such that
\begin{gather}
	\left(\grad q_h, \grad p_h^{n+1} \right)_{\Omega_e}
	-
	\left(\grad q_h,
		\left(
			p_h^{n+1} - \avg{p_h^{n+1}}
		\right) \, \n
	\right)_{\partial\Omega_e}
	-
	\left(q_h,
		\left(
			\avg{\grad p_h^{n+1}}
			+
			\tau
			\jump{p_h^{n+1}}
		\right)
		\cdot \n
	\right)_{\partial\Omega_e}
	\nonumber
	\\
	=
	\frac{\gamma_0}{\dt}
	\left(
		\grad q_h, \hat{\ve{u}}_h
	\right)_{\Omega_e}
	-
	\frac{\gamma_0}{\dt}
	\left(
		q_h,
		\avg{\hat{\ve{u}}_h
			} \cdot \n
	\right)_{\partial \Omega_e}
	\label{eqn:pressure_step_weak}
\end{gather}
for all $q_h\in\mathcal{V}_h^p$ and all elements $\Omega_e \in \Omega_h$,
where the stabilization parameter $\tau$ is chosen according
to~\cite{Hillewaert2013} omitted here for the sake
of brevity. Furthermore, we note that the consistent
boundary condition for the pressure Poisson
equation~\eqref{eqn:pressure_Neumann_BC} is rewritten introducing
the vorticity $\ve{\omega}_h^{n-i} \coloneqq \nabla \times \ve{u}_h^{n-i}$,
with $\ve{\omega}_h^{n-i} \in \mathcal{V}_h^u$, $i=0,\dots,m-1$ obtained via
element-local $L^2$ projections.
Additionally, boundary terms involving the intermediate velocity
$\hat{\ve{u}}_h$ need to be evaluated. To this end, we
rewrite~\eqref{eqn:convective_step} as
\begin{align}
		\hat{\ve{u}}_h
		=
		\ve{g}_{\hat{u}}\left(t_{n+1}\right)
		\coloneqq
		\frac{\dt}{\gamma_0}\ve{b}\left(t_{n+1}\right)
		-
			\frac{\dt}{\gamma_0}
			\,
			\left[\convectiveterm{\ve{u}_h}{\ve{u}_h}\right]^\#
		%
		+
		\left(
			\sum_{i=0}^{m-1}
			\frac{\alpha_i}{\gamma_0}
			\gu\left(t_{n-i}\right)
		\right)
		&& \text{on }\GD
		,
		\label{eqn:guhat_consistent_BC}
\end{align}
using $\ve{u}(t)=\gu(t)$ on $\GD$. Note here that enforcing~\eqref{eqn:guhat_consistent_BC}
is required for full consistency and stability in the small time step limit,
see~\citet{Fehn2017}. Altogether, boundary terms present
in~\eqref{eqn:pressure_step_weak} enforce
$\left(\grad p_h^{n+1}\right) \cdot \n = \hp\left(t_{n+1}\right)$ defined
in~\eqref{eqn:pressure_Neumann_BC} on $\GD$,
$p_h^{n+1}$ on $\GN$ via~\eqref{eqn:discrete_PPE_essential_pressure_BC},
and $\hat{\ve{u}}_h = \ve{g}_{\hat{u}}\left(t_{n+1}\right)$
according to~\eqref{eqn:guhat_consistent_BC} on $\GD$.

\paragraph{Velocity Projection Step}
The divergence-free velocity $\hhat{\ve{u}}_h$ is then computed
from a weak form corresponding to~\eqref{eqn:projection_step}, i.e,
find $\hhat{\ve{u}}_h \in \mathcal{V}_h^u$ such that
\begin{align}
	\frac{\gamma_0}{\dt}
	\left(
		\ve{v}_h,\hhat{\ve{u}}_h
	\right)_{\Omega_e}
	&=
	\frac{\gamma_0}{\dt}
	\left(
		\ve{v}_h,\hat{\ve{u}}_h
	\right)_{\Omega_e}
	+
	\left(
		\div \ve{v}_h, p_h^{n+1}
	\right)_{\Omega_e}
	-
	\left(
		\ve{v}_h, \avg{p_h^{n+1}} \, \n
	\right)_{\partial\Omega_e}
	&
	\forall \ve{v}_h\in\mathcal{V}_h^u,~\Omega_e \in \Omega_h
	,
	\label{eqn:projection_step_weak}
\end{align}
incorporating $p^{n+1}_h = \gp\left(t_{n+1}\right)$ on $\GN$~\eqref{eqn:discrete_PPE_essential_pressure_BC}.

\paragraph{Viscous Step}
The nonlinearities present in the viscous step~\eqref{eqn:viscous_step_with_convection}
are resolved by either temporal extrapolation or Picard iterations in the
viscosity and Newton's method for the convective terms.
Considering for a generalized Newtonian fluid,
we first update the apparent viscosity in a pointwise manner
based on the last velocity iterate $\ve{u}_h^{n+1, k}$,
\begin{align}
			\nu_h^{n+1,k+1}
		&=
			\eta\left(
				\dot{\gamma}\left(
					\grads \ve{u}_h^{n+1,k}
				\right)
			\right)
		&\text{in }\Omega_h
		.
		\label{eqn:viscosity_pointwise}
\end{align}
starting from the extrapolation $\ve{u}_h^\#$~\eqref{eqn:extrapolation} taken as initial guess.

\previouslyrevised{
	Alternatively, the viscosity might be projected onto a continuous space
	to reduce the memory footprint by storing data at the node points of a
	finite element basis instead of the integration points.
	This strategy, however, requires a global mass matrix solve
	and interpolation into quadrature points during element integration.
	Projecting the viscosity onto a discontinuous space allows faster element-local $L^2$ projections
	and poses less strict regularity requirements on $\nu(\ve{x}, t)$, but
	requires more memory.
	When using more integration points than there are polynomial support points (which is not the case here),
	a discontinuous viscosity space using nodal coefficient storage would also reduce the memory requirements, while still allowing element-local $L^2$ projections.
	The pointwise update employed herein, however,
	directly yields $\nu(\ve{x}, t) > 0$ at all integration points due to
	$\nu = \eta(\dot\gamma): \mathbb{R}^+ \mapsto \mathbb{R}^+\setminus\{0\}$,
	see also Eqn.~\eqref{eqn:viscosity_general}, which is, however, also
	not straight-forwardly guaranteeing positivity of the operator.
	All the mentioned options might thus be viable alternatives.
}

\previouslyrevised{Precomputing the viscosity} allows to split the potentially
very costly viscosity update from the remainder.
The last step in the splitting scheme being the weak counterpart
of~\eqref{eqn:viscous_step_with_convection} reads:
Find $\ve{u}_h^{n+1,k+1} \in \mathcal{V}_h^u$ such that
\begin{gather}
    \frac{\gamma_0}{\dt}
    \left(
    	\ve{v}_h,
    	\ve{u}_h^{n+1,k+1}
   	\right)_{\Omega_e}
    +
    \alpha_\mathrm{c}\,
	c_h^e\left(\ve{v}_h,\ve{u}_h^{n+1,k+1};\ve{u}_h^\star\right)_{\Omega_e}
    +
    v_h^e\left(\ve{v}_h, \ve{u}_h^{n+1,k+1}; \nu_h^{n+1,k+1}\right)
    \nonumber
    \\
    =
    \frac{\gamma_0}{\dt}
    \left(
    	\ve{v}_h,
    	\hhat{\ve{u}}_h
   	\right)_{\Omega_e}
    +
    \alpha_\mathrm{c}
    \sum_{i=0}^{m}
    \beta_i\,
    c_h^e\left(\ve{v}_h,\ve{u}_h^{n-i};\ve{u}_h^{n-i}\right)
    \label{eqn:viscous_step_weak}
\end{gather}
with the linearized implicit viscous term
$v_h^e\left(\ve{v}_h, \ve{u}_h; \nu_h\right)$~\eqref{eqn:viscous_term_weak}
and the convective term
$c_h^e\left(\ve{v}_h,\ve{u}_h;\ve{w}_h\right)$~\eqref{eqn:convective_term_weak}.
This gives rise to either a nonlinear problem if $\alpha_\mathrm{c}=1$
and $\ve{u}_h^\star = \ve{u}_h^{n+1,k+1}$ (requiring fixed-point iterations),
a linear implicit setup for $\alpha_\mathrm{c} = 1$ and
$\ve{u}_h^\star = \ve{u}_h^\#$~\eqref{eqn:extrapolation},
or a fully explicit treatment choosing $\alpha_\mathrm{c} = 0$.
The viscosity update~\eqref{eqn:viscosity_pointwise}
and the viscous step Eqn.~\eqref{eqn:viscous_step_weak} may be repeated, depending on the
linearization approach. In~\eqref{eqn:viscous_step_weak}, the boundary terms
are evaluated 
incorporating 
$\ve{u}_h^{n+1-i}=\gu\left(t_{n+1-i}\right)$, $i=0,\dots,m$ on $\GD$.

The viscous term $v_h^e\left(\ve{v}_h,\ve{u}_h;\nu_h\right)$ needs to account for a
spatially variable viscosity, differing from the Newtonian case.
Within this work, we consider the SIPG method to discretize
the stress divergence term, see, e.g., \cite{Arnold2000, Hesthaven2008} and others.
Revisiting the derivation as carried out for the Newtonian case in \cite{Fehn2021b},
the primal formulation results from
rewriting the second-order term as a system of first-order equations,
performing integration by parts multiple times and finally rewriting
the system in the primal variable and inserting numerical fluxes.
Introducing 
$\te{\tau}_{1,h} \coloneqq \nu_h \grad\ve{u}_h$ and
$\te{\sigma}_h \in [L^2(\Omega_h)]^{d\times d}$, we have
\begin{align}
		\left(
		\te{\sigma}_h,
		\te{\tau}_{1,h}
		\right)_{\Omega_e}
		&=
		\left(
		\te{\sigma}_h,
		\nu_h \grad\ve{u}_h
		\right)_{\Omega_e}
		=
		\left(
		\nu_h
		\te{\sigma}_h,
		\grad\ve{u}_h
		\right)_{\Omega_e}
		\nonumber\\
		&=
		-
		\left(
		\div\left(\nu_h
		\te{\sigma}_h\right),
		\ve{u}_h
		\right)_{\Omega_e}
		+
		\left(
		\nu_h \te{\sigma}_h,
		\ve{u}_h^\dag \otimes \n
		\right)_{\partial\Omega_e}
		\nonumber\\
		&=
		\left(
		\te{\sigma}_h,
		\nu_h \grad\ve{u}_h
		\right)_{\Omega_e}
		-
		\left(
		\te{\sigma}_h,
		\nu_h
		\left(\ve{u}_h - \ve{u}_h^\dag \right)
		\otimes \n
		\right)_{\partial\Omega_e}
		.
		\label{eqn:SIPG_tau_1_rewrite}
\end{align}
For the transpose gradient term we similarly define
$\te{\tau}_{2,h} \coloneqq \nu_h \left(\grad\ve{u}_h\right)^\top$ to get
\begin{align}
		\left(
		\te{\sigma}_h,
		\te{\tau}_{2,h}
		\right)_{\Omega_e}
		&=
		\left(
		\te{\sigma}_h,
		\nu_h \left(\grad\ve{u}_h\right)^\top
		\right)_{\Omega_e}
		=
		\left(
		\nu_h
		\te{\sigma}_h^\top,
		\grad\ve{u}_h
		\right)_{\Omega_e}
		\nonumber\\
		&=
		-
		\left(
		\div\left(\nu_h
		\te{\sigma}_h^\top\right),
		\ve{u}_h
		\right)_{\Omega_e}
		+
		\left(
		\nu_h \te{\sigma}_h^\top,
		\ve{u}_h^\dag \otimes \n
		\right)_{\partial\Omega_e}
		\nonumber\\
		&=
		\left(
		\te{\sigma}_h,
		\nu_h \left(\grad\ve{u}_h\right)^\top
		\right)_{\Omega_e}
		-
		\left(
		\te{\sigma}_h,
		\nu_h
		\left[\left(\ve{u}_h - \ve{u}_h^\dag \right)
		\otimes \n\right]^\top
		\right)_{\partial\Omega_e}
		.
		\label{eqn:SIPG_tau_2_rewrite}
\end{align}
Integrating the stress divergence by parts yields
\begin{align*}
		-
		\left(
		\ve{v}_h,
		\div\te{S}_h
		\right)_{\Omega_e}
		=
		-
		\left(
		\ve{v}_h
		,
		\div
		\left(
		\te{\tau}_{1,h} + \te{\tau}_{2,h}
		\right)
		\right)_{\Omega_e}
		=
		\left(
		\grad \ve{v}_h
		,
		\te{\tau}_{1,h}
		+
		\te{\tau}_{2,h}
		\right)_{\Omega_e}
		-
		\left(
		\ve{v}_h
		,
		\left(
		\te{\tau}_{1,h}^\dag
		+
		\te{\tau}_{2,h}^\dag
		\right) \cdot \n
		\right)_{\partial\Omega_e}
		,
\end{align*}
from which we obtain the primal formulation inserting Eqns.~\eqref{eqn:SIPG_tau_1_rewrite}
and~\eqref{eqn:SIPG_tau_2_rewrite} and setting $\te{\sigma}_h=\grad\ve{v}_h$, which gives
\begin{align}
		\left(
		\grad\ve{v}_h
		,
		2 \, \nu_h \grads \ve{u}_h
		\right)_{\Omega_e}
		-
		\left(
		\grad\ve{v}_h
		,
		2\,
		\nu_h
		\left[
		\left(\ve{u}_h - \ve{u}_h^\dag\right)\otimes\n
		\right]^\mathrm{S}
		\right)_{\partial\Omega_e}
		-
		\left(
		\ve{v}_h
		,
		\left(
		\te{\tau}_{1,h}^\dag
		+
		\te{\tau}_{2,h}^\dag
		\right) \cdot \n
		\right)_{\partial\Omega_e}
		\eqqcolon
		\,\,
		v_h^e(\ve{v}_h,\ve{u}_h;\nu_h)
		.
		\label{eqn:viscous_term_weak}
\end{align}
The numerical fluxes within the SIPG method are defined as
\begin{align*}
		\ve{u}_h^\dag \coloneqq \avg{\ve{u}_h}
		,
		\qquad
		\te{\tau}_{1,h}^\dag
		\coloneqq
		\avg{\nu_h\grad\ve{u}_h}
		-
		\tau \jump{\ve{u}_h}
		,
		\qquad
		\te{\tau}_{2,h}^\dag
		\coloneqq
		\left(\te{\tau}_{1,h}^\dag \right)^\top
		,
\end{align*}
where the stabilization parameter $\tau$ for tensor-product finite elements
is defined according to~\cite{Hillewaert2013}.
Boundary conditions are enforced weakly by inserting boundary
data into the respective numerical flux functions and adopting the mirror
principle, for details see~\citet{Fehn2017, Hesthaven2008}.
The final result respecting variable viscosities is hence closely related to the
classic formulation for the Newtonian case~\cite{Fehn2021b}.

\subsection{Coupled Formulation}

Multiplying the balance of linear momentum, continuity equations
and rheological law  of the coupled
formulation~\eqref{eqn:momentum_balance_time_discrete}--\eqref{eqn:continuity_equation_time_discrete}
by test functions $\ve{v}_h \in \mathcal{V}_h^u$ and $q_h \in \mathcal{V}_h^p$
and integrating by parts leads to the
weak form of the (potentially still nonlinear) problem of finding
$\left(\ve{u}_h^{n+1}, p_h^{n+1}\right)\in
\mathcal{V}_h^u \times \mathcal{V}_h^p$ such that
\begin{align}
		\frac{\gamma_0}{\dt}
		\left(
		\ve{v}_h
		,
		\ve{u}_h^{n+1}
		\right)_{\Omega_e}
		-
		\sum_{i=0}^{m-1}
		\frac{\alpha_i}{\dt}
		\left(
		\ve{v}_h
		,
		\ve{u}_h^{n-i}
		\right)_{\Omega_e}
		+
		c_h^e\left(\ve{v}_h,\ve{u}_h^{n+1}; \ve{u}^\star_h\right)
		+
		v_h^e\left(\ve{v}_h,\ve{u}^{n+1}_h; \nu_h^{n+1}(\ve{u}_h^\star)\right)
		& \nonumber \\[-1.5ex]
		-
		\left(
		\div \ve{v}_h, p_h^{n+1}
		\right)_{\Omega_e}
		+
		\left(
		\ve{v}_h , \avg{p_h^{n+1}}
		\right)_{\partial\Omega_e}
		&=
		\left(
		\ve{v}_h,\ve{b}\left(t_{n+1}\right)
		\right)_{\Omega_e}
		\label{eqn:mom_bal_weak}
		,
		\\
		\left(
		\grad q_h, \ve{u}_h^{n+1}
		\right)_{\Omega_e}
		-
		\left(
		q_h, \avg{\ve{u}_h^{n+1}} \cdot \n
		\right)_{\partial \Omega_e}
		&= 0
		,
		\label{eqn:conti_weak}
\end{align}
for all $\left(\ve{v}_h, q_h\right)
\in \mathcal{V}_h^u \times \mathcal{V}_h^p$
and all elements $\Omega_e\in\Omega_h$. Similarly,
the convective and viscous terms can be considered
implicitly based on
$\ve{u}^\star = \ve{u}^{n+1}$,
linearly implicit based on an extrapolation
$\ve{u}^\star = \ve{u}^\#$~\eqref{eqn:extrapolation},
or completely explicit (not shown here).
The traction boundary terms in the present form do not allow enforcing the
full traction vector $\ve{h}$ due to the pressure gradient term in the
momentum balance equation~\eqref{eqn:mom_bal_weak} being integrated by parts.
Since the focus within this work lies on the practically relevant case
of imposing $\hu$ and $g_p$, we refrain from reformulating
the weak form for the sake of brevity.

\subsection{Consistent Stabilization}

The $L^2$-conforming DG formulations are further
enhanced via consistent divergence and continuity penalty
terms, weakly enforcing the incompressibility
constraint and inter-element continuity of the normal velocity.
Independent of choosing the coupled solution approach or the
splitting scheme, the obtained velocity field $\ve{u}_h^{n+1}$ is
postprocessed at the end of each time step to find
$\tilde{\ve{u}}_h\in\mathcal{V}_h^u$~\cite{Fehn2018b} such that
\begin{align}
		\left(
		\ve{v}_h , \tilde{\ve{u}}_h
		\right)_{\Omega_e}
		+
		\left(
		\div\ve{v}_h , \tau_\mathrm{div}^e \div\tilde{\ve{u}}_h
		\right)_{\Omega_e}
		+
		\left(
		\ve{v}_h \cdot \n, \tau_\mathrm{cont}^e \ojump{\tilde{\ve{u}}_h} \cdot \n
		\right)_{\partial\Omega_e}
		=
		\left(
		\ve{v}_h , \ve{u}_h^{n+1}
		\right)_{\Omega_e}
		\label{eqn:penalty_step}
\end{align}
for all $\ve{v}_h\in\mathcal{V}_h^u$ and all elements $\Omega_e \in \Omega_h$.
The divergence and continuity penalty parameters, $\tau_\mathrm{div}^e$ and
$\tau_\mathrm{cont}^e$, are chosen according to~\cite{Fehn2018b} and the oriented
jump term $\ojump{\tilde{\ve{u}}_h}$ incorporates boundary data
$\tilde{\ve{u}}_h = \gu$ on $\GD$. Afterwards, the postprocessed velocity is
then considered as $\ve{u}^{n+1}_h$ without introducing further changes to
the notation.

\section{Solution Algorithms}
\label{sec:solution_algorithms}

The fully discrete schemes for the solution of incompressible
generalized Newtonian fluids corresponding to the various linearization
variants and coupled/projection solvers for the velocity-pressure system
adopting DG discretizations are summarized in the following.
For both schemes, the initial guesses for linear and nonlinear solvers
are taken as the $m$-th order extrapolations, e.g.,
\begin{align}
		\ve{u}_h^{n+1,k=0} = \sum_{i=0}^{m-1} \beta_i \ve{u}_h^{n-i}
		,
		\qquad
		p_h^{n+1,k=0} = \sum_{i=0}^{m-1} \beta_i p_h^{n-i}
		,
		\qquad
		\nu_h^{n+1,k=0} = \sum_{i=0}^{m-1} \beta_i \nu_h^{n-i}
		,
		\label{eqn:newton_init_guess}
\end{align}
for the velocity, pressure and viscosity where applicable.

\subsection{Projection Solver}

The splitting scheme leading to a projection solver incorporating
a \previouslyrevised{Newton-like} solver directly treating the
convective nonlinearity via Newton's scheme and the potentially
nonlinear viscous step via a Picard iteration are summarized in Alg.~\ref{alg:splitting_scheme}.

To simplify the solver, we ignore the nonlinearities stemming
from the viscosity and adopt a Picard linearization
$\nu_h^{n+1,k} = \eta(\dot{\gamma}(\ve{u}^{n+1,k}))$.
The systems corresponding to the viscous step to be solved at each
Newton step $k=0,\dots,N_\mathrm{nl}$ are defined in the velocity space,
such that the convergence criterion of the nonlinear solver
is based on the actual residual of the viscous step denoted
as ${\ve{r}}_h^{\mathrm{visc},k+1}$, corresponding
to~\eqref{eqn:viscous_step_with_convection}.
Convergence of the nonlinear solver is reached as soon as
the relative
$\|\ve{r}_h^{\mathrm{visc},k+1}\|
\leq \epsilon_\mathrm{nl}^\mathrm{rel} \, \|\ve{r}_h^{\mathrm{visc},0}\|$
or absolute
$\|\ve{r}_h^{\mathrm{visc},k+1}\|\leq \epsilon_\mathrm{nl}^\mathrm{abs}$
convergence criterion is reached.
Additionally, the viscosity fixed-point iteration is accelerated via
Aitken acceleration~\cite{Irons1969}, which recovers quadratic convergence
under suitable conditions~\cite{Verzhbitskii2011}. Within the acceleration scheme,
we employ the full nonlinear/linearized residual to compute the adaptive
relaxation parameter, circumventing false convergence at stagnation points.

\begin{algorithm}
	\begin{algorithmic}[1]
		\caption{Projection solver with optionally nonlinear viscous step}
		\label{alg:splitting_scheme}
		\Function{ProjectionSolver}{$\{\ve{u}_h^{n-i}$, $p_h^{n-i}$,
			$\nu_h^{n-i}\}_{i=0,\dots,m-1}$, $\dt$,
			$\epsilon_\mathrm{nl}^\mathrm{rel}$,
			$\epsilon_\mathrm{nl}^\mathrm{abs}$, $N_\mathrm{nl}$}
		\State initialize iterates by $m$-th order
		extrapolation~\eqref{eqn:newton_init_guess}
		\State \textbf{convective step:} compute intermediate velocity
		$\hat{\ve{u}}_h$ via~\eqref{eqn:convective_step_weak}
		\State \textbf{vorticity projection:} compute vorticity
		$\ve{\omega}_h^n \coloneqq \grad \times \ve{u}_h^n$ via
		element-local $L^2$ projection
		\State \textbf{pressure step:} compute the pressure $p_h^{n+1}$
		via~\eqref{eqn:pressure_step_weak}
		\State \textbf{velocity projection step:} compute weakly
		divergence-free $\hhat{\ve{u}}_h$ via~\eqref{eqn:projection_step_weak}
		\State $k = 0$
		\Comment{initialize counter}
		\While {
			$k < N_\mathrm{nl}$
			and
			(
			$||\ve{r}_h^{\mathrm{visc},k+1}||
			>
			\epsilon_\mathrm{nl}^\mathrm{rel} \, ||\ve{r}_h^{\mathrm{visc},0}||$
			or
			$||\ve{r}_h^{\mathrm{visc},k+1}||
			>
			\epsilon_\mathrm{nl}^\mathrm{abs}$
			)
		}
		\If{$k = 0$ or viscosity implicit}
		\Comment{optional Picard update of the viscosity}
		\State update viscosity $\nu_h^{n+1,k+1}$
		via~\eqref{eqn:viscosity_pointwise}
		\Else
		\State $\nu_h^{n+1,k+1} \gets \nu_h^{n+1,k}$
		\EndIf
		\State \textbf{viscous step:} update $\ve{u}_h^{n+1,k+1}$ from
		linearized~\eqref{eqn:viscous_step_weak}
		\Comment{Newton step or linear problem solve}
		\If{convective term not implicit and viscosity linearized implicit}
		\State \textbf{break}
		\Comment{break loop if convective and viscous terms are linearized}
		\EndIf
		\State $k\gets k +1$
		\Comment{update iteration counter}
		\EndWhile
		\State postprocess $\ve{u}_h^{n+1,k+1}$ and $p_h^{n+1,k+1}$
		via divergence and continuity penalty step~\eqref{eqn:penalty_step}
		\State \Return $\ve{u}_h^{n+1, k+1}$,
		$p_h^{n+1,k+1}$,
		$\nu_h^{n+1,k+1}$, $k-1$
		\EndFunction
	\end{algorithmic}
\end{algorithm}

\subsection{Coupled Solver}
Taking a closer look at system~\eqref{eqn:mom_bal_weak}--\eqref{eqn:conti_weak},
we see that when adopting Newton's method and considering an implicit
velocity in the rheological law, $\nu^{n+1} = \eta(\dot{\gamma}(\grads \ve{u}^{n+1}))$,
the velocity block may become quite intricate depending on the
rheological law considered.
Denoting the increments within a Newton scheme as $\delta \ve{u}_h^k$,
and $\delta p_h^k$, which are used to update the iterates
as $\ve{u}^{n+1,k+1}_h = \ve{u}^{n+1,k}_h + \delta \ve{u}_h^k$ and similarly
for $p^{n+1,k}_h$, the systems to be solved at each
Newton step $k=0,\dots,N_\mathrm{nl}$ can be written as
\begin{align}
	\begin{pmatrix}
		\te{A}_\mathrm{nl}\left(\ve{u}_h^{n+1,k},\nu_h^{n+1,k}\right)
		& \te{B}^\top
		\\
		\te{B} & \te{0}^{\phantom{\top}}
	\end{pmatrix}
	\begin{pmatrix}
		\delta\ve{u}_h^k
		\\
		\delta p_h^k
	\end{pmatrix}
	=
	\begin{pmatrix*}[l]
		-\ve{r}_h^{u,k}\left(\ve{u}_h^{n+1,k}, p_h^{n+1,k}, \nu_h^{n+1,k}\right)
		\\
		-\ve{r}_h^{p,k}\left(\ve{u}_h^{n+1,k}\right)
	\end{pmatrix*}
	.
	\label{eqn:newton_nonlinear_convection}
\end{align}
Also here, we
ignore the nonlinearities stemming from the nonlinear
viscosity and perform a Picard scheme for the viscosity
$\nu_h^{n+1,k} = \eta(\dot{\gamma}(\ve{u}^{n+1,k}))$
via pointwise updates, which is then followed by a standard Newton step of
the velocity-pressure system with linearized viscosity
until the velocity-pressure residual of the updated solution
$\ve{r}_h^{k+1} \coloneqq (\ve{r}_h^{u,k+1}, \ve{r}_h^{p,k+1})^\top$
fulfills the relative or absolute convergence criterion,
$\|\ve{r}_h^{k+1}\|\leq \epsilon_\mathrm{nl}^\mathrm{rel} \, \|\ve{r}_h^0\|$
or
$\|\ve{r}_h^{k+1}\|\leq \epsilon_\mathrm{nl}^\mathrm{abs}$
respectively.
Similar to the splitting scheme, the viscosity fixed-point iteration is accelerated via
Aitken acceleration~\cite{Irons1969} as well. Also within this acceleration scheme,
we employ the full nonlinear/linearized residual to compute the adaptive
relaxation parameter, to avoid false convergence at stagnation points.

The chosen approach
allows reusing standard preconditioners considering
for spatially variable (but linear) viscosity, avoiding more involved
block-preconditioners, essentially shifting some of the intricacy to the wrapping
nonlinear solver. Treating the convective term explicitly or linearizing the convective
velocity leads to a system linear in the velocity-velocity block, such that
an Oseen problem of the form
\begin{align}
	\begin{pmatrix*}[l]
		\te{A}
		& \te{B}^\top \\ \te{B} & \te{0}
	\end{pmatrix*}
	\begin{pmatrix}
		\ve{u}_h^{n+1,k+1}
		\\
		p_h^{n+1,k+1}
	\end{pmatrix}
	=
	\begin{pmatrix*}
		\ve{f}_h
		\\
		\ve{g}_h
	\end{pmatrix*}
	,
	\label{eqn:oseen_linearized_convection}
\end{align}
needs to be solved. In~\eqref{eqn:oseen_linearized_convection},
the left-hand side is linear in $\ve{u}_h^{n+1,k+1}$, and linearized
via $\nu_h^{n+1,k}$ within a Picard scheme. The right-hand side
$\left(\ve{f}_h, \te{g}_h\right)^\top$ incorporates the body force term,
numerical flux terms on the domain boundaries and potentially a fully explicit
convective term. The related algorithm to be executed at each time step is
summarized in Alg.~\ref{alg:newton_coupled_all}, covering the variants
of the convective term treated i) implicitly, ii) linearly implicit, or
iii) explicitly and treating the viscosity i) implicitly or ii) linearly
implicit.
\begin{algorithm}
	\begin{algorithmic}[1]
		\caption{Coupled solver with viscosity Picard iteration}
		\label{alg:newton_coupled_all}
		\Function{CoupledSolver}{$\{\ve{u}_h^{n-i}$, $p_h^{n-i}$,
			$\nu_h^{n-i}\}_{i=0,\dots,m-1}$, $\dt$,
			$\epsilon_\mathrm{nl}^\mathrm{rel}$,
			$\epsilon_\mathrm{nl}^\mathrm{abs}$, $N_\mathrm{nl}$}
		\State initialize iterates by $m$-th order
		       extrapolation~\eqref{eqn:newton_init_guess}
		\State enforce Dirichlet boundary conditions on iterates
		\State $k = 0$
		\Comment{initialize counter}
		\While {
			$k < N_\mathrm{nl}$
			and
			(
			$||\ve{r}_h^{k+1}||
			>
			\epsilon_\mathrm{nl}^\mathrm{rel} \, ||\ve{r}_h^0||$
			or
			$||\ve{r}_h^{k+1}||
			>
			\epsilon_\mathrm{nl}^\mathrm{abs}$
			)
		}
		\If{$k = 0$ or viscosity implicit}
		\Comment{optional Picard update of the viscosity}
			\State update viscosity $\nu_h^{n+1,k+1}$
			via~\eqref{eqn:viscosity_pointwise}
		\Else
		    \State $\nu_h^{n+1,k+1} \gets \nu_h^{n+1,k}$
		\EndIf
		\If{convective term implicit}
			\State solve for velocity-pressure Newton
			update via~\eqref{eqn:newton_nonlinear_convection}
			\State $\ve{u}_h^{n+1,k+1}
			\gets \ve{u}_h^{n+1,k} + \delta \ve{u}_h^k$,
			$p_h^{n+1,k+1} \gets p_h^{n+1,k} + \delta p_h^k$
			\Comment{update velocity and pressure iterates}
		\Else
			\State compute $\ve{u}_h^{n+1,k+1}$ and $p_h^{n+1,k+1}$
			solving the Oseen problem~\eqref{eqn:oseen_linearized_convection}
			\If{viscosity linearized implicit}
				\State \textbf{break}
				\Comment{break loop if convective and
					     viscous terms are linearized}
			\EndIf
		\EndIf
		\State $k\gets k +1$
		\Comment{update iteration counter}
		\EndWhile
		\State postprocess $\ve{u}_h^{n+1,k+1}$ and $p_h^{n+1,k+1}$
		via divergence and continuity penalty step~\eqref{eqn:penalty_step}
		\State \Return $\ve{u}_h^{n+1, k+1}$,
					   $p_h^{n+1,k+1}$,
					   $\nu_h^{n+1,k+1}$, $k-1$
		\EndFunction
	\end{algorithmic}
\end{algorithm}

\subsection{Matrix-free Preconditioners}
\label{sec:matrix-free}

The linear systems arising within the coupled and projection approaches are
solved via iterative Krylov methods. In the projection scheme, the convective
step, the pressure step, and the velocity projection step lead to symmetric
systems, while the viscous step potentially results in a non-symmetric system
depending on the linearizatoin variant.
The velocity-velocity block in the coupled approach follows a similar rationale,
where explicit treatment of the convective term leads to symmetry,
while treating the convective term linearly implicit or implicitly leads to non-symmetric
systems. For the coupled solver, we apply the generalized minimal
residual method~(GMRES)~\cite{Saad2003}.
A conjugate gradient method is used for the pressure step and
for the viscous step depending on the linearization variant.
$L^2$-conforming DG discretizations of the
convective~\eqref{eqn:convective_step_weak} and the velocity
projection~\eqref{eqn:projection_step_weak} steps lead to block-diagonal
mass matrices, such that these steps
can be treated in a cell-wise fashion with tensor-product variants of the inverse~\cite{Kronbichler2016}.

Within the Krylov methods, the action of the discretized operator on a vector
is realized in a matrix-free fashion and realized via numerical quadrature,
exploiting tensor product structure of the shape functions and the quadrature
rules via sum factorization techniques
(cf.~\cite{KronbichlerKormann2012, KronbichlerKormann2019, Vos2010}).
Furthermore, SIMD vectorization is utilized over batches of
elements, see~\cite{KronbichlerKormann2012} for details.

For preconditioning the viscous terms, a multigrid strategy relying on our previous
developments~\cite{KronbichlerWall2018, Fehn2020, Fehn2021b, Munch2023} is chosen.
We construct a preconditioner following the $hp$-multigrid preconditioning strategy
with matrix-free smoothers similar to Fehn~et~al.~\cite{Fehn2020},
where first the conversion from a discontinuous to a continuous polynomial space is
performed, whereafter the polynomial degree is lowered recursively from $k$ to $k-1$,
after which $h$-coarsening is performed. The individual $h$-levels are created by
refining an initial coarse grid resolving the geometry's topology, which are, however,
typically not able to resolve the intricate physical solution within the domain and can
hence only be used as coarse levels within a multigrid hierarchy. This strategy is denoted as
$cph$-multigrid in~\cite{Fehn2020}.
On the coarsest level of the multigrid preconditioner,
a low and fixed number of algebraic multigrid (AMG) cycles from the
\texttt{Trilinos ML} package~\cite{Heroux2012, Gee2006}
or a direct solver if the system size falls below 2000 degrees of freedom (DoFs) is used.
Furthermore, we consider double precision arithmetic in the outer Krylov
solver, while the preconditioner operates in single
precision. However, the AMG coarse grid solvers used herein operate in double precision
only. Fine-scale errors introduced by this mixed-precision strategy are
well-captured by the multigrid smoothers, see~\cite{KronbichlerKormann2012,
KronbichlerLjungkvist2019}.

We employ standard multigrid V-cycles (see, e.g.,~\cite{Brandt1977, Hackbusch1985,
Trottenberg2001}), and a Chebyshev-accelerated Jacobi scheme~\cite{KronbichlerWall2018,
Fehn2020, Adams2003} as smoother. The construction of this smoother incorporates
precomputing the inverse of the matrix diagonal on each level, while the residual evaluation
in the Jacobi-type iteration is based on the level operator application. Based on previous
investigations on matrix-free preconditioning of challenging elliptic operators~\cite{Schussnig2025,
Davydov2020}, we store the scalar viscosity field evaluated in the integration points of
each level to avoid reevaluating these potentially costly functions. Whenever the
linearization vector is updated, the quadrature point data of the multigrid hierarchy is
updated on all levels using the multigrid transfer operator.

The Navier--Stokes equations in their coupled form~\eqref{eqn:oseen_linearized_convection}
are preconditioned via a block-triangular preconditioner~\cite{Benzi2005} of the form
\begin{align*}
	\mathcal{P}^{-1}
	\coloneqq
	\begin{pmatrix*}[l]
		\te{A}
		& \te{B}^\top \\ \te{0} & \te{S}
	\end{pmatrix*}^{-1}
	=
	\begin{pmatrix*}[l]
		\te{A}^{-1}
		& \te{0} \\ \te{0} & \id
	\end{pmatrix*}
	\begin{pmatrix*}
		\te{0}
		& -\te{B}^\top \\ \te{0} & \id
	\end{pmatrix*}
	\begin{pmatrix*}[l]
		\id
		& \te{0} \\ \te{0} & \te{S}^{-1}
	\end{pmatrix*}
	,
	\quad
	\mathrm{where}
	\quad
	\te{S}
	=
	-
	\te{B}\te{A}^{-1}\te{B}^\top
	.
\end{align*}
Due to the saddle point structure of system~\eqref{eqn:oseen_linearized_convection},
a GMRES method is required, which converges in two iterations using exact inverses
of $\te{A}$ and the negative Schur complement $\te{S}$~\cite{Benzi2005}.
The specific form of $\te{A}$ and hence $\te{S}$ critically impacts GMRES convergence.
The problem and solver parameters, that is, the time step and grid size, the viscosity
and its gradient, and the convective term influence the optimal choice for
approximating $\te{S}$. Aiming for incompressible flows characterized by low to
medium Reynolds numbers and medium viscosity contrasts,
we consider
the Cahouet--Chabard preconditioner~\cite{Cahouet1988}, namely,
\begin{align}
	\te{S}
	\approx
	\te{M}_p(\nu) - \nicefrac{1}{\Delta t}\,\te{L}_p
	.
	\label{eqn:schur_complement_cahouet_chabard}
\end{align}
The mass matrix is scaled with the inverse viscosity, while
$\te{L}_p$ corresponds to the SIPG discretization of the Poisson operator
as in the pressure step~\eqref{eqn:pressure_step_weak}.
The corresponding element-wise contributions are hence of the form
\begin{gather*}
	%
	\left( q_h, \nicefrac{1}{\nu_h} \, p_h^{n+1} \right)_{\Omega_e}
	+
	\nicefrac{1}{\Delta t}
	\left(\grad q_h, \grad p_h^{n+1} \right)_{\Omega_e}
	\\
	-
	\nicefrac{1}{\Delta t}
	\left(\grad q_h,
		\left(
			p_h^{n+1} - \avg{p_h^{n+1}}
		\right) \, \n
	\right)_{\partial\Omega_e}
	-
	\nicefrac{1}{\Delta t}
	\left(q_h,
		\left(
			\avg{\grad p_h^{n+1}}
			+
			\tau
			\jump{p_h^{n+1}}
		\right)
		\cdot \n
	\right)_{\partial\Omega_e}
	.
\end{gather*}
Note here again that this Schur complement approximation breaks down for
i) extreme contrasts in $\nu$ and
ii) dominant convection, which we further investigate in Sec.~\ref{sec:numerical_results}.

\section{Numerical Results}
\label{sec:numerical_results}

The methods presented in this work are implemented in the software project
\href{https://github.com/exadg/exadg}{\texttt{ExaDG}}~\cite{ExaDGgithub} (see \cite{Fehn2021b}
for a comprehensive overview and \cite{Arndt2020} for the exa-scale project as a whole),
which implements numerical solvers for many PDE model problems in computational fluid and
structural dynamics based on the \texttt{deal.II}~\cite{dealII96} finite element
library and in particular its matrix-free infrastructure~\cite{KronbichlerKormann2012, KronbichlerKormann2019}.

The following numerical tests are designed to
i) verify the method and its consistent implementation using a smooth
analytical solution, see Sec.~\ref{sec:numerical_examples_man_sol},
ii) demonstrate the preconditioner performance for increasing viscosity contrasts,
see Sec.~\ref{sec:numerical_examples_cavity}, and
iii) showcase performance in engineering-size problems in an application to the
backward-facing step benchmark problem featuring a non-smooth solution,
see Sec.~\ref{sec:numerical_examples_backward_facing_step}.

Performance evaluations are run on an AMD EPYC 9254 processor
with 24 physical cores (2.9 GHz base frequency) per socket with two sockets per node and a
$2\times 12$-channel DDR5-4800 memory interface.
A single node features 384 GB of memory, where
the sum of L2 and L3 caches amounts to 305 MB, which
corresponds to $\approx38 \times 10^6$ floating point double-precision numbers.
The full width of the AVX-512 instruction set (8
double-precision or 16 single-precision floating point numbers) is used in a
vectorization-across-cells strategy~\cite{KronbichlerKormann2019}. The
GNU compiler version 13.2.0 with flags ``-O3 -march=native'' and
Intel OneAPI MPI, version 2021.12 are used. The full node is
utilized with 48 threads
with MPI-only parallelization.

\previouslyrevised{
In what follows we will refer to the maximum element Courant number, which we define based on the
polynomial degree of the velocity $k_u$, the velocity field $\ve{u}_h$ and the element size $h_e$,
as the maximum over all $N_e$ elements with $N_q$ quadrature points~\cite{Karniadakis2013}:
\begin{align}
		\mathrm{Cr}
		\coloneqq
		\max_{e=1,\dots,N_e}
			\mathrm{Cr}_e
		\quad
		\text{with}
		\quad
			\mathrm{Cr}_e
			\coloneqq
			\max_{q=1,\dots, N_q}
			\left.\frac{\Delta t \, \|\ve{u}\|}{\nicefrac{h_e}{(k_u)^r}}\right\rvert_{q,e}
		,
		\label{eqn:courant_definition}
\end{align}
with the parameter is chosen as $r=1.5$
(see, e.g.,~\cite{Fehn2018b, Fehn2021b} for a discussion regarding this choice).
}

\subsection{Manufactured Solution}
\label{sec:numerical_examples_man_sol}
We verify the expected convergence
rates in space and time on the unit square $\Omega = (0,1)^2$
with pure Dirichlet conditions.
\previouslyrevised{The pressure level is shifted to match the prescribed solution since the
pressure is only defined up to a constant in this setting.}
The relative errors are defined as
\begin{gather*}
		e^u_\mathrm{rel}
		\coloneqq
		\frac{\|\ve{u}(\ve{x}, T) - \ve{u}_h(T)\|_{L^2(\Omega_h)}}
			 {\|\ve{u}(\ve{x}, T)\|_{L^2(\Omega_h)}}
		,
		\quad
		e^p_\mathrm{rel}
		\coloneqq
		\frac{\|p(\ve{x}, T) - p_h(\ve{x}, T)\|_{L^2(\Omega_h)}}
			 {\|p(\ve{x}, T)\|_{L^2(\Omega_h)}}
		.
\end{gather*}
We design the body force vector $\ve{b}$ such that the solution
to~\eqref{eqn:momentum_balance_strong_form}--\eqref{eqn:continuity_equation}
in two space dimensions is given as
\begin{gather*}
	u_1(\ve{x}, t) = \cos(t) \, \sin(x_1) \, \cos(x_2)
	,
	\quad
	u_2(\ve{x}, t) = -\cos(t) \, \cos(x_1) \, \sin(x_2)
	,
	\quad
	p(\ve{x}, t) = \cos(t) \, \cos (x_1 \, x_2)
	,
\end{gather*}
and $\nu=\eta(\dot{\gamma})$ \previouslyrevised{following Eqn.}~\eqref{eqn:viscosity_general} accordingly.
For all tests here, the explicit versions of the schemes are used with linear solver tolerances set to
$\epsilon_\mathrm{lin}^\mathrm{abs} = 10^{-12}$ and
$\epsilon_\mathrm{lin}^\mathrm{rel} = 10^{-6}$.

We first verify spatial convergence rates for $k_u=2,3,4$ and $k_p = k_u-1$ by a sequence of mesh refinements.
The rheological parameters are $\kappa=\nicefrac{a}{2}=2\,b=1$, $\lambda=1~\text{s}$ $\rho=1~\text{kg/m}^3$
and $\eta_0 = 2 \, \eta_\infty = 0.1~\text{m}^2/\text{s}$.
Fig.~\ref{fig:convergence_space} reveals optimal ($k+1$) convergence rates in both the relative velocity and
pressure $L^2$ norms choosing $\Delta t=10^{-5}~\text{s}$ and $T=0.1~\text{s}$ in the BDF-2 scheme.
\previouslyrevised{With these settings, the maximum encountered element Courant number~\eqref{eqn:courant_definition} and Reynolds number are
$\mathrm{Cr} \lesssim 0.035$ and $\mathrm{Re} \lesssim 200$.}
The coupled solution scheme
and the projection solver deliver identical results up to the temporal discretization error.
Reducing the time step size to $\Delta t = 10^{-6}~\text{s}$ lowers the error achieved as $\nicefrac{h}{L}\rightarrow0$
for the splitting scheme as well, such that the linear solver tolerance
can be ruled out as source of error stagnation around $5\times10^{-8}$ for $\Delta t=10^{-5}~\text{s}$
for both velocity and pressure relative $L^2$ norms.
Interestingly, keeping the linear solver tolerance unchanged, the coupled solution scheme can reach lower errors
in both velocity and pressure. Hence, we conclude that the error stagnation observed stems from the projection itself.
The splitting error of the scheme scales with the time step size, but is higher than the temporal discretization error
of the coupled scheme using BDF-2.

\pgfplotstableread[comment chars={c}]{./results/spatial_convergence/spatial_convergence_carreau_coupled_p4.txt}\DataSpatialCarreauCoupledPFour
\pgfplotstableread[comment chars={c}]{./results/spatial_convergence/spatial_convergence_carreau_splitting_p4.txt}\DataSpatialCarreauSplittingPFour
\pgfplotstableread[comment chars={c}]{./results/spatial_convergence/spatial_convergence_carreau_coupled_p3.txt}\DataSpatialCarreauCoupledPThree
\pgfplotstableread[comment chars={c}]{./results/spatial_convergence/spatial_convergence_carreau_splitting_p3.txt}\DataSpatialCarreauSplittingPThree
\pgfplotstableread[comment chars={c}]{./results/spatial_convergence/spatial_convergence_carreau_coupled_p2.txt}\DataSpatialCarreauCoupledPTwo
\pgfplotstableread[comment chars={c}]{./results/spatial_convergence/spatial_convergence_carreau_splitting_p2.txt}\DataSpatialCarreauSplittingPTwo
\pgfplotsset{width=0.5\linewidth}
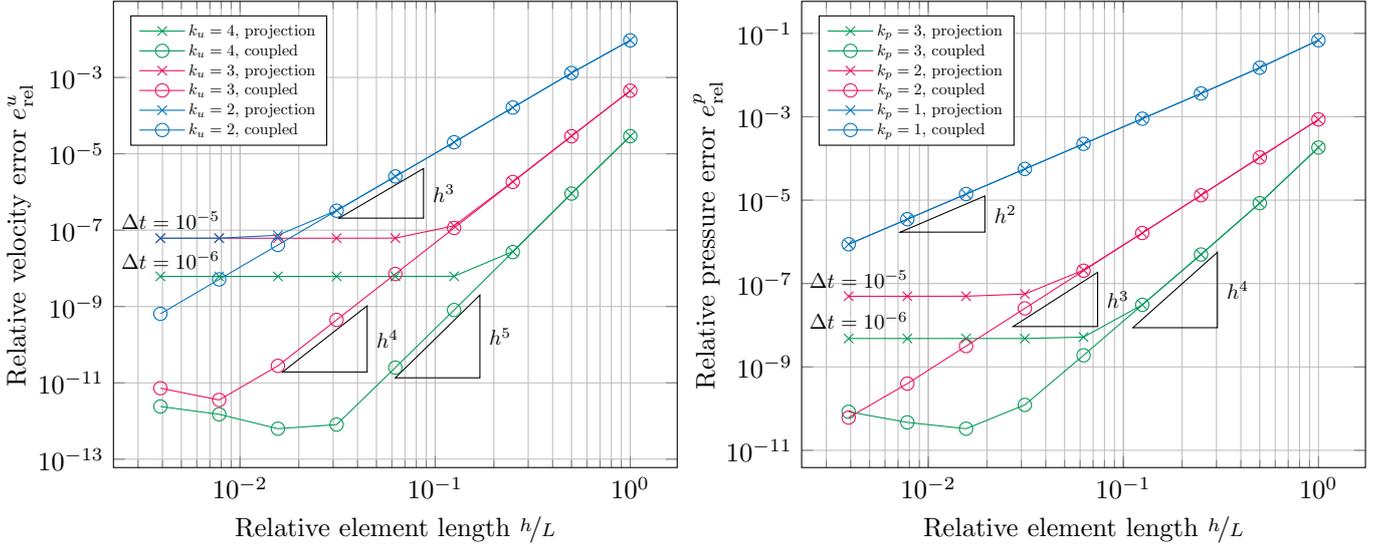
\begin{figure}[h!]
	\begin{tikzpicture}
		\begin{loglogaxis}[
			xlabel={Relative element length $\nicefrac{h}{L}$},
			ylabel={{Relative velocity error $e^u_\mathrm{rel}$}},
			grid=both,
			legend entries={
							{$k_u=4$, projection},
							{$k_u=4$, coupled},
							{$k_u=3$, projection},
				            {$k_u=3$, coupled},
				            {$k_u=2$, projection},
				            {$k_u=2$, coupled},
				            },
			legend style={
				nodes={scale=0.6, transform shape}},
				legend pos=north west,
				font = \normalsize,
				mark options = {scale = 0.5},
				legend cell align={left}
			]
			\addplot [ForestGreen,      mark=x,      mark size=5pt] table [x=h, y=u] {\DataSpatialCarreauSplittingPFour};
			\addplot [ForestGreen,      mark=o,      mark size=5pt] table [x=h, y=u] {\DataSpatialCarreauCoupledPFour};
			\addplot [OrangeRed,      mark=x,      mark size=5pt] table [x=h, y=u] {\DataSpatialCarreauSplittingPThree};
			\addplot [OrangeRed,      mark=o, mark size=5pt] table [x=h, y=u] {\DataSpatialCarreauCoupledPThree};
			\addplot [RoyalBlue,     mark=x,      mark size=5pt] table [x=h, y=u] {\DataSpatialCarreauSplittingPTwo};
			\addplot [RoyalBlue,     mark=o, mark size=5pt] table [x=h, y=u] {\DataSpatialCarreauCoupledPTwo};

			\logLogSlopeTriangle{0.55}{0.15}{0.535}{3}{black}{\scalebox{0.8}{$h^3$}}
			\logLogSlopeTriangle{0.45}{0.15}{0.205}{4}{black}{\scalebox{0.8}{$h^4$}}
			\logLogSlopeTriangle{0.65}{0.15}{0.1925}{5}{black}{\scalebox{0.8}{$h^5$}}
		\end{loglogaxis}
		\node at (0.75,3.275){\scalebox{0.8}{$\Delta t = 10^{-5}$}};
		\node at (0.75,2.75){\scalebox{0.8}{$\Delta t = 10^{-6}$}};
	\end{tikzpicture}
	\begin{tikzpicture}
		\begin{loglogaxis}[
			xlabel={Relative element length $\nicefrac{h}{L}$},
			ylabel={{Relative pressure error $e^p_\mathrm{rel}$}},
			grid=both,
			legend entries={
							{$k_p=3$, projection},
							{$k_p=3$, coupled},
							{$k_p=2$, projection},
							{$k_p=2$, coupled},
							{$k_p=1$, projection},
							{$k_p=1$, coupled},
							},
			legend style={
				nodes={scale=0.6, transform shape}},
			legend pos=north west,
			font = \normalsize,
			mark options = {scale = 0.5},
			legend cell align={left}
			]
			\addplot [ForestGreen,      mark=x,      mark size=5pt] table [x=h, y=p] {\DataSpatialCarreauSplittingPFour};
			\addplot [ForestGreen,      mark=o,      mark size=5pt] table [x=h, y=p] {\DataSpatialCarreauCoupledPFour};
			\addplot [OrangeRed,  mark=x,      mark size=5pt] table [x=h, y=p] {\DataSpatialCarreauSplittingPThree};
			\addplot [OrangeRed,  mark=o, mark size=5pt] table [x=h, y=p] {\DataSpatialCarreauCoupledPThree};
			\addplot [RoyalBlue, mark=x,      mark size=5pt] table [x=h, y=p] {\DataSpatialCarreauSplittingPTwo};
			\addplot [RoyalBlue, mark=o, mark size=5pt] table [x=h, y=p] {\DataSpatialCarreauCoupledPTwo};

			\logLogSlopeTriangle{0.325}{0.15}{0.505}{2}{black}{\scalebox{0.8}{$h^2$}}
			\logLogSlopeTriangle{0.525}{0.15}{0.3025}{3}{black}{\scalebox{0.8}{$h^3$}}
			\logLogSlopeTriangle{0.7375}{0.15}{0.3}{4.175}{black}{\scalebox{0.8}{$h^4$}}
		\end{loglogaxis}
		\node at (0.75,2.5){\scalebox{0.8}{$\Delta t = 10^{-5}$}};
		\node at (0.75,1.95){\scalebox{0.8}{$\Delta t = 10^{-6}$}};
	\end{tikzpicture}
	\caption{Spatial convergence study employing inf-sup stable finite element pairs
		     with velocity degree $k_u$ and pressure degree $k_p = k_u-1$ in the
		     projection and coupled solvers.
		     Optimal orders of accuracy in the relative $L^2$ errors are achieved up
		     to linear solver tolerance.}
	\label{fig:convergence_space}
\end{figure}

Second, we verify the order of the BDF time integrator by reducing the time step size $\Delta t$.
The end time is $T=10~\text{s}$, while the spatial discretization via an $8\times8$ grid with polynomials of
degree $k_u = k_p+1 = 4$ is fixed. The rheological parameters are
$\kappa=\nicefrac{a}{2}=2\,b=1$, $\lambda = 1~\text{s}$, $\rho=1000~\text{kg/m}^3$
and $\eta_0 = 50 \, \eta_\infty = 10^{-6}~\text{m}^2/\text{s}$
in the range of blood. To rule out stability issues for large time step
sizes stemming from the convective term, we consider only the Stokes problem for the
temporal convergence study.
\previouslyrevised{The maximum encountered element Courant~\eqref{eqn:courant_definition} number is
	$\mathrm{Cr} \approx 64$.}
Investigating the results in Fig.~\ref{fig:convergence_time}, two observations are made.

The first observation concerns the stability issues of the BDF-3 time integrator
even in this simplified scenario and for practically relevant parameters.
While BDF-1 and BDF-2 time integrators of the coupled and splitting solver yield
almost identical results for all time steps up to the spatial discretization error,
the BDF-3 scheme shows decreased stability for small time step sizes.
The onset of these instabilities is related to the linear/nonlinear solver settings,
where a stricter tolerance triggers the instabilities earlier. This effect is independent
of the fluid model, as it is also present in the Newtonian case shown in
Fig.~\ref{fig:convergence_time}. These results are not surprising, as there
exists no proof of unconditional stability for
BDF-3~\cite{Karniadakis1991, Guermond2003a, Guermond2006, Huang2025stability}
to the best of our knowledge.
While in many practical scenarios the scheme
has been reported to perform well, based on the results shown in
Fig.~\ref{fig:convergence_time} we can generally not recommend BDF-3,
as the error behavior is in general not known.

\pgfplotstableread[comment chars={c}]{./results/temporal_convergence/temporal_convergence_carreau_coupled_bdf1_fine_extended.txt}\DataTemporalCarreauCoupledBDFone
\pgfplotstableread[comment chars={c}]{./results/temporal_convergence/temporal_convergence_carreau_splitting_bdf1_fine_extended.txt}\DataTemporalCarreauSplittingBDFone
\pgfplotstableread[comment chars={c}]{./results/temporal_convergence/temporal_convergence_carreau_coupled_bdf2_fine_extended.txt}\DataTemporalCarreauCoupledBDFtwo
\pgfplotstableread[comment chars={c}]{./results/temporal_convergence/temporal_convergence_carreau_splitting_bdf2_fine_extended.txt}\DataTemporalCarreauSplittingBDFtwo
\pgfplotstableread[comment chars={c}]{./results/temporal_convergence/temporal_convergence_carreau_coupled_bdf3_fine.txt}\DataTemporalCarreauCoupledBDFthree

\pgfplotstableread[comment chars={c}]{./results/temporal_convergence/temporal_convergence_carreau_splitting_bdf3_fine_lowtol_Newtonian.txt}\DataTemporalCarreauSplittingBDFthreeLowTolNewtonian
\pgfplotstableread[comment chars={c}]{./results/temporal_convergence/temporal_convergence_carreau_splitting_bdf3_fine_hightol_Newtonian.txt}\DataTemporalCarreauSplittingBDFthreeHighTolNewtonian
\pgfplotstableread[comment chars={c}]{./results/temporal_convergence/temporal_convergence_carreau_splitting_bdf3_fine_lowtol_genNewtonian.txt}\DataTemporalCarreauSplittingBDFthreeLowTolGenNewtonian
\pgfplotstableread[comment chars={c}]{./results/temporal_convergence/temporal_convergence_carreau_splitting_bdf3_fine_hightol_genNewtonian.txt}\DataTemporalCarreauSplittingBDFthreeHighTolGenNewtonian
\pgfplotsset{width=0.5\linewidth}
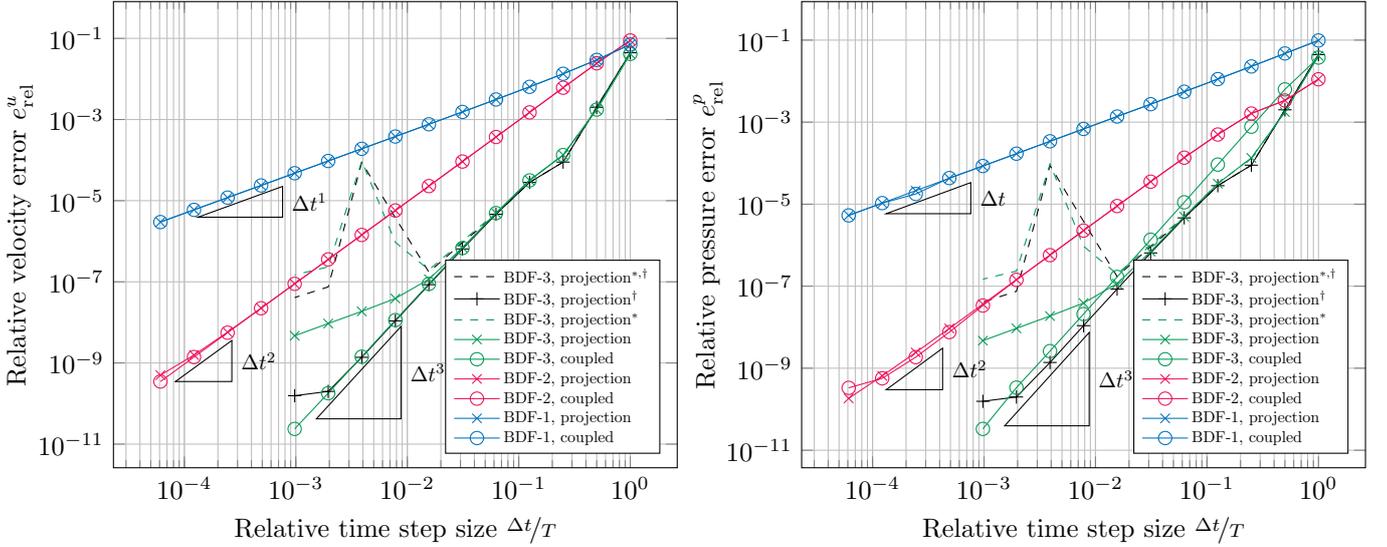
\begin{figure}[h!]
	\begin{tikzpicture}
		\begin{loglogaxis}[
			xlabel={Relative time step size $\nicefrac{\Delta t}{T}$},
			ylabel={{Relative velocity error $e^u_\mathrm{rel}$}},
			grid=both,
			legend entries={{BDF-3, projection$^{*,\dag}$},
							{BDF-3, projection$^\dag$},
							{BDF-3, projection$^*$},
							{BDF-3, projection},
							{BDF-3, coupled},
							{BDF-2, projection},
							{BDF-2, coupled},
							{BDF-1, projection},
							{BDF-1, coupled}
							},
			legend style={
			nodes={scale=0.6, transform shape}},
			legend pos=south east,
			font = \normalsize,
			mark options = {scale = 0.5},
			legend cell align={left}
			]
		\addplot [black,      mark=none, dashed, mark size=5pt] table [x=dt, y=u] {\DataTemporalCarreauSplittingBDFthreeLowTolNewtonian};
		\addplot [black,      mark=+,      mark size=5pt] table [x=dt, y=u] {\DataTemporalCarreauSplittingBDFthreeHighTolNewtonian};

		\addplot [ForestGreen,      mark=none, dashed, mark options=solid,     mark size=5pt] table [x=dt, y=u] {\DataTemporalCarreauSplittingBDFthreeLowTolGenNewtonian};
		\addplot [ForestGreen,      mark=x,      mark size=5pt] table [x=dt, y=u] {\DataTemporalCarreauSplittingBDFthreeHighTolGenNewtonian};

		\addplot [ForestGreen,      mark=o,      mark size=5pt] table [x=dt, y=u] {\DataTemporalCarreauCoupledBDFthree};
		\addplot [OrangeRed,      mark=x,      mark size=5pt] table [x=dt, y=u] {\DataTemporalCarreauSplittingBDFtwo};
		\addplot [OrangeRed,      mark=o, mark size=5pt] table [x=dt, y=u] {\DataTemporalCarreauCoupledBDFtwo};
		\addplot [RoyalBlue,     mark=x,      mark size=5pt] table [x=dt, y=u] {\DataTemporalCarreauSplittingBDFone};
		\addplot [RoyalBlue,     mark=o, mark size=5pt] table [x=dt, y=u] {\DataTemporalCarreauCoupledBDFone};

		\logLogSlopeTriangle{0.30}{0.15}{0.5375}{1}{black}{\scalebox{0.8}{$\Delta t^1$}}
		\logLogSlopeTriangle{0.21}{0.1}{0.185}{2}{black}{\scalebox{0.8}{$\Delta t^2$}}
		\logLogSlopeTriangle{0.51}{0.15}{0.105}{3}{black}{\scalebox{0.8}{$\Delta t^3$}}
	\end{loglogaxis}
\end{tikzpicture}
\begin{tikzpicture}
	\begin{loglogaxis}[
		xlabel={Relative time step size $\nicefrac{\Delta t}{T}$},
		ylabel={{Relative pressure error $e^p_\mathrm{rel}$}},
		grid=both,
		legend entries={{BDF-3, projection$^{*,\dag}$},
						{BDF-3, projection$^{\dag}$},
						{BDF-3, projection$^*$},
						{BDF-3, projection},
						{BDF-3, coupled},
						{BDF-2, projection},
						{BDF-2, coupled},
						{BDF-1, projection},
						{BDF-1, coupled},
						},
			legend style={
			nodes={scale=0.6, transform shape}},
			legend pos=south east,
			font = \normalsize,
			mark options = {scale = 0.5},
		legend cell align={left}
		]
		\addplot [black,      mark=none, dashed,     mark size=5pt] table [x=dt, y=u] {\DataTemporalCarreauSplittingBDFthreeLowTolNewtonian};
		\addplot [black,      mark=+, mark size=5pt] table [x=dt, y=u] 	{\DataTemporalCarreauSplittingBDFthreeHighTolNewtonian};

		\addplot [ForestGreen,      mark=none, dashed,     mark size=5pt] table [x=dt, y=u] 	{\DataTemporalCarreauSplittingBDFthreeLowTolGenNewtonian};
		\addplot [ForestGreen,      mark=x,      mark size=5pt] table [x=dt, y=u] 	{\DataTemporalCarreauSplittingBDFthreeHighTolGenNewtonian};

		\addplot [ForestGreen,      mark=o,      mark size=5pt] table [x=dt, y=p] {\DataTemporalCarreauCoupledBDFthree};
		\addplot [OrangeRed,      mark=x,      mark size=5pt] table [x=dt, y=p] {\DataTemporalCarreauSplittingBDFtwo};
		\addplot [OrangeRed,      mark=o, mark size=5pt] table [x=dt, y=p] {\DataTemporalCarreauCoupledBDFtwo};
		\addplot [RoyalBlue,     mark=x,      mark size=5pt] table [x=dt, y=p] {\DataTemporalCarreauSplittingBDFone};
		\addplot [RoyalBlue,     mark=o, mark size=5pt] table [x=dt, y=p] {\DataTemporalCarreauCoupledBDFone};

		\logLogSlopeTriangle{0.30}{0.15}{0.545}{1}{black}{\scalebox{0.8}{$\Delta t$}}
		\logLogSlopeTriangle{0.25}{0.1}{0.1675}{2}{black}{\scalebox{0.8}{$\Delta t^2$}}
		\logLogSlopeTriangle{0.51}{0.15}{0.09}{3}{black}{\scalebox{0.8}{$\Delta t^3$}}
	\end{loglogaxis}
\end{tikzpicture}
\caption{Temporal convergence study employing uniform time steps and BDF-$m$ schemes
	     $m=1,2,3$. Optimal orders of accuracy in the relative $L^2$ errors are achieved
	     with the coupled solver. The projection solver yields identical results up to $m=3$.
	     The known unstable behavior is also observed for the Newtonian splitting scheme ($\dag$)
	     and is found to be more drastic for lower linear solver tolerance
	     ($*$, $\epsilon_\mathrm{lin}^\mathrm{abs} = 10^{-16}$,
	     $\epsilon_\mathrm{lin}^\mathrm{rel} = 10^{-8}$).}
\label{fig:convergence_time}
\end{figure}

The second observation regarding Fig.~\ref{fig:convergence_time} concerns the decrease
to a linear convergence rate from the optimal ones for BDF-$m$, $m=2,3$.
Note, however, that this reduction is not seen for the Newtonian case.
The decreased convergence order
observed for the generalized Newtonian case is not connected to the
pressure boundary condition~\eqref{eqn:pressure_Neumann_BC},
as even a solution with $\ve{n} \cdot \grad p\rvert_{\partial \Omega} = 0$
leads to the reduction. Furthermore, one cannot recover optimal convergence
\previouslychanged{rates by including the divergence of the viscous term on the right-hand side of the PPE
as done in~\cite{Karamanos2000}.}
However, optimal convergence rates in the velocity for the generalized Newtonian
case can be recovered by inserting the exact pressure solution after solving the PPE,
thereby identifying the pressure as the source of error.
\previouslychanged{The splitting scheme itself
leads to the reduction for the generalized Newtonian case, as also observed in~\cite{Karamanos2000}.
The adopted scheme based on the method presented by~\citet{Karniadakis1991}}
can be recast as a velocity-projection scheme, hence suffers similarly from pressure
boundary layers, see, e.g.~\cite{Deteix2018, Deteix2019} for possible remedies.

The error from which on the rates are reduced depends on the viscosity and its gradient.
This can be demonstrated by decreasing the
upper viscosity bound from $\eta_0=5\times10^{-2}$,
while keeping the lower bound fixed at $\eta_\infty = 5\times10^{-2}$
as shown in Fig.~\ref{fig:convergence_time_2} for the BDF-2 time integrator.
For the BDF-3 integrator, a similar effect is observed and omitted here for brevity.
Nonetheless, for a large parameter range and related applications the present scheme
does indeed deliver accurate results, when non-Newtonian effects are comparably mild. This is the case
when $\mathcal{O}(\eta_0/\eta_\infty)\leq100$ and medium to high Reynolds numbers,
\previouslyrevised{approximately $ \mathrm{Re}\geq 100$,}
are considered. In such scenarios, the expected error is small, as, e.g., in blood flows within large arteries
or turbulent flows as considered by~\cite{Blackburn2025}.

\pgfplotstableread[comment chars={c}]{./results/temporal_convergence_new/nu0_5em2_CS_bdf2.txt}\DataBDFtwoCSnupowerminustwo
\pgfplotstableread[comment chars={c}]{./results/temporal_convergence_new/nu0_5em2_DS_bdf2.txt}\DataBDFtwoDSnupowerminustwo
\pgfplotstableread[comment chars={c}]{./results/temporal_convergence_new/nu0_5em3_DS_bdf2.txt}\DataBDFtwoDSnupowerminusthree
\pgfplotstableread[comment chars={c}]{./results/temporal_convergence_new/nu0_5em4_DS_bdf2.txt}\DataBDFtwoDSnupowerminusfour
\pgfplotstableread[comment chars={c}]{./results/temporal_convergence_new/nu0_5em5_DS_bdf2.txt}\DataBDFtwoDSnupowerminusfive

\pgfplotsset{width=0.5\linewidth}
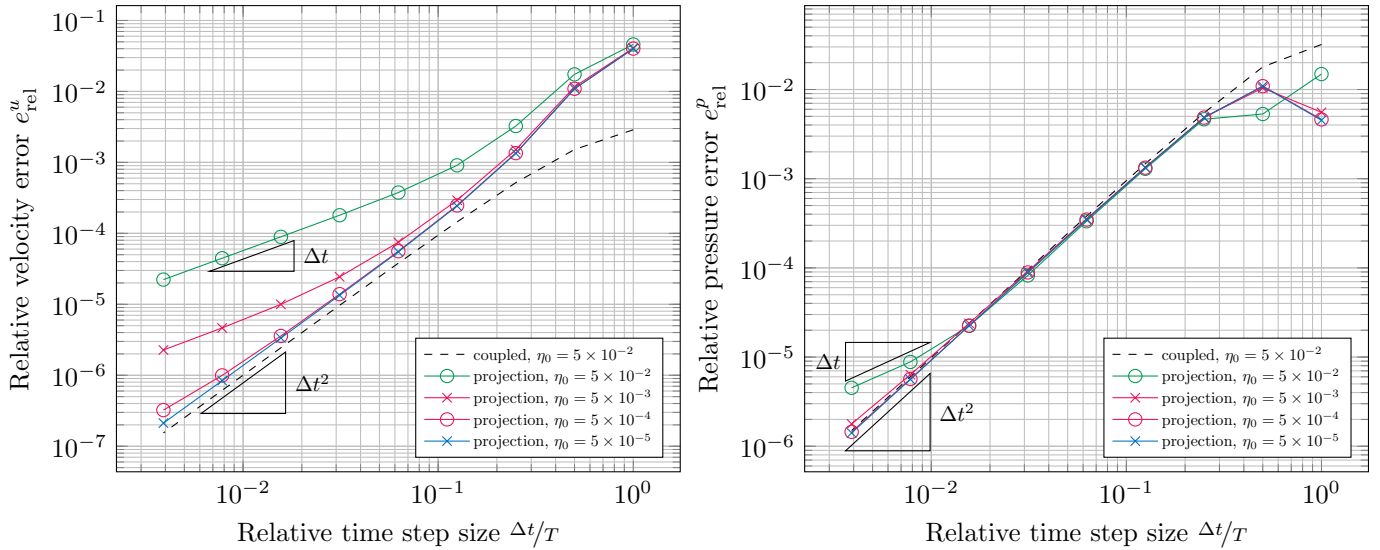
\begin{figure}[h!]
	\begin{tikzpicture}
		\begin{loglogaxis}[
			xlabel={Relative time step size $\nicefrac{\Delta t}{T}$},
			ylabel={{Relative velocity error $e^u_\mathrm{rel}$}},
			grid=both,
			legend entries={
				{coupled, $\eta_0 = 5\times10^{-2}$},
				{projection, $\eta_0 = 5\times10^{-2}$},
				{projection, $\eta_0 = 5\times10^{-3}$},
				{projection, $\eta_0 = 5\times10^{-4}$},
				{projection, $\eta_0 = 5\times10^{-5}$}
				},
			legend style={
			nodes={scale=0.6, transform shape}},
			legend pos=south east,
			font = \normalsize,
			mark options = {scale = 0.5},
			legend cell align={left}
			]
			\addplot [black,      mark=none, dashed, mark size=5pt] table [x=dt, y=u] {\DataBDFtwoCSnupowerminustwo};

			\addplot [ForestGreen,      mark=o,      mark size=5pt] table [x=dt, y=u] {\DataBDFtwoDSnupowerminustwo};
			\addplot [OrangeRed,      mark=x,      mark size=5pt] table [x=dt, y=u] {\DataBDFtwoDSnupowerminusthree};
			\addplot [OrangeRed,      mark=o, mark size=5pt] table [x=dt, y=u] {\DataBDFtwoDSnupowerminusfour};
			\addplot [RoyalBlue,     mark=x,      mark size=5pt] table [x=dt, y=u] {\DataBDFtwoDSnupowerminusfive};

			\logLogSlopeTriangle{0.315}{0.15}{0.43}{1}{black}{\scalebox{0.8}{$\Delta t$}}
			\logLogSlopeTriangle{0.3}{0.15}{0.125}{2}{black}{\scalebox{0.8}{$\Delta t^2$}}
\end{loglogaxis}
\end{tikzpicture}
\begin{tikzpicture}
	\begin{loglogaxis}[
		xlabel={Relative time step size $\nicefrac{\Delta t}{T}$},
		ylabel={{Relative pressure error $e^p_\mathrm{rel}$}},
		grid=both,
		legend entries={
			{coupled, $\eta_0 = 5\times10^{-2}$},
			{projection, $\eta_0 = 5\times10^{-2}$},
			{projection, $\eta_0 = 5\times10^{-3}$},
			{projection, $\eta_0 = 5\times10^{-4}$},
			{projection, $\eta_0 = 5\times10^{-5}$}
		},
		legend style={
			nodes={scale=0.6, transform shape}},
		legend pos=south east,
		font = \normalsize,
		mark options = {scale = 0.5},
		legend cell align={left}
		]
		\addplot [black,      mark=none, dashed, mark size=5pt] table [x=dt, y=p] {\DataBDFtwoCSnupowerminustwo};

		\addplot [ForestGreen,      mark=o,      mark size=5pt] table [x=dt, y=p] {\DataBDFtwoDSnupowerminustwo};
		\addplot [OrangeRed,      mark=x,      mark size=5pt] table [x=dt, y=p] {\DataBDFtwoDSnupowerminusthree};
		\addplot [OrangeRed,      mark=o, mark size=5pt] table [x=dt, y=p] {\DataBDFtwoDSnupowerminusfour};
		\addplot [RoyalBlue,     mark=x,      mark size=5pt] table [x=dt, y=p] {\DataBDFtwoDSnupowerminusfive};

		\logLogSlopeTriangle{0.0725}{-0.15}{0.2775}{1}{black}{\scalebox{0.8}{\hspace*{-6.5mm}$\Delta t$}}
		\logLogSlopeTriangle{0.2225}{0.15}{0.045}{2}{black}{\scalebox{0.8}{$\Delta t^2$}}
	\end{loglogaxis}
\end{tikzpicture}
\caption{Temporal convergence study employing uniform time steps and BDF-$2$ schemes
under variation of the upper viscosity limit $\eta_0$, keeping $\eta_\infty=5\times10^{-2}~\text{m}^2/\text{s}$ fixed.
Optimal orders of accuracy in the relative $L^2$ errors are achieved
with the coupled solver. The projection solver yields optimal rates for the Newtonian case,
but reduces earlier to linear convergence for high viscosity contrasts.}
\label{fig:convergence_time_2}
\end{figure}

In summary, the spatial convergence results show optimal behavior, while the temporal convergence is
optimal up to a certain point depending on the difference in the upper and lower viscosity limits.
For the generalized Newtonian case, an error of order $\mathcal{O}(\Delta t)$
stemming from the PPE boundary condition can dominate.

\subsection{Lid-Driven Cavity}
\label{sec:numerical_examples_cavity}
The next numerical example focuses on the preconditioner
performance under increasing viscosity contrasts of generalized Newtonian models.
We solve the lid-driven cavity benchmark problem in $\Omega=(0,1)^3$, fixing some of the rheological
parameters to $\kappa = \nicefrac{a}{2} = 2\,b = 1$, $\lambda=10~\text{s}$, while
we vary the upper and lower viscosity bounds $\eta_0$ and $\eta_\infty$.
The domain is discretized into $32\times32\times32$ elements with
the velocity and pressure degrees set to $k_u = 4$ and $k_p = 3$, respectively.
This yields $14.4\times10^6$ velocity and pressure DoFs.
The BDF-2 time integration scheme with a fixed time step size of $\Delta t= 0.01~\text{s}$ is used
to advance the solution from $t=0$ to $T=0.2~\text{s}$ in 20 time steps.

The fluid is initially at rest and accelerated by smoothly
ramping up the prescribed velocity at the lid (at $x_2 = 1$) in space and time via
\begin{gather*}
	\gu(\ve{x},t) = (g_{u_1}, 0, 0)^\top
	,
	\quad
	\text{where}
	\quad
	g_{u_1}(\ve{x},t)
	\coloneqq
	\zeta_t(t)
	\,
	\zeta_x(x_1)
	\,
	\zeta_x(x_3)
	\\
	\text{and}
	\quad
	\zeta_t(t)
	\coloneqq
	\begin{cases}
		\sin^2(5\pi\,t) & t < 0.1,\\
		1.0 & \text{else}
		,
	\end{cases}
	,
	\quad
	\zeta_x(x)
	\coloneqq
	\begin{cases}
		1-\cos^4(5\pi\,x) & x < 0.1,\\
		1-\cos^4(5\pi\,(x-1)) & x > 0.9,\\
		1 & \text{else}.
	\end{cases}
\end{gather*}
This choice avoids the pressure singularity at $x=y=1$, cf.~\cite{Frutos2016, Pacheco2021_c}.
\previouslyrevised{The resulting maximum observed element Courant number~\eqref{eqn:courant_definition} is thus $\mathrm{Cr}\approx 2.56$.}

The nonlinear solver tolerance is set to $\epsilon_\mathrm{nl}^\mathrm{rel} = 10^{-6}$ and
$\epsilon_\mathrm{nl}^\mathrm{abs} = 10^{-12}$ (if applicable),
while the linear solver tolerance targeted is $\epsilon_\mathrm{lin}^\mathrm{abs} = 10^{-12}$ and
either $\epsilon_\mathrm{lin}^\mathrm{rel} = 10^{-6}$ for individual steps, or
$\epsilon_\mathrm{lin}^\mathrm{rel} = 10^{-3}$ for systems solved within a nonlinear solver.
\previouslyrevised{These settings render the Newton solver inexact,
	which is a viable option typically performing well for large nonlinear systems
	as further demonstrated in Sec.~\ref{sec:numerical_examples_backward_facing_step}.
}

We neglect the convective term in this example, which enables a competitive use of the Cahouet--Chabard
preconditioner~\eqref{eqn:schur_complement_cahouet_chabard}.
The overall preconditioner settings are chosen such that only the toughest problems
featuring the highest viscosity contrasts lead to divergence for the coupled solver,
while for other parameter combinations significantly cheaper options would have been available.
Hence, the throughput results shown here do not show the highest achievable performance,
a topic which is systematically assessed by a performance comparison in
Sec.~\ref{sec:numerical_examples_backward_facing_step} below.

\begin{figure}[h!]
	\centering
	\begin{subfigure}{.49\textwidth}
		\centering
		\includegraphics[width=1.0\linewidth, draft=\draftSec]{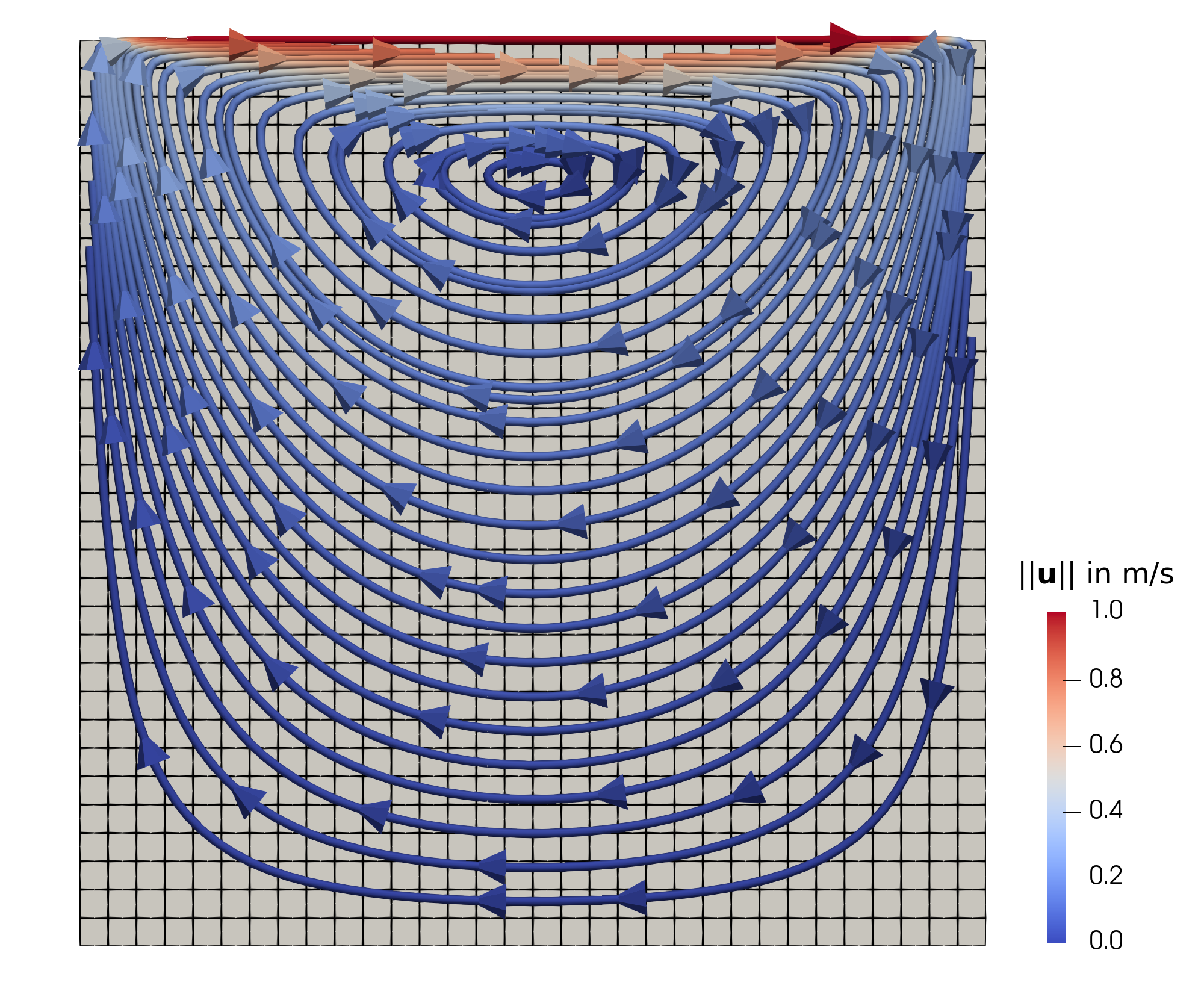}
		\caption{Velocity $\ve{u}$}
	\end{subfigure}
	\begin{subfigure}{.49\textwidth}
		\centering
		\includegraphics[width=1.0\linewidth, draft=\draftSec]{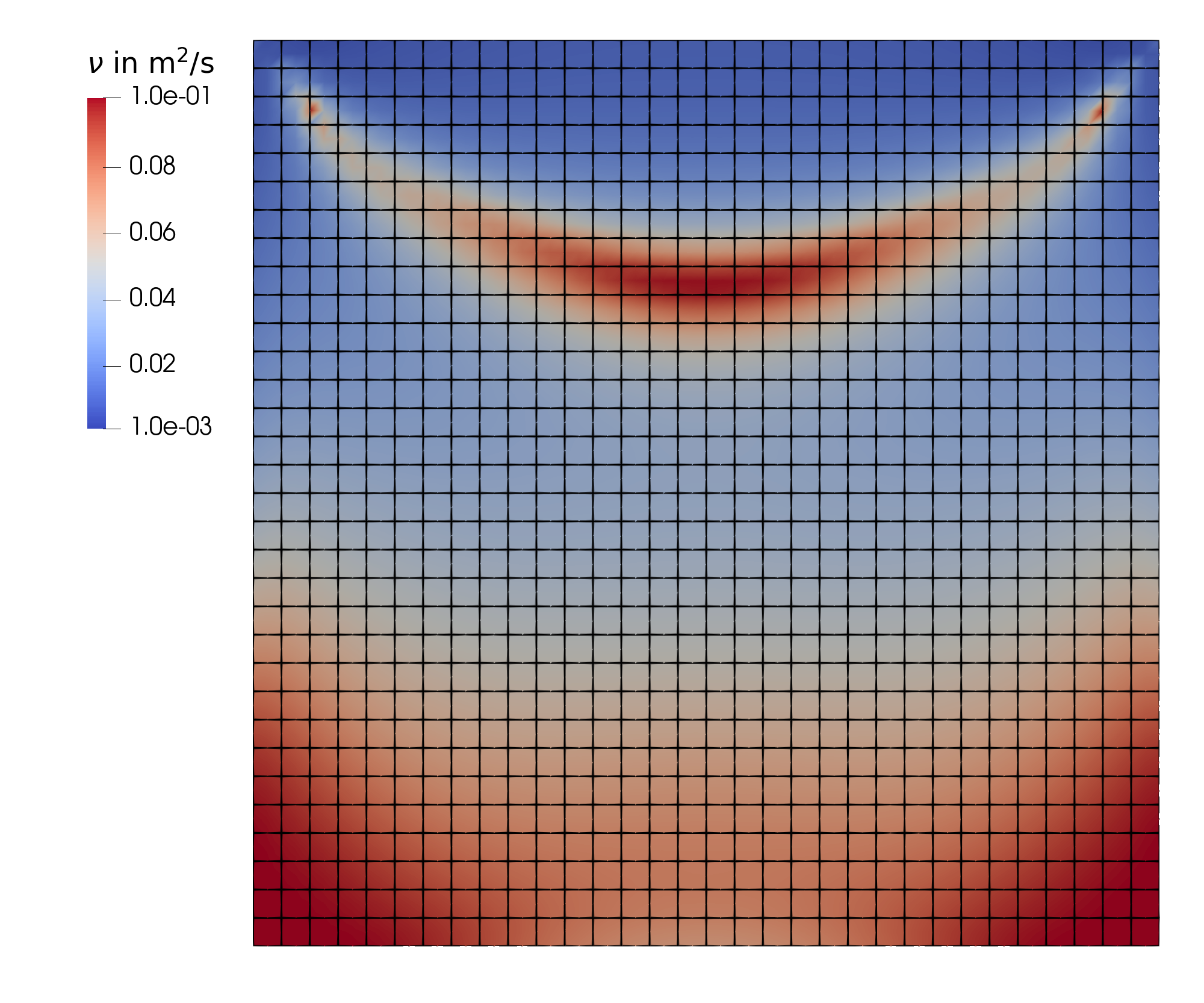}
		\caption{Viscosity $\nu$}
	\end{subfigure}
	\caption{Exemplary steady-state solution on the $xy$ symmetry plane of the cavity using $\eta_0 = 100 \, \eta_\infty = 10^{-1}~\mathrm{m}^2\mathrm{/s}$.}
	\label{fig:cavity_2D_results}
\end{figure}

A representative solution of a two-dimensional version of the problem with
$\eta_\infty = 10^{-3}~\text{m}^2/\text{s}$, and $\eta_0 - \eta_\infty = 100 \, \eta_\infty$ is shown in Fig.~\ref{fig:cavity_2D_results}.
One observes a circulatory flow and---depending on the
rheological properties---a band of increased apparent visocity
beneath the lid. On the lid and near the leading and trailing edges of the lid,
the viscosity reduces due to the locally increased shear rate, while at the bottom of the
cavity, the apparent viscosity reaches its maximum value. Such a setup is thus able to
reproduce the effects seen in practically relevant settings incorporating generealized
Newtonian fluids in engineering applications.
We note that compared to the multi-sinker benchmark problem in earth
mantle convection (see, e.g., \cite{May2014, Rudi2017, Clevenger2021}), the present setup
does not possess pronounced localized viscosity contrasts. However, we believe that the
current scenario represents a realistic behavior in the flow problems at hand,
despite the lower demands on preconditioning.

\subsubsection{\previouslyrevised{Constant viscosity}}

First, we vary $\nu=\eta_\infty=\eta_0 = 10^{-9},10^{-8},\dots,1.0~\text{m}^2/\text{s}$ in a
constant viscosity Stokes problem to establish a baseline for comparison.
The viscosity range leads to a pressure boundary layer depending on $\nu$ when using
the projection scheme. This drawback is also present in the original method~\cite{Karniadakis1991},
and the extension to variable viscosities is not expected (or intended) to remedy it.
The boundary layer thickness scales as
$\mathcal{O}(\sqrt{\nu \Delta t})$, therefore the results here only show that the arising systems
can be solved for a large range of parameters, while the resulting solutions might, in fact, be inaccurate.

Tab.~\ref{tab:cavity_results_Newtonian} lists the iteration counts of the
coupled solver with Cahouet--Chabard
preconditioner \previouslyrevised{as in Eqn.}~\eqref{eqn:schur_complement_cahouet_chabard}. Overall, we observe
bounded and low iteration counts for all involved problems in the coupled
solution scheme and the splitting method.
At the two ends of the considered viscosity range the iteration counts are lower than
in the middle. This is caused by the Schur complement approximation being slightly worse
for the medium parameter range and the block inverses being approximated only.
Using the present settings, a multigrid preconditioner is necessary for the penalty step~\eqref{eqn:penalty_step}
to converge robustly, which is connected to the pressure boundary layer and
related inaccurate velocity field.
This step then also contributes significantly to the
overall compute time. \previouslyrevised{For lower viscosity or smaller
time step sizes, the preconditioner can be significantly simplified,
solving only an element-wise block-diagonal system of equations
where the coefficients come from the mass operator.}

Despite the restrictions on the test and solver setups mentioned above,
one can appreciate an approximately 1.5--2$\times$ speed-up of the
splitting scheme over the coupled approach. Furthermore,
the linearized scheme can
significantly improve the throughput depending on the problem at hand. The linearized versions of the coupled
and splitting schemes deliver roughly 2--6$\times$ the throughput of their
nonlinear counterparts, which of course depends on the linear and nonlinear solver
settings and on the time step size chosen.

\begin{table}[h!]
	\centering
	\caption{Throughput and average number of linear iterations required to achieve
		a residual reduction by $10^6$ for the Stokes problem with constant viscosity
		$\nu = \eta_0 = \eta_\infty$ \previouslyrevised{(n.a. = not applicable)}.}
	{
		\scriptsize
		\label{tab:cavity_results_Newtonian}
		\begin{tabular}{||c || c || c | c | c | c | c ||}
			\hline
			&&&&&&\\[-1.25ex]
			\multirow{3}{*}{\vspace*{2mm}Solver}
			&\multicolumn{1}{|c||}{Viscosity}
			&\multicolumn{1}{|c|}{Lin. iter.}
			&\multicolumn{1}{|c|}{Lin. iter.}
			&\multicolumn{1}{|c|}{Lin. iter.}
			&\multicolumn{1}{|c|}{Lin. iter.}
			&\multicolumn{1}{|c||}{Throughput}
			\\
			& [m$^2$/s] & coupled [1] & pressure [1]
			& viscous [1] & penalty [1] & [$10^4$ DoFs/s/core/time step]
			\\[1.25ex]
			\hline\hline
			&&&&&&
			\\[-1.25ex]
			\multirow{10}{*}{\vspace*{-18mm}\rotatebox{90}{coupled approach}}
			& $10^{-9}$ & \s 4.5 & \previouslyrevised{n.a.} & \previouslyrevised{n.a.} & 0.0 & 40.2 \\[1.5ex]
			& $10^{-8}$ & 11.4   & \previouslyrevised{n.a.} & \previouslyrevised{n.a.} & 1.1 & 23.0 \\[1.5ex]
			& $10^{-7}$ & 16.4   & \previouslyrevised{n.a.} & \previouslyrevised{n.a.} & 1.9 & 18.3 \\[1.5ex]
			& $10^{-6}$ & 19.4   & \previouslyrevised{n.a.} & \previouslyrevised{n.a.} & 2.8 & 15.7 \\[1.5ex]
			& $10^{-5}$ & 19.5   & \previouslyrevised{n.a.} & \previouslyrevised{n.a.} & 2.9 & 16.8 \\[1.5ex]
			& $10^{-4}$ & 18.8   & \previouslyrevised{n.a.} & \previouslyrevised{n.a.} & 2.8 & 16.0 \\[1.5ex]
			& $10^{-3}$ & 14.6   & \previouslyrevised{n.a.} & \previouslyrevised{n.a.} & 2.6 & 18.8 \\[1.5ex]
			& $10^{-2}$ & 12.8   & \previouslyrevised{n.a.} & \previouslyrevised{n.a.} & 3.3 & 20.4 \\[1.5ex]
			& $10^{-1}$ & 12.0   & \previouslyrevised{n.a.} & \previouslyrevised{n.a.} & 3.8 & 20.9 \\[1.5ex]
			& $1.0$     & \s 9.7 & \previouslyrevised{n.a.} & \previouslyrevised{n.a.} & 3.9 & 23.3
			\\[0.75ex]
			\hline
			\hline
			&&&&&&\\[-0.75ex]
			\multirow{10}{*}{\vspace*{-18mm}\rotatebox{90}{splitting/projection}}
			& $10^{-9}$ & \previouslyrevised{n.a.} & \s 2.5 & \s 1.5 & 0.0 & 56.3 \\[1.5ex]
			& $10^{-8}$ & \previouslyrevised{n.a.} & \s 3.1 & \s 2.2 & 1.1 & 45.1 \\[1.5ex]
			& $10^{-7}$ & \previouslyrevised{n.a.} & \s 3.7 & \s 3.0 & 1.9 & 40.8 \\[1.5ex]
			& $10^{-6}$ & \previouslyrevised{n.a.} & \s 3.8 & \s 4.0 & 2.8 & 35.9 \\[1.5ex]
			& $10^{-5}$ & \previouslyrevised{n.a.} & \s 3.8 & \s 5.3 & 2.9 & 33.2 \\[1.5ex]
			& $10^{-4}$ & \previouslyrevised{n.a.} & \s 3.8 & \s 5.9 & 2.8 & 31.9 \\[1.5ex]
			& $10^{-3}$ & \previouslyrevised{n.a.} & \s 3.8 & \s 6.1 & 2.9 & 32.1 \\[1.5ex]
			& $10^{-2}$ & \previouslyrevised{n.a.} & \s 3.8 & \s 6.5 & 3.8 & 31.8 \\[1.5ex]
			& $10^{-1}$ & \previouslyrevised{n.a.} & \s 3.7 & \s 6.2 & 4.2 & 32.4 \\[1.5ex]
			& $1.0$     & \previouslyrevised{n.a.} & \s 3.6 & \s 5.7 & 4.6 & 31.2 \\[0.75ex]
			\hline
		\end{tabular}
	}
\end{table}

\subsubsection{\previouslyrevised{Variable viscosity}}

In the second test, we set $\eta_\infty=10^{-6}~\text{m}^2/\text{s}$ and increase the viscosity
contrast $\eta_0-\eta_\infty$. Tab.~\ref{tab:cavity_results_genNewtonian} lists the
number of iterations required to reach convergence in the nonlinear and linear versions
of the coupled and splitting schemes. With increasing viscosity contrast, the nonlinear
and linear iteration counts increase. At the end of the viscosity spectrum, the
nonlinear solvers cannot reach the end of the desired time interval. The linearized
versions of the coupled and splitting schemes require a higher number of iterations
in the linear solvers, since the tolerance settings are more stringent. Notably, the
linearized time integration schemes allow the solver to remain robust until the end of the time interval, even
for the highest viscosity contrast observed in this benchmark.
Similarly to the constant-viscosity case, the throughput of the splitting scheme
is in general 1.25--2$\times$ higher. Again, the accuracy might be limited for the high viscosity
cases, but the focus in this example lies on iteration counts.

\begin{table}[h!]
	\centering
	\caption{Throughput, nonlinear and linear iterations of the coupled solver and viscous
		step within the projection scheme required to achieve convergence for the various
		linearization variants for fixed lower viscosity limit
		$\eta_\infty = 10^{-6}~\text{m}^2\text{/s}$ and increasing upper limit $\eta_0$
		\previouslyrevised{(n.a. = not applicable, div. = diverged)}.}
	{
		\scriptsize
		\label{tab:cavity_results_genNewtonian}
		\begin{tabular}{||c || c || c | c | c || c | c ||}
			\hline
			&&\multicolumn{3}{|c||}{\,}&\multicolumn{2}{|c||}{\,}\\[-1.25ex]
			\multirow{4}{*}{\vspace*{2mm}Solver}
			&
			\multirow{4}{*}{\vspace*{2mm}$\eta_0 - \eta_\infty$}
			&
			\multicolumn{3}{|c||}{Implicit viscosity, $\nu = \nu(\ve{u}^{n+1, k+1})$}
			&
			\multicolumn{2}{|c||}{Linearized viscosity, $\nu = \nu(\ve{u}^{n+1,k})$}
			\\
			\cline{3-7}
			&&&&&&
			\\[-2ex]
			&
			&\multicolumn{1}{|c|}{Nonlin. iter.}
			&\multicolumn{1}{|c|}{Lin. iter.}
			&\multicolumn{1}{|c||}{Throughput}
			&\multicolumn{1}{|c|}{Lin. iter.}
			&\multicolumn{1}{|c||}{Throughput}
			\\
			&
			[m$^2$/s]
			&
			[1]&(avg.) [1]&[$10^4$ DoFs/s/core/time step]&[1]&[$10^4 $ DoFs/s/core/time step]
			\\[1.25ex]
			\hline\hline
			&&&&&&
			\\[-1.25ex]
			\multirow{6}{*}{\vspace*{-18mm}\rotatebox{90}{coupled approach}}
			&0&\previouslyrevised{n.a.}&\previouslyrevised{n.a.}&\previouslyrevised{n.a.} &19.4&16.0\\[1.5ex]
			&$10^{-6}$&\s 3.4 & 5.6& 7.1  &13.5&14.4\\[1.5ex]
			&$10^{-5}$&\s 4.8 & 5.5& 5.3  &15.1&13.5\\[1.5ex]
			&$10^{-4}$&\s 5.2 & 5.3& 5.1  &13.8&14.0\\[1.5ex]
			&$10^{-3}$&\s 7.0 & 5.6& 3.8  &16.0&12.8\\[1.5ex]
			&$10^{-2}$&  11.4 & 8.4& 2.0  &19.1&12.0\\[1.5ex]
			&$10^{-1}$&  14.7 & 7.9& 1.6  &21.7&10.9\\[1.5ex]
			&1        &\s div.&div.& div. &17.1&12.4\\[0.75ex]
			\hline
			\hline
			&&&&&&\\[-0.75ex]
			\multirow{6}{*}{\vspace*{-18mm}\rotatebox{90}{splitting/projection}}
			&0&\previouslyrevised{n.a.}&\previouslyrevised{n.a.}&\previouslyrevised{n.a.}  &\s 4.0& 36.3 \\[1.5ex]
			&$10^{-6}$&\s 2.9 &2.8 &  10.1 &\s 3.0& 21.5 \\[1.5ex]
			&$10^{-5}$&\s 3.9 &3.1 &\s 7.8 &\s 3.6& 22.0 \\[1.5ex]
			&$10^{-4}$&\s 4.4 &2.9 &\s 7.2 &\s 3.0& 21.3 \\[1.5ex]
			&$10^{-3}$&\s 6.6 &3.6 &\s 5.1 &\s 3.9& 20.8 \\[1.5ex]
			&$10^{-2}$&  10.0 &7.0 &\s 2.8 &\s 4.9& 20.1 \\[1.5ex]
			&$10^{-1}$&  10.9 &5.6 &\s 2.6 &\s 6.3& 16.6 \\[1.5ex]
			&1        &\s div.&div.&\s div.&\s 6.8& 12.4 \\[0.75ex]
			\hline
		\end{tabular}
	}
\end{table}

\subsection{Backward-Facing Step}
\label{sec:numerical_examples_backward_facing_step}

This numerical example demonstrates the performance of the various discretization
variants adopting the classical backward-facing step benchmark problem with
hemodynamics-inspired parameters.
The geometry is chosen according to~\cite{Choi2005}, where the
channel height at the inlet is $H=5.2~\text{mm}$,
the step height is $s = 0.9423\,H$, the
distance from the inlet to the step is $L_1 = 10\,H$, and
the distance from the step to the outlet is $L_2=20\,H$.
Although the two-dimensional geometry is extruded in the direction normal to the step
to generate a three-dimensional geometry, the flow remains 2D for sufficiently low Reynolds numbers.
Regarding boundary conditions, the
inflow velocity profile for $x_2 \in [0,H]$ is set to
\begin{gather*}
		\gu(\ve{x},t) = (g_{u_1}, 0, 0)^\top
		,
		\quad
		\text{where}
		\quad
		g_{u_1}(\ve{x},t)
		\coloneqq
		\zeta_t(t)
		\,
		\zeta_x(x_2)
		\,
		u_\mathrm{in,max}
		\\
		\quad
		\text{and}
		\quad
		\zeta_t(t)
		\coloneqq
		\begin{cases}
			\sin^2\left(\frac{\pi\,t}{2\,t_\mathrm{ramp}}\right) & t < t_\mathrm{ramp},\\
			1.0 & \text{else},
		\end{cases}
		,
		\quad
		\zeta_x(x)
		\coloneqq
		1 - \left(\frac{x}{\nicefrac{H}{2}}\right)^2
		,
\end{gather*}
with $t_\mathrm{ramp} = 0.3~\text{s}$ and $u_\mathrm{in,max} \approx 0.662~\text{m/s}$ to give
$\text{Re}=300$, defining $\text{Re} = \nicefrac{2}{\nu_\infty} \, Q_\mathrm{max}$
with $Q_\mathrm{max} \coloneqq \nicefrac{2}{3} \, H \, u_\mathrm{in,max}$.
The outlet features zero Neumann boundary conditions, while on the step and the upper and lower walls,
no-slip conditions, $\ve{u}=\ve{0}$, are enforced. Periodic boundary conditions are
enforced on the remaining parts of the boundary.
The rheological parameters considered correspond to blood at a hematocrit
value of 45\% and are taken from~\citet{Kwon2008}. Physiological parameters for a
Carreau model ($a = 2\,\kappa = 1$) are summarized in Tab.~\ref{tab:blood_parameters}.
These parameters lie in the range yielding optimal convergence rates down to practically relevant tolerances
according to the results presented in Sec.~\ref{sec:numerical_examples_man_sol}.
\begin{table}[h!]
	\centering
	\caption{Blood parameters corresponding to 45\% hematocrit for the Carreau
		model~\eqref{eqn:viscosity_general}, taken from~\cite{Kwon2008}.}
	{
		\scriptsize
		\label{tab:blood_parameters}
		\begin{tabular}{||c | c | c | c | c | c | c ||}
			\hline
			&&&&&&\\[-1.25ex]
			$\rho$~[kg/m$^3$]
			& $\eta_\infty$~[m$^2$/s]
			& $\eta_0$~[m$^2$/s]
			& $\kappa$~[-]
			& $a$~[-]
			& $n$~[-]
			& $\lambda$~[s]
			\\[1.25ex]
			\hline\hline
			&&&&&&
			\\[-1.25ex]
			1050 & $3.29\times10^{-6}$ & $153.33\times10^{-6}$ & 2 & 1 & 0.479 & 39.41 \\[1.5ex]
			\hline
		\end{tabular}
	}
\end{table}

\begin{figure}[h!]
	\centering
	\includegraphics[width=1.0\linewidth, draft=\draftSec]{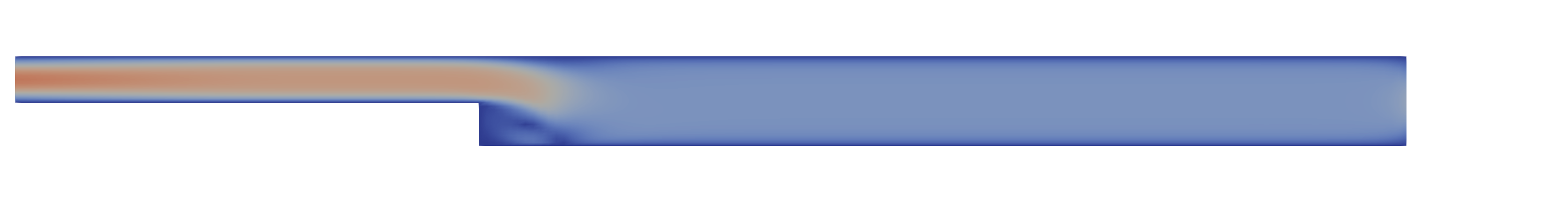}
	\vspace{-15mm}

	\includegraphics[width=1.0\linewidth, draft=\draftSec]{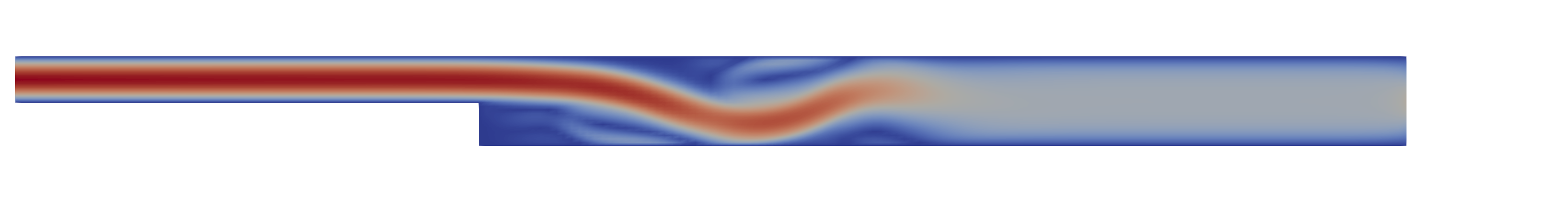}
	\vspace{-15mm}

	\includegraphics[width=1.0\linewidth, draft=\draftSec]{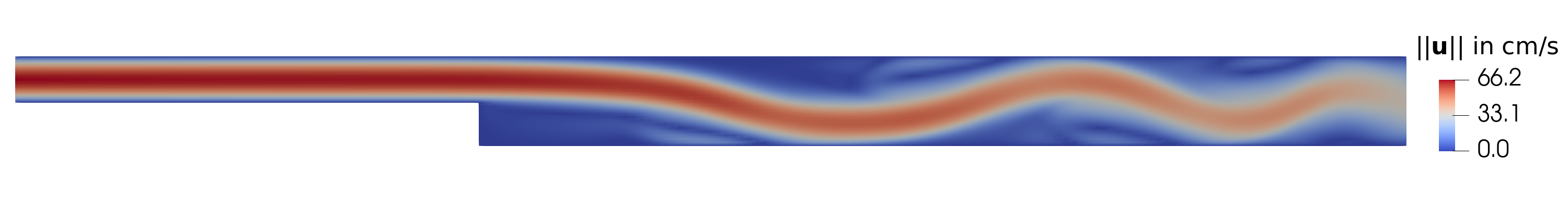}
	\vspace{-15mm}

	\includegraphics[width=1.0\linewidth, draft=\draftSec]{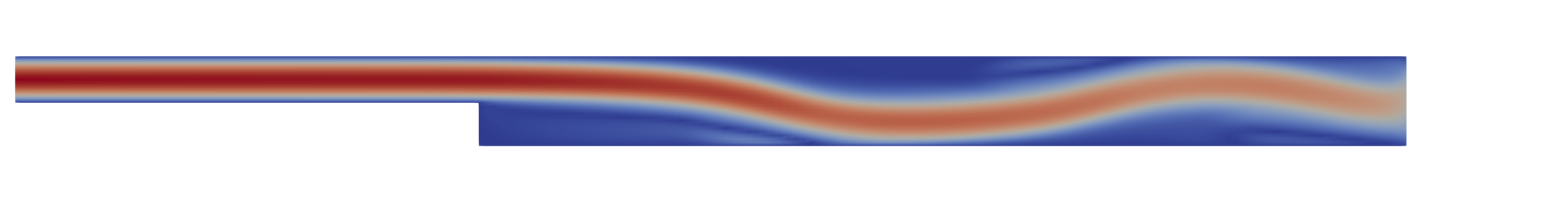}
	\vspace{-15mm}

	\includegraphics[width=1.0\linewidth, draft=\draftSec]{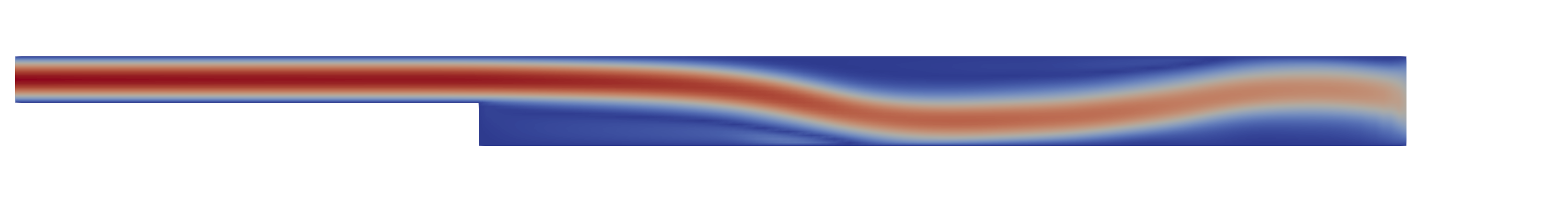}
	\caption{Velocity magnitude $\ve{u}$ at $t = 0.2, 0.4, 0.6, 0.8$ and $1.0$~s for $\mathrm{Re}=300$ in
		the backward-facing step benchmark with smoothly ramped inlet velocity.
		Legend applies to all solution snapshots.}
	\label{fig:backward_facing_step_snapshots}
\end{figure}

Given these parameters, a smooth behavior is observed during the ramp-up phase of the inflow profile as
shown in Fig.~\ref{fig:backward_facing_step_snapshots} for $t\in(0,1]$.
To reduce computational costs, the time interval considered in the numerical experiments
presented within this section is chosen as $t\in(0,0.3]$,
where the solution is not yet fully developed,
but iteration counts are in a similar range as when considering $t\in(0,1]$.
The nonlinear solver tolerance, if applicable, is set to $\epsilon_\mathrm{nl}^\mathrm{rel} = 10^{-3}$ and
$\epsilon_\mathrm{nl}^\mathrm{abs} = 10^{-12}$.
The linear solver tolerance targeted is $\epsilon_\mathrm{lin}^\mathrm{abs} = 10^{-12}$ and
$\epsilon_\mathrm{lin}^\mathrm{rel} = 10^{-3}$ for linear problems.
For linear problems within a nonlinear solver, we set
$\epsilon_\mathrm{lin}^\mathrm{abs} = \iota\times10^{-12}$ and
$\epsilon_\mathrm{lin}^\mathrm{rel} = \iota\times10^{-1}$,
choosing $\iota \in \{1, 0.5, \dots, 10^{-3}\}$
such that the time to solution is minimal for each solver variant individually.

\subsubsection{\previouslyrevised{Comparing linearization variants}}

To compare the linearization and time integration schemes, we aim to obtain the
optimal Courant number $\mathrm{Cr}_\mathrm{opt}$ for each variant such that the overall wall time is minimized.
$\mathrm{Cr}_\mathrm{opt}$ is obtained up to a
tolerance of $0.25$ by selecting $\mathrm{Cr} \in \{0.25,0.5,\dots,10\}$.
The Courant number enters the time step selection procedure together with the
polynomial degree of the velocity $k_u$, the velocity field $\ve{u}_h$ and the element size $h_e$.
A BDF-2 scheme with time step selection according to an element-local CFL condition~\cite{Karniadakis2013}
\previouslyrevised{
based on~\eqref{eqn:courant_definition} reads
\begin{align}
	\Delta t
	=
	\min_{e = 1,\dots,N_e}
	\{
		\Delta t_e
	\}
	,
	\quad
	\text{where}
	\quad
	\Delta t_e
	\coloneqq
	\min_{q=1, \dots, N_q}
	\frac{\mathrm{Cr}}{(k_u)^{r}}
	\left.\frac{h_e}{\|\ve{u}_h\|}\right\rvert_{q,e}
	,
	\label{eqn:CFL_timestep_selection}
\end{align}
with $\mathrm{Cr}$ being the target value. Therein, we again choose $r=1.5$
and compute the velocity-to-mesh-size ratio as
$\|\ve{u}_h\|\,h_e^{-1} = \|\te{J}^{-1}\,\ve{u}^h\|_{q,e}$,
evaluated at the $q=1, \dots, N_q$ quadrature points of element $e$, with
$\te{J}$ denoting the Jacobian of the finite element mapping.
}

The spatial resolutions with velocity degrees $k_u=2,3,4$
are selected such that the processor caches are saturated for all methods, which was verified for $\approx15\times10^6$ DoFs.
This leads to the hexahedra-based discretizations
as detailed in Tab.~\ref{tab:backward_facing_step_resolutions}, which lie within the practically relevant range
for laminar flows.
The cell count and resulting number of
DoFs are adapted by extending the computational domain in the direction perpendicular to the step,
since the solution is constant in this direction.

Due to the re-entrant corner, the solution is
non-smooth on the lip of the backward-facing step such that higher-order convergence rates can in general not be expected.
Nonetheless, we consider higher-order approximation in the current experiment for the following reasons:
\begin{enumerate}[label=\roman*)]
	\item Higher-order discretizations can still pay off in practical scenarios even in the pre-asymptotic range
	with non-smooth solutions as the current one if they deliver a similar DoF-to-error ratio
	(see, e.g., \cite{Fehn2018}).
	\item Contrary to matrix-based methods, matrix-free methods typically deliver higher
	throughput per DoF for moderately high polynomial degrees (e.g., $k_u = 2,\dots,8$),
	motivating their use from a performance perspective.
\end{enumerate}

\begin{table}[h!]
	\centering
	\caption{DoF counts as obtained after spatial discretization
		of the backward facing step benchmark geometry with varying cell counts to reach a similar
		number of DoFs for various velocity and pressure polynomial degrees, $k_u$ and $k_p$.}
	{
		\scriptsize
		\label{tab:backward_facing_step_resolutions}
		\begin{tabular}{||c | c | c | c | c | c ||}
			\hline
			&&&&&\\[-1.25ex]
			$k_u$
			& $k_p$
			& Number of cells
			& Velocity DoFs
			& Pressure DoFs
			& Total DoFs
			\\[1.25ex]
			\hline\hline
			&&&&&
			\\[-1.25ex]
			2 & 1 &  $166.40\times10^4$ & $13.48\times10^6$ & $1.33\times10^6$ & $14.81\times10^6$\\
			3 & 2 & \s$71.68\times10^4$ & $13.76\times10^6$ & $1.57\times10^6$ & $15.70\times10^6$\\
			4 & 3 & \s$32.64\times10^4$ & $12.24\times10^6$ & $2.09\times10^6$ & $14.33\times10^6$
			\\[1.5ex]
			\hline
		\end{tabular}
	}
\end{table}

A simulation is considered diverged, if at any point any linear or nonlinear solver requires
more than 100 iterations to achieve the required tolerance or the simulation takes more than 3 hours overall.
The resulting best performing runs are summarized in Tab.~\ref{tab:backward_facing_step_comparison}, where the following observations are made:
\begin{enumerate}[label=\roman*)]
	\item The individual linearization variants of the coupled and splitting schemes
	achieve a similar maximum stable Courant number $\mathrm{Cr}_\mathrm{max}$ for
	the most part, while some differing
	trends are also observed. Similar observations hold for $\mathrm{Cr}_\mathrm{opt}$.
	\item Cr$_\mathrm{opt}$ is not necessarily equal Cr$_\mathrm{max}$.
	Good initial guesses for (non-)linear solvers ease solving the (non-)linear systems and
	conditional stability of the linearized schemes hinders increasing Cr beyond specific limits
	for each variant. An optimally performing Cr$_\mathrm{opt}$ is typically found somewhere in between.
	\item Increasing the polynomial order from 2 to 3, the time to solution always increases,
	while for the higher orders 3 and 4, there is no uniform trend, with the solution time
	even decreasing in some cases. This is related to slightly smaller problem size,
	potentially richer $hp$-multigrid preconditioner, differing spatial discretizations
	and increased throughput in operator application for
	increased polynomial degrees.
	\item Explicit treatment of the convective term significantly restricts Cr$_\mathrm{max}$,
	rendering these variants less robust. However, such cases in fact allow for cheaper preconditioners
	with lower costs per iteration, potentially reaching similar or
	even lower overall time-to-solution despite a larger number of time steps.
\end{enumerate}

\begin{table}[h!]
	\centering
	\caption{Optimal Courant number Cr$_\mathrm{opt}$ yielding shortest time to solution,
		maximum Courant number Cr$_\mathrm{max}$ still converging,
		throughput per time step and time to solution reached
		for solver and linearization variants adopting
		various strategies of (semi-)implicit time integration of convective and viscous terms.
		The $\times$ symbol \previouslyrevised{indicates} that with the respective settings,
		none of the runs converged fast enough.
		}
	{
		\scriptsize
		\label{tab:backward_facing_step_comparison}
		\begin{tabular}{||c || c | c | c | c | c || c | c | c || c | c | c || c | c | c ||}
			\hline
			&\multicolumn{3}{|c|}{\,}
			&\multicolumn{2}{|c||}{\,}
			&\multicolumn{3}{|c||}{\,}
			&\multicolumn{3}{|c||}{\,}
			&\multicolumn{3}{|c||}{\,}
			\\[-1.25ex]
			\multirow{4}{*}{\vspace*{-0mm}\rotatebox{90}{Solver}}
			&\multicolumn{3}{|c|}{convective term}
			&\multicolumn{2}{|c||}{viscous term}
			&\multicolumn{3}{|c||}{$\mathrm{Cr}_\mathrm{opt}$ ($\mathrm{Cr}_\mathrm{max}$) identified}
			&\multicolumn{3}{|c||}{Throughput}
			&\multicolumn{3}{|c||}{Time to solution}
			\\
			\cline{2-6}
			&&&&&&
			\multicolumn{3}{|c||}{{}}
			&
			\multicolumn{3}{|c||}{{}}
			&
			\multicolumn{3}{|c||}{{}}
			\\[-2ex]
			&\multirow{1}{*}{Impl.}
			&\multirow{1}{*}{Lin.}
			&\multirow{1}{*}{Expl.}
			&\multirow{1}{*}{Impl.}
			&\multirow{1}{*}{Lin.}
			&\multicolumn{3}{|c||}{[-]}
			&\multicolumn{3}{|c||}{[$10^5$~DoFs/s/core/time step]}
			&\multicolumn{3}{|c||}{[s]}
			\\
			\cline{7-15}
			&&&&&&&&&&&&&&
			\\[-1.75ex]
			&&impl.&&&impl.
			&$k_u=2$&$k_u=3$&$k_u=4$
			&$k_u=2$&$k_u=3$&$k_u=4$
			&$k_u=2$&$k_u=3$&$k_u=4$
			\\
			\hline\hline
			&&&&&&&&&&&&&&
			\\[-1.25ex]
			\multirow{6}{*}{\vspace*{-8mm}\rotatebox{90}{\hspace*{-1.8cm}coupled approach}}
			& $\bullet$ & &
			& $\bullet$ &
			& 1.75
			& 1.75
			& 1.50
			& 1.24
			& 1.28
			& 1.05
			& \s584
			& \s981
			& 1240
			\\
			&&&&&&($\geq10$)&($7.75$)&($5.00$)&&&&&&
			\\[1.5ex]
			&  & $\bullet$ &
			& $\bullet$ &
			& $4.75$
			& $8.50$
			& $7.00$
			& 0.55
			& 0.18
			& 0.21
			& \s507
			& 1570
			& 1380
			\\
			&&&&&&($5.50$)&($9.75$)&($7.00$)&&&&&&
			\\[1.5ex]
			&  &  & $\bullet$
			& $\bullet$ &
			& 0.25
			& 1.00
			& $\times$
			& 1.72
			& 1.62
			& $\times$
			& 2850
			& 2790
			& $\times$
			\\
			&&&&&&(0.25)&(1.00)&($\times$)&&&&&&
			\\[1.5ex]
			\cline{2-15}
			&&&&&&&&&&&&&&
			\\[-0.75ex]
			& $\bullet$ &  &
			&  & $\bullet$
			& 1.25
			& 1.25
			& 1.25
			& 1.34
			& 1.51
			& 0.94
			& \s749
			& 1160
			& 1640
			\\
			&&&&&&($4.50$)&($4.25$)&($2.75$)&&&&&&
			\\[1.5ex]
			&  & $\bullet$ &
			&  & $\bullet$
			& 4.50
			& 1.75
			& 2.25
			& 0.85
			& 1.65
			& 0.86
			& \s347
			& \s759
			& 1020
			\\
			&&&&&&($5.25$)&($5.00$)&($4.00$)&&&&&&
			\\[1.5ex]
			&  &  & $\bullet$
			&  & $\bullet$
			& 0.25
			& 0.50
			& 0.25
			& 3.88
			& 3.21
			& 2.39
			& 1260
			& 1460
			& 3190
			\\
			&&&&&&(0.25)&(0.50)&(0.25)&&&&&&
			\\[0.75ex]
			\hline\hline
			&&&&&&&&&&&&&&
			\\[-0.75ex]
			\multirow{6}{*}{\rotatebox{90}{\hspace*{-2.4cm}splitting/projection}}
			& $\bullet$ &  &
			& $\bullet$ &
			& 1.25
			& 1.75
			& 1.25
			& 2.23
			& 1.73
			& 1.43
			& \s449
			& \s722
			& 1060
			\\
			&&&&&&($\geq 10$)&($8.75$)&($5.75$)&&&&&&
			\\[1.5ex]
			&  & $\bullet$ &
			& $\bullet$ &
			& 5.00
			& 10.0
			& 9.25
			& 0.69
			& 0.31
			& 0.35
			& \s389
			& 783
			& 652
			\\
			&&&&&&($5.50$)&($\geq10$)&($9.25$)&&&&&&
			\\[1.5ex]
			&  &  & $\bullet$
			& $\bullet$ &
			& 0.25
			& 0.50
			& 0.25
			& 3.47
			& 2.65
			& 1.82
			& 1410
			& 1680
			& 4180
			\\
			&&&&&&(0.25)&(1.00)&(0.25)&&&&&&
			\\[1.5ex]
			\cline{2-15}
			&&&&&&&&&&&&&&
			\\[-0.75ex]
			& $\bullet$ &  &
			&  & $\bullet$
			& 1.50
			& 1.50
			& 1.25
			& 2.16
			& 2.25
			& 1.56
			& \s385
			& \s645
			& \s993
			\\
			&&&&&&($\geq10$)&($9.25$)&($5.25$)&&&&&&
			\\[1.5ex]
			&  & $\bullet$ &
			&  & $\bullet$
			& 4.75
			& 9.50
			& 9.75
			& 1.95
			& 0.70
			& 0.76
			& \s142
			& \s360
			& \s288
			\\
			&&&&&&($5.50$)&($\geq10$)&($9.75$)&&&&&&
			\\[1.5ex]
			&  &  & $\bullet$
			&  & $\bullet$
			& 0.25
			& 0.50
			& 0.25
			& 6.53
			& 5.03
			& 3.48
			& \s747
			& \s889
			& 2190
			\\
			&&&&&&(0.25)&(0.50)&(0.25)&&&&&&
			\\[0.75ex]
			\hline
		\end{tabular}
	}
\end{table}

Based on Tab.~\ref{tab:backward_facing_step_comparison},
the speed-up of the splitting scheme over the coupled formulation
is given in Tab.~\ref{tab:backward_facing_step_speedup}.
The relative speeds of each variant taking the fastest variant as baseline,
namely the splitting scheme with linearly implicit convective
and viscous terms, are given in
Tab.~\ref{tab:backward_facing_step_relative_speed}.
We observe that the coupled scheme always requires more time to solution independent of the
linearization variant chosen, and the speed-up of splitting over the coupled scheme
lies between 1.3 and 3.54, where, additionally,
the coupled scheme did not always converge fast enough.
Comparing to the fastest variant overall, the relative compute time lies between 1.0 and 29.44,
which indicates convergence issues using some linearizations and Courant numbers.
Leaving out the variants with explicit convection,
we obtain a range between 1.0 and 11.55---still a significant difference.

Restricting these comparisons to the individual polynomial degrees used, we obtain values ranging from
1.0 to 20.07 for $k_u=2$, 1.0 to 7.75 for $k_u =3$, and 1.0 to 14.51 for $k_u=4$, where convergence problems
are observed for explicit convection.
Comparing to the fully implicit variant of the coupled and splitting solver,
the relative compute time ranges from 0.59 to 5.46 for the coupled solver and
0.32 to 9.31 for the splitting scheme, respectively.
These relative compute times observed keeping the polynomial degree fixed demonstrate that the linearization variants
impact the solve time for each polynomial degree, leading to a wide range for $k_u=2,3,4$, which also holds true
comparing variants for the coupled and splitting schemes separately.

\begin{table}[h!]
	\centering
	\caption{Speed-up of the splitting scheme over the coupled solver for various
		strategies of (semi-)implicit time integration of convective and viscous terms.
		The \previouslyrevised{$\times$ symbol indicates} that with the respective settings,
		the coupled solver did not converge fast enough.
	}
	{
		\scriptsize
		\label{tab:backward_facing_step_speedup}
		\begin{tabular}{|| c | c | c | c | c || c | c | c ||}
			\hline
			\multicolumn{3}{||c|}{\,}
			&\multicolumn{2}{|c||}{\,}
			&\multicolumn{3}{|c||}{\,}
			\\[-1.75ex]
			\multicolumn{3}{||c|}{convective term}
			&\multicolumn{2}{|c||}{viscous term}
			&\multicolumn{3}{|c||}{Speed-up splitting}
			\\
			\cline{0-4}
			&&&&&
			\multicolumn{3}{|c||}{{}}
			\\[-2ex]
			\multirow{1}{*}{Impl.}
			&\multirow{1}{*}{Lin.}
			&\multirow{1}{*}{Expl.}
			&\multirow{1}{*}{Impl.}
			&\multirow{1}{*}{Lin.}
			&\multicolumn{3}{|c||}{over coupled scheme~[-]}
			\\
			\cline{6-8}
			&&&&&&&
			\\[-1.75ex]
			&impl.&&&impl.
			&$k_u=2$&$k_u=3$&$k_u=4$
			\\
			\hline\hline
			&&&&&&&
			\\[-1.25ex]
			$\bullet$ & &
			& $\bullet$ &
			& 1.30
			& 1.36
			& 1.17
			\\[1.5ex]
			& $\bullet$ &
			& $\bullet$ &
			& 1.30
			& 2.00
			& 2.11
			\\[1.5ex]
			&  & $\bullet$
			& $\bullet$ &
			& 2.02
			& 1.66
			& $\times$
			\\[1.5ex]
			\cline{0-7}
			&&&&&&&
			\\[-0.75ex]
			$\bullet$ &  &
			&  & $\bullet$
			& 1.95
			& 1.80
			& 1.65
			\\[1.5ex]
			& $\bullet$ &
			&  & $\bullet$
			& 2.44
			& 2.11
			& 3.54
			\\[1.5ex]
			&  & $\bullet$
			&  & $\bullet$
			& 1.69
			& 1.64
			& 1.46
			\\[0.75ex]
			\hline
		\end{tabular}
	}
\end{table}

%
%
\begin{table}
	\centering
	\caption{Relative time to solution for various
		strategies of (semi-)implicit time integration of convective and viscous terms.
		The $\times$ symbol \previouslyrevised{indicates} that with the respective settings,
		the coupled solver did not converge fast enough.
	}
	{
		\scriptsize
		\label{tab:backward_facing_step_relative_speed}
		\begin{tabular}{||c || c | c | c | c | c || c | c | c || c | c | c || c | c | c ||}
			\hline
			&\multicolumn{3}{|c|}{\,}
			&\multicolumn{2}{|c||}{\,}
			&\multicolumn{3}{|c||}{\,}
			&\multicolumn{3}{|c||}{\,}
			&\multicolumn{3}{|c||}{\,}
			\\[-1.25ex]
			\multirow{4}{*}{\vspace*{2mm}\rotatebox{90}{Solver}}
			&\multicolumn{3}{|c|}{convective term}
			&\multicolumn{2}{|c||}{viscous term}
			&\multicolumn{3}{|c||}{Rel. time to solution}
			&\multicolumn{3}{|c||}{Rel. time to solution}
			&\multicolumn{3}{|c||}{Rel. time to solution}
			\\
			\cline{2-6}
			&&&&&&
			\multicolumn{3}{|c||}{{}}
			&
			\multicolumn{3}{|c||}{{}}
			&
			\multicolumn{3}{|c||}{{}}
			\\[-2ex]
			&\multirow{1}{*}{Impl.}
			&\multirow{1}{*}{Lin.}
			&\multirow{1}{*}{Expl.}
			&\multirow{1}{*}{Impl.}
			&\multirow{1}{*}{Lin.}
			&\multicolumn{3}{|c||}{(overall) [-]}
			&\multicolumn{3}{|c||}{(per degree) [-]}
			&\multicolumn{3}{|c||}{(per scheme) [-]}
			\\
			\cline{7-15}
			&&&&&&&&&&&&&&
			\\[-1.25ex]
			&&impl.&&&impl.
			&$k_u=2$&$k_u=3$&$k_u=4$
			&$k_u=2$&$k_u=3$&$k_u=4$
			&$k_u=2$&$k_u=3$&$k_u=4$
			\\
			\hline\hline
			&&&&&&&&&&&&&&
			\\[-1.25ex]
			\multirow{6}{*}{\vspace*{-8mm}\rotatebox{90}{\hspace*{-0.3cm}coupled approach}}
			& $\bullet$ & &
			& $\bullet$ &
			& \s4.11
			& \s6.91
			& \s8.73
			& \s4.11
			& \s2.73
			& \s4.31
			& $\mathbf{1.00}$
			& 1.68
			& 2.12
			\\[1.5ex]
			&  & $\bullet$ &
			& $\bullet$ &
			& \s3.57
			& 11.06
			& \s9.72
			& \s3.57
			& \s4.36
			& \s4.79
			& 0.87
			& 2.67
			& 2.36
			\\[1.5ex]
			&  &  & $\bullet$
			& $\bullet$ &
			& 20.07
			& 19.64
			& $\times$
			& 20.07
			& \s7.75
			& $\times$
			& 4.88
			& 4.78
			& $\times$
			\\[1.5ex]
			\cline{2-15}
			&&&&&&&&&&&&&&
			\\[-0.75ex]
			& $\bullet$ &  &
			&  & $\bullet$
			& \s5.27
			& \s8.17
			& 11.55
			& \s5.27
			& \s3.22
			& \s5.69
			& 1.28
			& 1.99
			& 2.81
			\\[1.5ex]
			&  & $\bullet$ &
			&  & $\bullet$
			& \s2.44
			& \s5.35
			& \s7.18
			& \s2.44
			& \s2.11
			& \s3.54
			& 0.59
			& 1.30
			& 1.75
			\\[1.5ex]
			&  &  & $\bullet$
			&  & $\bullet$
			& \s8.87
			& 10.28
			& 22.46
			& \s8.87
			& \s4.06
			& 11.08
			& 2.16
			& 2.50
			& 5.46
			\\[0.75ex]
			\hline\hline
			&&&&&&&&&&&&&&
			\\[-0.75ex]
			\multirow{6}{*}{\rotatebox{90}{\hspace*{-0.6cm}splitting/projection}}
			& $\bullet$ &  &
			& $\bullet$ &
			& \s3.16
			& \s5.08
			& \s7.46
			& \s3.16
			& \s2.01
			& \s3.68
			& $\mathbf{1.00}$
			& 1.61
			& 2.36
			\\[1.5ex]
			&  & $\bullet$ &
			& $\bullet$ &
			& \s2.74
			& \s5.51
			& \s4.59
			& \s2.74
			& \s2.18
			& \s2.26
			& 0.87
			& 1.74
			& 1.45
			\\[1.5ex]
			&  &  & $\bullet$
			& $\bullet$ &
			& \s9.93
			& 11.83
			& 29.44
			& \s9.93
			& \s4.67
			& 14.51
			& 3.14
			& 3.74
			& 9.31
			\\[1.5ex]
			\cline{2-15}
			&&&&&&&&&&&&&&
			\\[-0.75ex]
			& $\bullet$ &  &
			&  & $\bullet$
			& \s2.71
			& \s4.54
			& \s6.99
			& \s2.71
			& \s1.79
			& \s3.45
			& 0.86
			& 1.44
			& 2.21
			\\[1.5ex]
			&  & $\bullet$ &
			&  & $\bullet$
			& \s$\mathbf{1.00}$
			& \s2.54
			& \s2.03
			& \s$\mathbf{1.00}$
			& \s$\mathbf{1.00}$
			& \s$\mathbf{1.00}$
			& 0.32
			& 0.80
			& 0.64
			\\[1.5ex]
			&  &  & $\bullet$
			&  & $\bullet$
			& \s5.26
			& \s6.26
			& 15.42
			& \s5.26
			& \s2.47
			& \s7.60
			& 1.66
			& 1.98
			& 4.88
			\\[0.75ex]
			\hline
		\end{tabular}
	}
\end{table}

\subsubsection{\previouslyrevised{Nonlinear and linear solver performance}}
\label{sec:nonlin_lin_solver_performance}

\begin{figure}
	\centering
	\begin{subfigure}{.245\textwidth}
		\centering
		\includegraphics[width=\linewidth, draft=\draftSec]{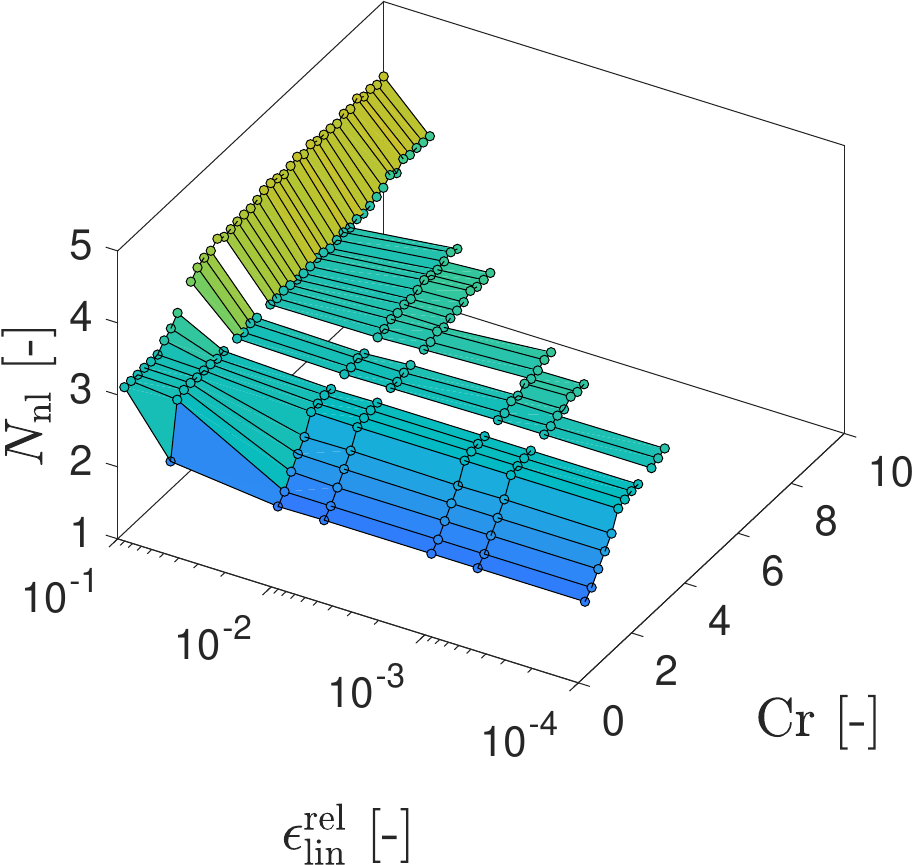}
		\caption{$\ve{u}_h^{n+1}\cdot\nabla\ve{u}_h^{n+1}$, $\nu_h = \nu_h^{n+1}$}
	\end{subfigure}
	\begin{subfigure}{.245\textwidth}
		\centering
		\includegraphics[width=\linewidth, draft=\draftSec]{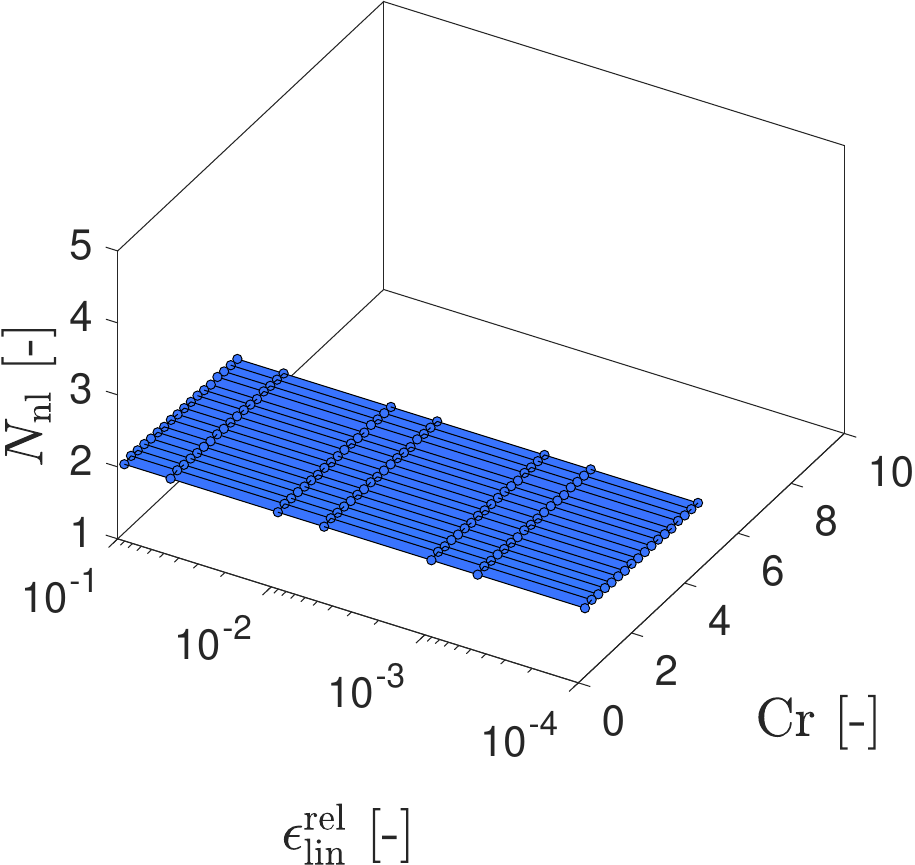}
		\caption{$\ve{u}_h^{n+1}\cdot\nabla\ve{u}_h^{n+1}$, $\nu_h = \nu_h^\#$}
	\end{subfigure}
	\begin{subfigure}{.245\textwidth}
		\centering
		\includegraphics[width=\linewidth, draft=\draftSec]{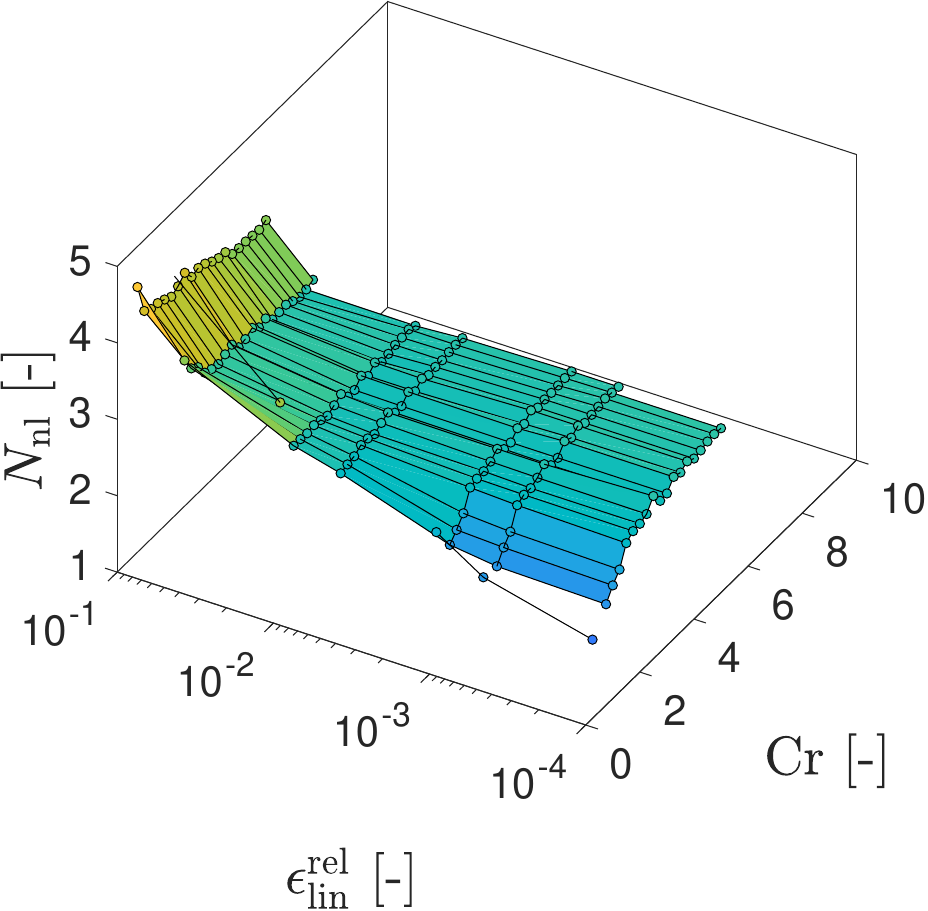}
		\caption{$\ve{u}_h^\#\cdot\nabla\ve{u}_h^{n+1}$, $\nu_h = \nu_h^{n+1}$}
	\end{subfigure}
	\begin{subfigure}{.24\textwidth}
		\centering
		\includegraphics[width=\linewidth, draft=\draftSec]{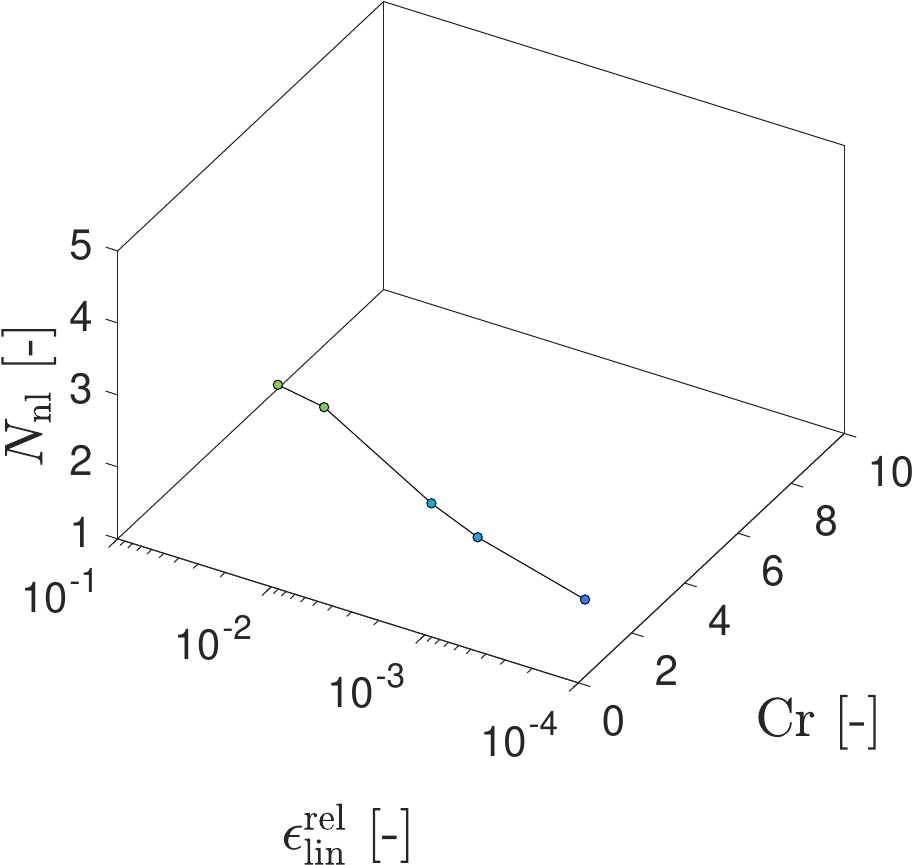}
		\caption{
			$\left[\convectiveterm{\ve{u}}{\ve{u}}\right]^\#$, $\nu_h = \nu_h^{n+1}$}
	\end{subfigure}
	\caption{Nonlinear solver iterations $N_\mathrm{nl}$ for various linearization variants
	of the coupled solution scheme with $k_u = 2$ over the target Courant number Cr and inner linear Krylov
	solver tolerance $\epsilon_\mathrm{lin}^\mathrm{rel}$
	(and accordingly scaled absolute inner Krylov solver tolerance). \previouslyrevised{Points omitted indicate solver divergence.}}
	\label{fig:backward_facing_step_nonlinear_solver_behavior_coupled}
\end{figure}

\begin{figure}
	\centering
	\begin{subfigure}{.245\textwidth}
		\centering
		\includegraphics[width=\linewidth, draft=\draftSec]{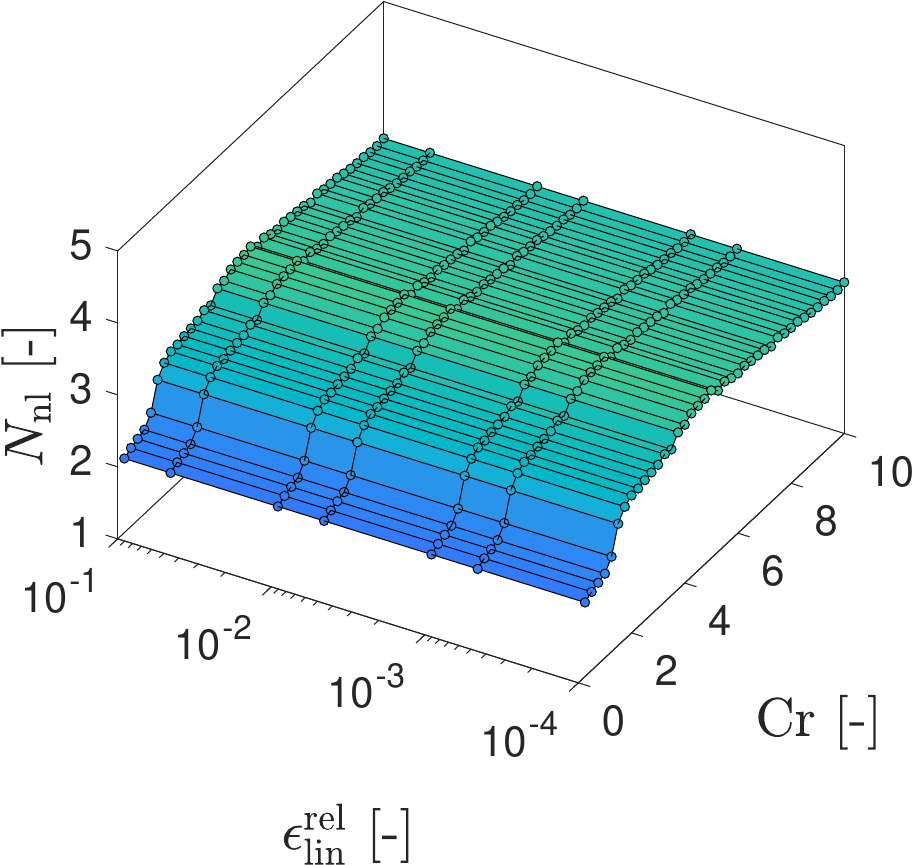}
		\caption{$\ve{u}_h^{n+1}\cdot\nabla\ve{u}_h^{n+1}$, $\nu_h = \nu_h^{n+1}$}
	\end{subfigure}
	\begin{subfigure}{.245\textwidth}
		\centering
		\includegraphics[width=\linewidth, draft=\draftSec]{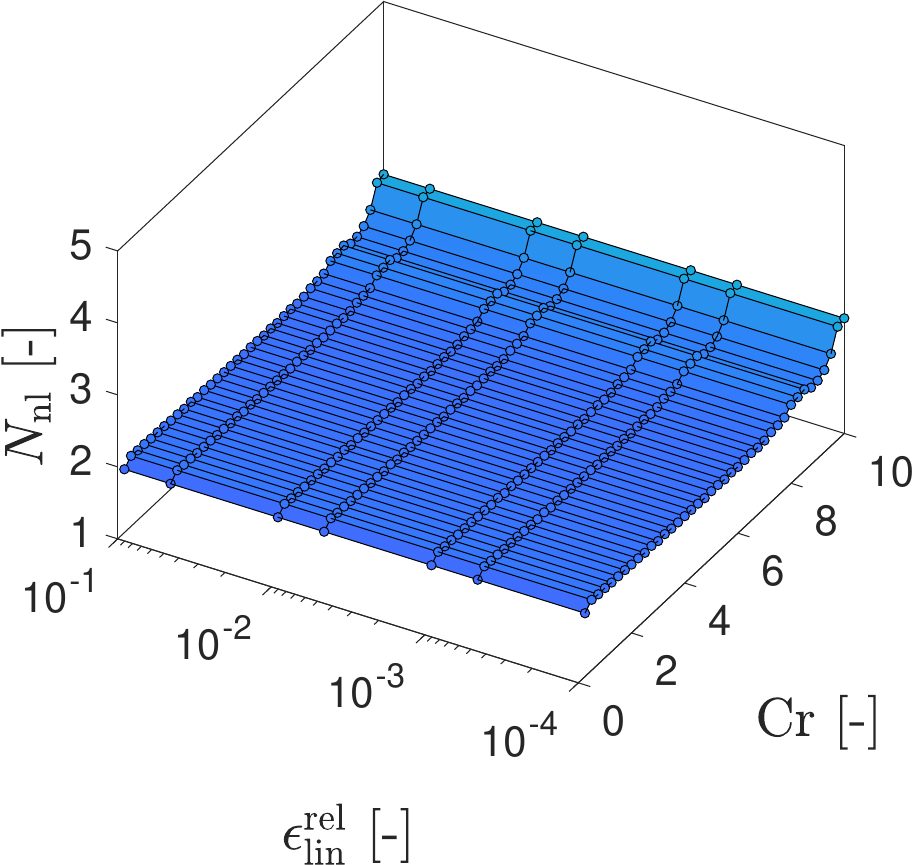}
		\caption{$\ve{u}_h^{n+1}\cdot\nabla\ve{u}_h^{n+1}$, $\nu_h = \nu_h^\#$}
	\end{subfigure}
	\begin{subfigure}{.245\textwidth}
		\centering
		\includegraphics[width=\linewidth, draft=\draftSec]{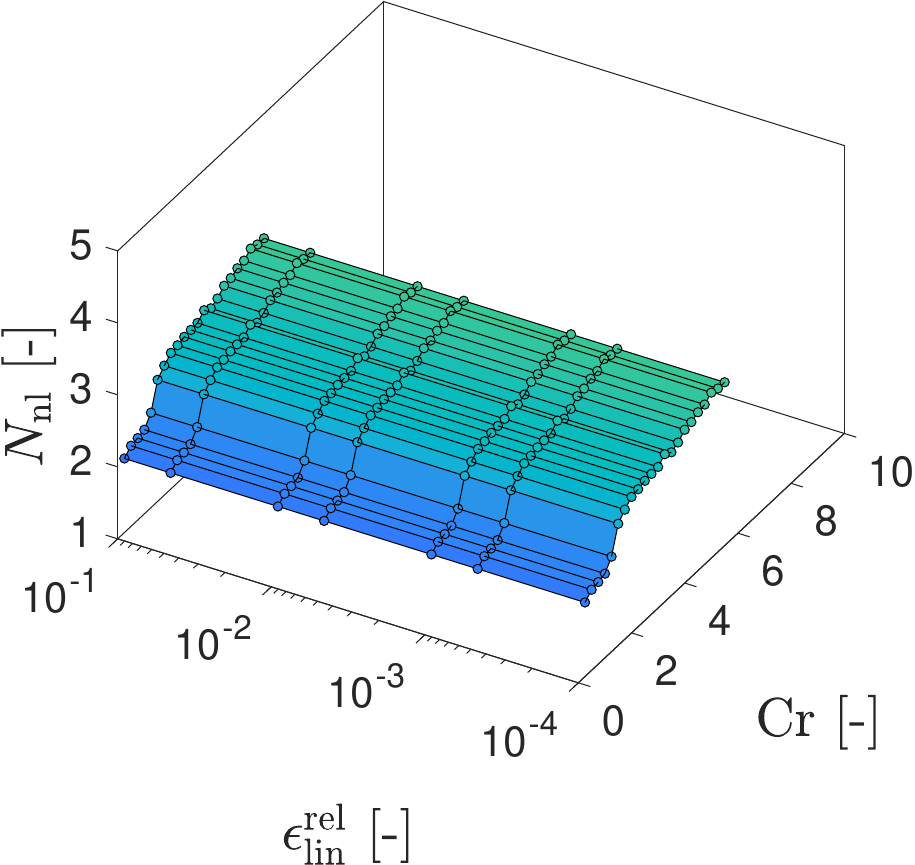}
		\caption{$\ve{u}_h^\#\cdot\nabla\ve{u}_h^{n+1}$, $\nu_h = \nu_h^{n+1}$}
	\end{subfigure}
	\begin{subfigure}{.24\textwidth}
		\centering
		\includegraphics[width=\linewidth, draft=\draftSec]{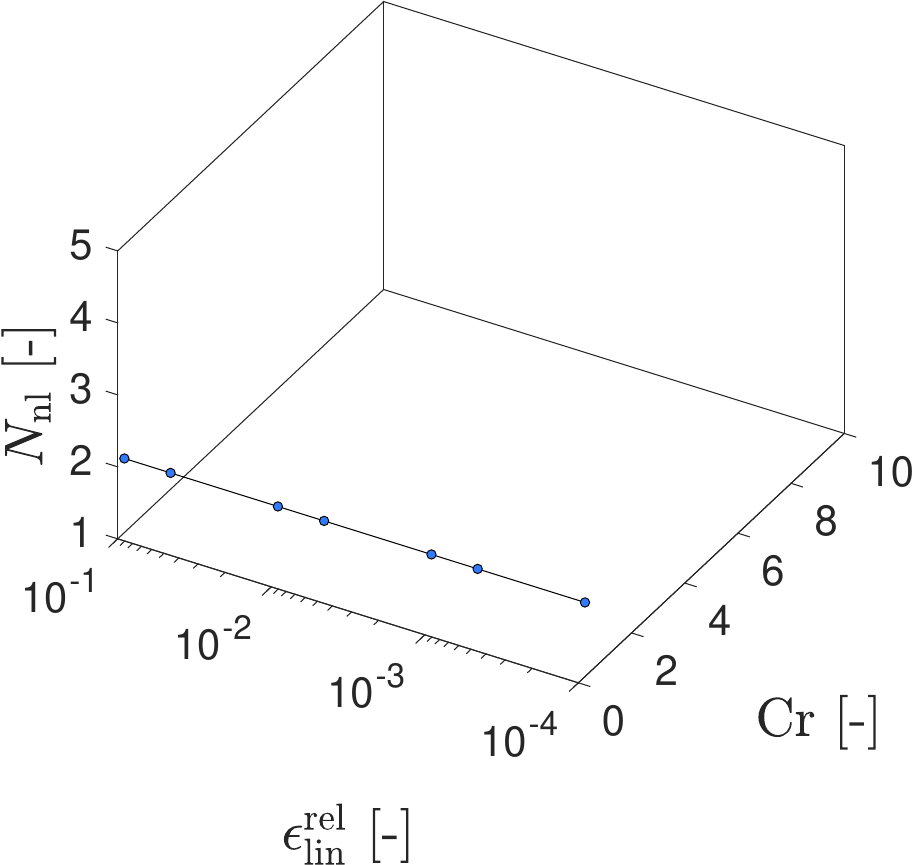}
		\caption{
			$\left[\convectiveterm{\ve{u}}{\ve{u}}\right]^\#$,
			$\nu_h = \nu_h^{n+1}$}
	\end{subfigure}
	\caption{Nonlinear solver iterations $N_\mathrm{nl}$ for various linearization variants
	of the projection scheme with $k_u = 2$  over the target Courant number Cr and inner linear Krylov
	solver tolerance $\epsilon_\mathrm{lin}^\mathrm{rel}$
	(and accordingly scaled absolute inner Krylov solver tolerance). \previouslyrevised{Points omitted indicate solver divergence.}}
	\label{fig:backward_facing_step_nonlinear_solver_behavior_splitting}
\end{figure}

Condensing the numerical experiments in
Tabs.~\ref{tab:backward_facing_step_comparison}--\ref{tab:backward_facing_step_relative_speed}
inevitably leaves out information regarding the (non-)linear solvers and their behavior.
To further address this, let us first investigate resulting iteration counts of
the nonlinear solvers for each nonlinear variant.
Figs.~\ref{fig:backward_facing_step_nonlinear_solver_behavior_coupled}
and~\ref{fig:backward_facing_step_nonlinear_solver_behavior_splitting}
depict results for the coupled and projection solver,
where all polynomial degrees $k_u=2,3,4$ show essentially similar trends.
As already observed in Tab.~\ref{tab:backward_facing_step_comparison},
higher polynomial degrees tend to require lower Cr numbers for stability.
For the coupled solver, using an implicit treatment of the viscosity increases the nonlinear iteration count,
if the inner linear solver tolerance is not chosen strict enough.
The convective term alone is not as sensitive to the inner solver tolerance, as seen from
Fig.~\ref{fig:backward_facing_step_nonlinear_solver_behavior_coupled}(b).
For the splitting solver, we do not observe such a dependency on the inner Krylov solver tolerance,
see Fig.~\ref{fig:backward_facing_step_nonlinear_solver_behavior_splitting}.
Both Figs.~\ref{fig:backward_facing_step_nonlinear_solver_behavior_coupled}
and~\ref{fig:backward_facing_step_nonlinear_solver_behavior_splitting} indicate
that the inner Krylov solver tolerance chosen from
$\epsilon_\mathrm{lin}^\mathrm{rel} \in \{\epsilon_\mathrm{nl},
\min\{ 0.1, 100 \times \epsilon_\mathrm{nl}^\mathrm{rel} \}\}$
and similarly chosen absolute tolerances
might not increase the nonlinear solver iteration counts excessively, with the splitting solver
being much more forgiving in this regard.
\previouslyrevised{
	In general, solving the linear systems within a nonlinear solver with lower accuracy
	renders the nonlinear solver inexact.
	Such a strategy is a known viable option, typically performing well for large nonlinear systems
	as is also reflected in the results in Fig.~\ref{fig:backward_facing_step_nonlinear_solver_behavior_coupled}.
}

\begin{figure}[h!]
	\centering
	\begin{subfigure}{.245\textwidth}
		\centering
		\includegraphics[width=\linewidth, draft=\draftSec]{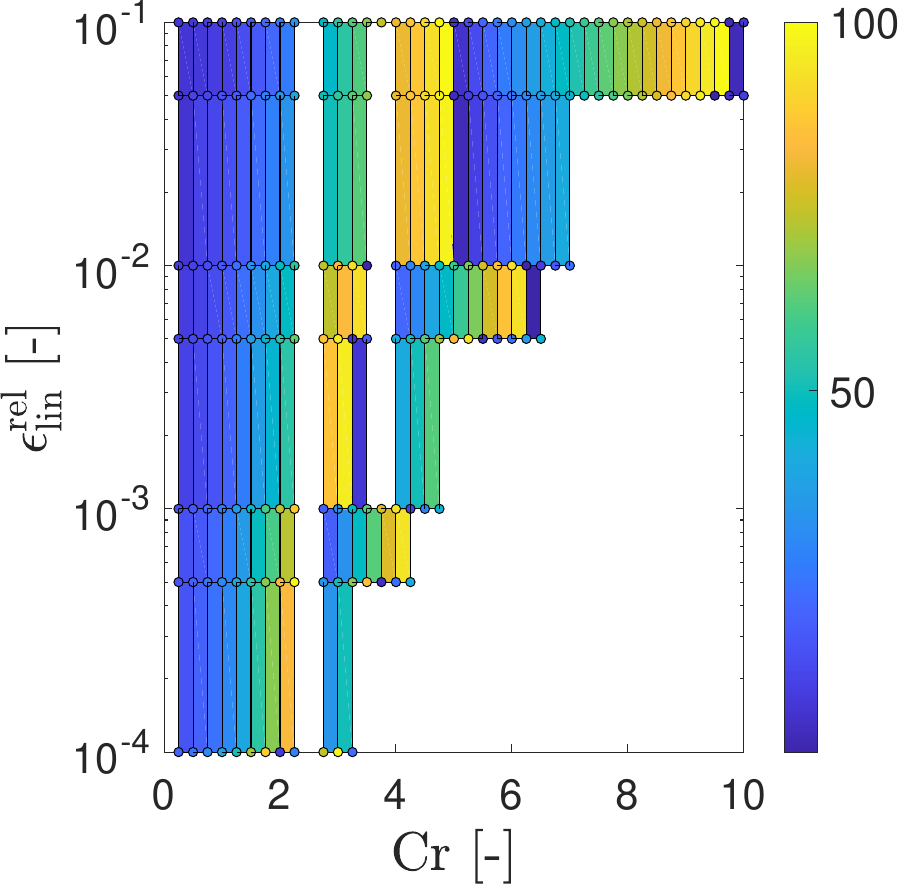}
		\caption{$\ve{u}_h^{n+1}\cdot\nabla\ve{u}_h^{n+1}$, $\nu_h = \nu_h^{n+1}$}
	\end{subfigure}
	\begin{subfigure}{.245\textwidth}
		\centering
		\includegraphics[width=\linewidth, draft=\draftSec]{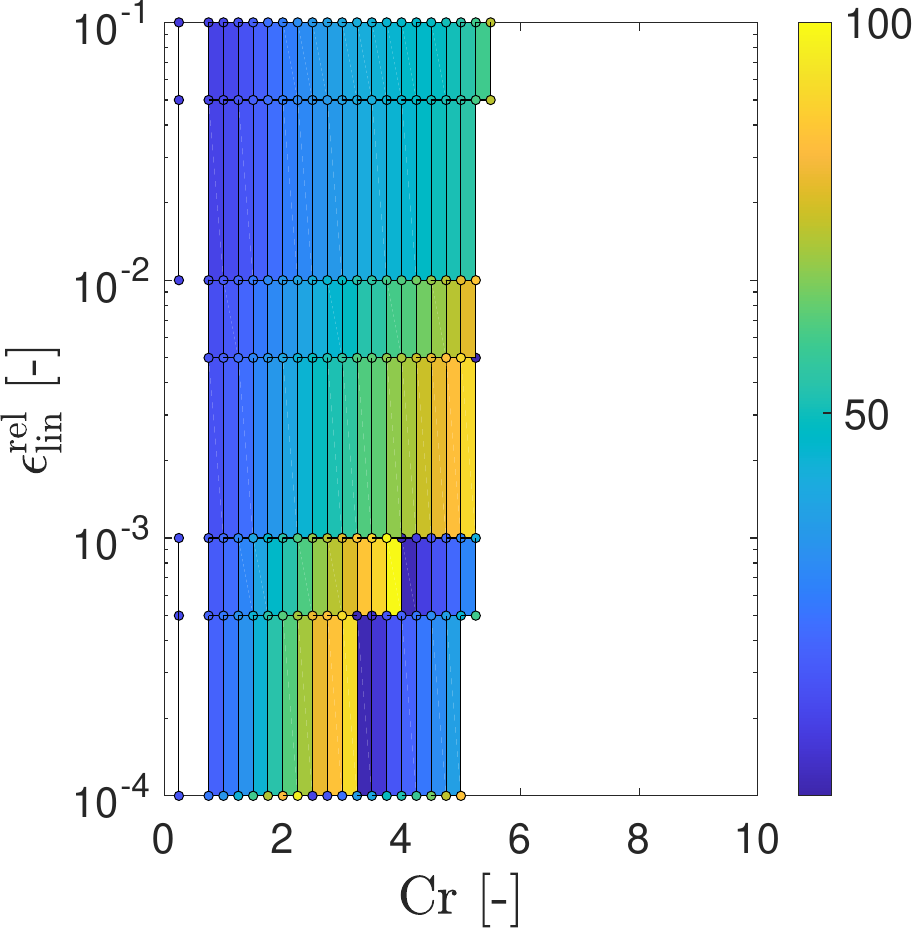}
		\caption{$\ve{u}_h^\#\cdot\nabla\ve{u}_h^{n+1}$, $\nu_h = \nu_h^{n+1}$}
	\end{subfigure}
	\begin{subfigure}{.245\textwidth}
		\centering
		\includegraphics[width=\linewidth, draft=\draftSec]{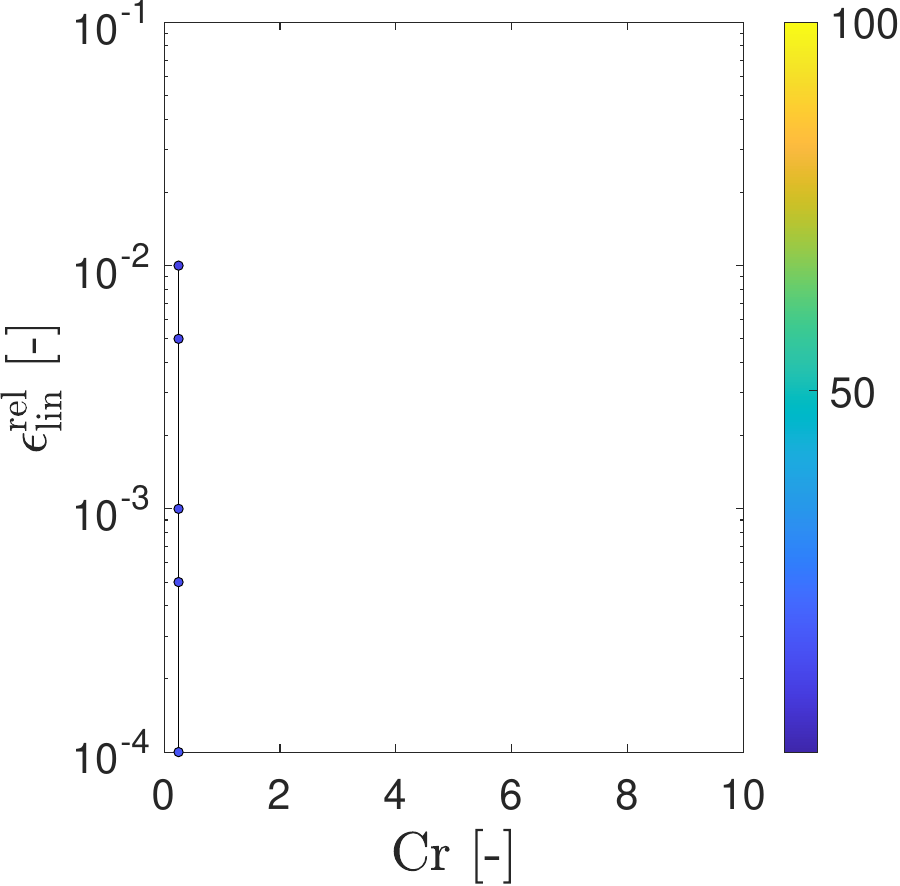}
		\caption{$\left[\convectiveterm{\ve{u}}{\ve{u}}\right]^\#$, $\nu_h = \nu_h^{n+1}$}
	\end{subfigure}
	\\
	\begin{subfigure}{.245\textwidth}
		\centering
		\includegraphics[width=\linewidth, draft=\draftSec]{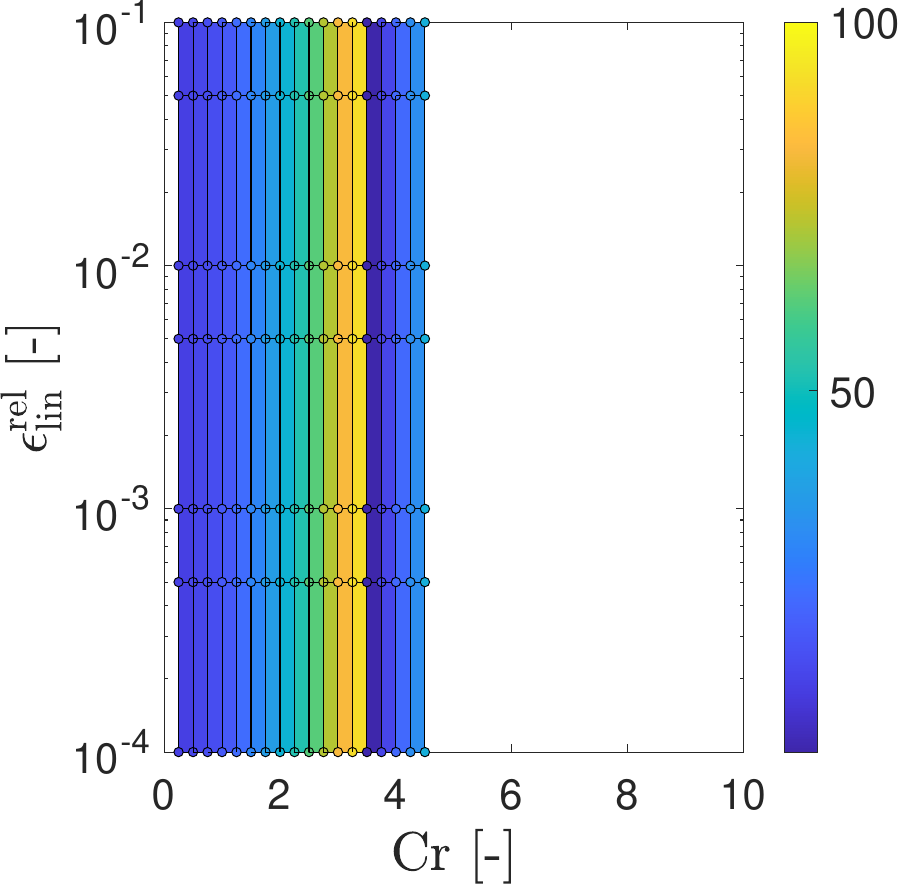}
		\caption{$\ve{u}_h^{n+1}\cdot\nabla\ve{u}_h^{n+1}$, $\nu_h = \nu_h^\#$}
	\end{subfigure}
	\begin{subfigure}{.245\textwidth}
		\centering
		\includegraphics[width=\linewidth, draft=\draftSec]{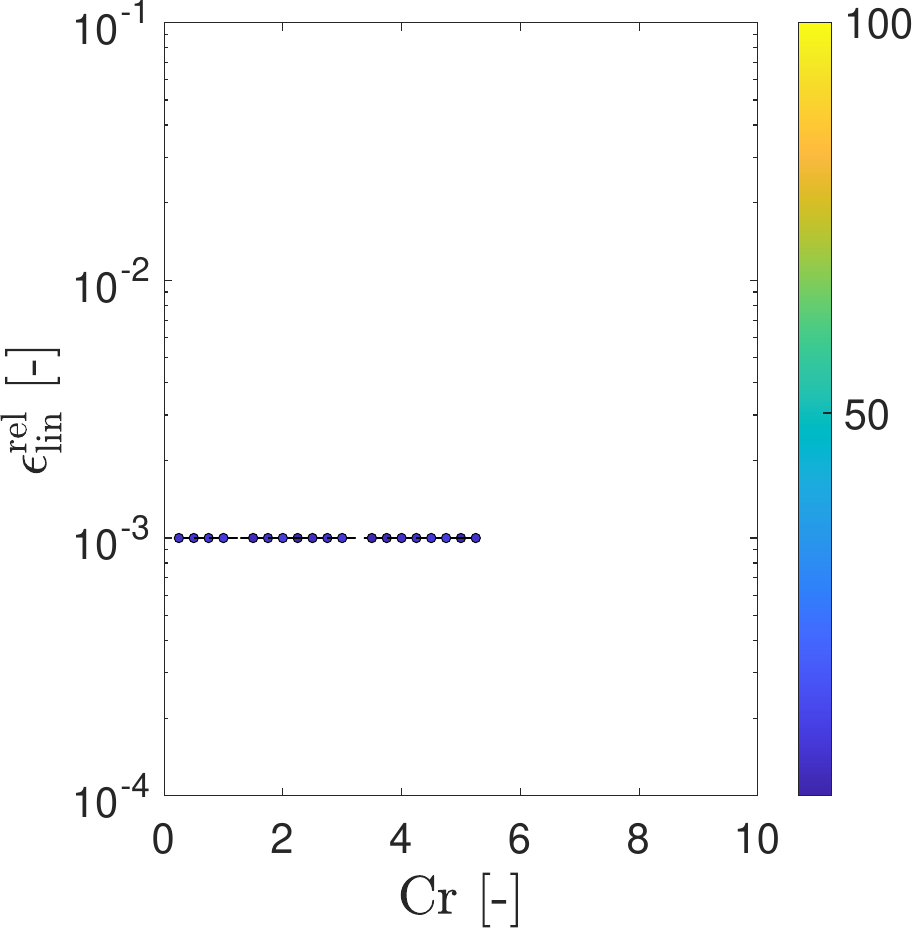}
		\caption{$\ve{u}_h^\#\cdot\nabla\ve{u}_h^{n+1}$, $\nu_h = \nu_h^\#$}
	\end{subfigure}
	\begin{subfigure}{.245\textwidth}
		\centering
		\includegraphics[width=\linewidth, draft=\draftSec]{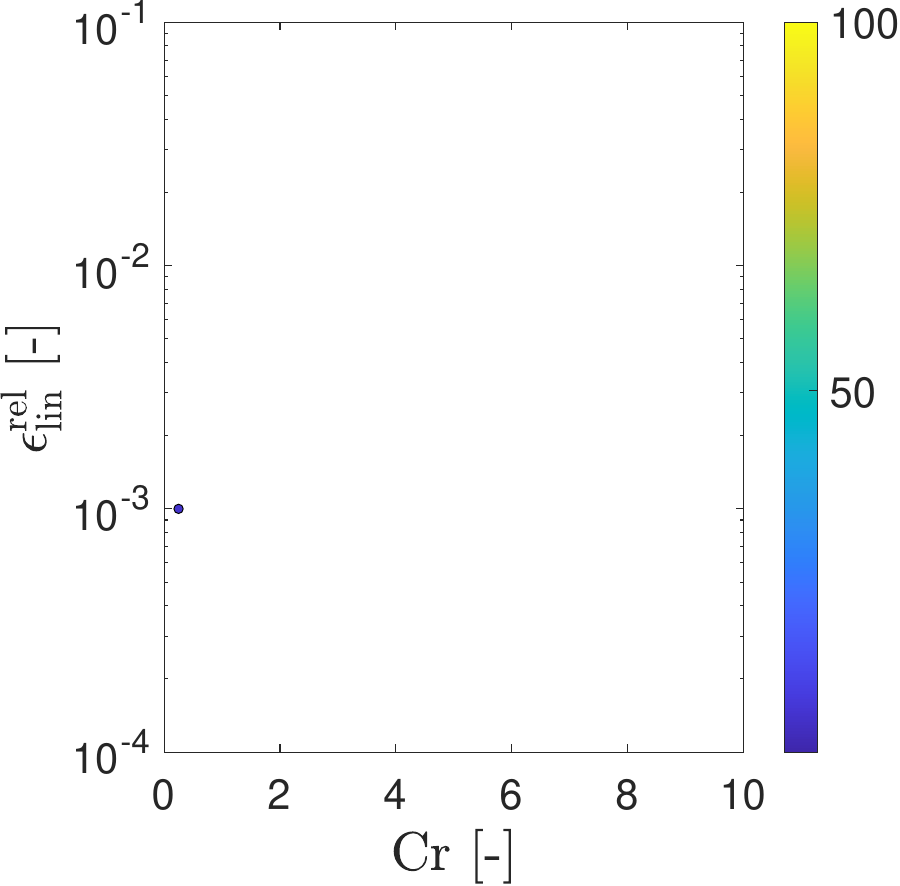}
		\caption{$\left[\convectiveterm{\ve{u}}{\ve{u}}\right]^\#$, $\nu_h = \nu_h^\#$}
	\end{subfigure}
	\caption{Linear solver iterations of the coupled system for various linearization variants
		of the coupled solution scheme with $k_u = 2$  over the target Courant number Cr and inner linear Krylov
		solver tolerance $\epsilon_\mathrm{lin}^\mathrm{rel}$
		(and accordingly scaled absolute inner Krylov solver tolerance).
		\textit{Linear} problems are solved with $\epsilon_\mathrm{lin}^\mathrm{rel}=10^{-3}$ only.
		\previouslyrevised{Points omitted indicate solver divergence, \revised{but linear schemes (e) and (f) consider only $\epsilon_\mathrm{lin}^\mathrm{rel} = 10^{-3}$.}}
		}
	\label{fig:backward_facing_step_linear_solver_behavior_coupled}
\end{figure}

The second aspect not visible when condensing the numerical parameter sweeps
into Tab.~\ref{tab:backward_facing_step_comparison} is the linear solver performance
under variation of the linear solver tolerance and targeted Courant number.
Varying the time step size and outer solver tolerance naturally impacts the approximation quality.
In the present tests, every solver took more than roughly 80 time steps to complete the interval.
Given the rather short time interval and the second-order time integration scheme,
all results are assumed acceptable if converged, while the fastest variants where verified by visual inspection.
For the presented data regarding the
convective (linearized) implicit coupled solver to be fully comparable to the splitting scheme,
reasonable linear iteration counts are required.
The average iteration counts per time step over all time steps in
Fig.~\ref{fig:backward_facing_step_linear_solver_behavior_coupled} suggest
that the Schur complement approximation performs well in some cases,
while for other parameter combinations, the approximation is not accurate enough
since it neglects the convective term completely.
We observe that for implicitly and linearly implicit treated convective term many Cr
and solver tolerances can be chosen, while an explicit convective term yields temporal instabilities
due to the range of Cr chosen as $[0.25,\dots,10]$.

For the variants treating the convective term at least linearly implicit, we observe
``waves'' of increasing iteration counts.
This might be due to various target Courant number resulting in better or worse initial
guesses to the instationary solution. The higher-order extrapolations taken as initial
guesses for the linear solvers might under-/overshoot depending on the target Courant number.
The linearization of the viscous term also impacts the maximum reachable Courant number.
Also here, the admissible time step range is reduced as for the
linearized implicit convective term.
A possible reason for this behavior lies in the initial guesses of the viscosity
significantly degrading in quality when $\mathrm{Cr} > 5$.
Note that all the \textit{linear} problems are solved with a tolerance of
$\epsilon_\mathrm{lin}^\mathrm{rel}=10^{-3}$ only. No parameter sweeps of
$\epsilon_\mathrm{lin}^\mathrm{rel}$ are executed for these cases.

Initial tests with the classical PCD preconditioner~\cite{Elman2014, Kay2003}
(designed for the constant viscosity case; no results shown in this work) have shown worse behavior than the
Cahouet--Chabard approach, which \textit{does} account for variable viscosity.
Preconditioning the Navier--Stokes equations accounting for variable viscosity and
medium to high Reynolds numbers remains challenging---which is exactly the motivation for
splitting schemes in the first place.
However, preconditioners for the coupled system might be subject to improvements in the future,
potentially impacting some of the conclusions drawn here.
Nonetheless, the present results demonstrate that an off-the-shelf
block-preconditioner is only to some degree competitive with the splitting scheme
and might suffer from decreased robustness with respect to the physical and numerical parameters.

\begin{figure}
	\centering
	\begin{subfigure}{.245\textwidth}
		\centering
		\includegraphics[width=\linewidth, draft=\draftSec]{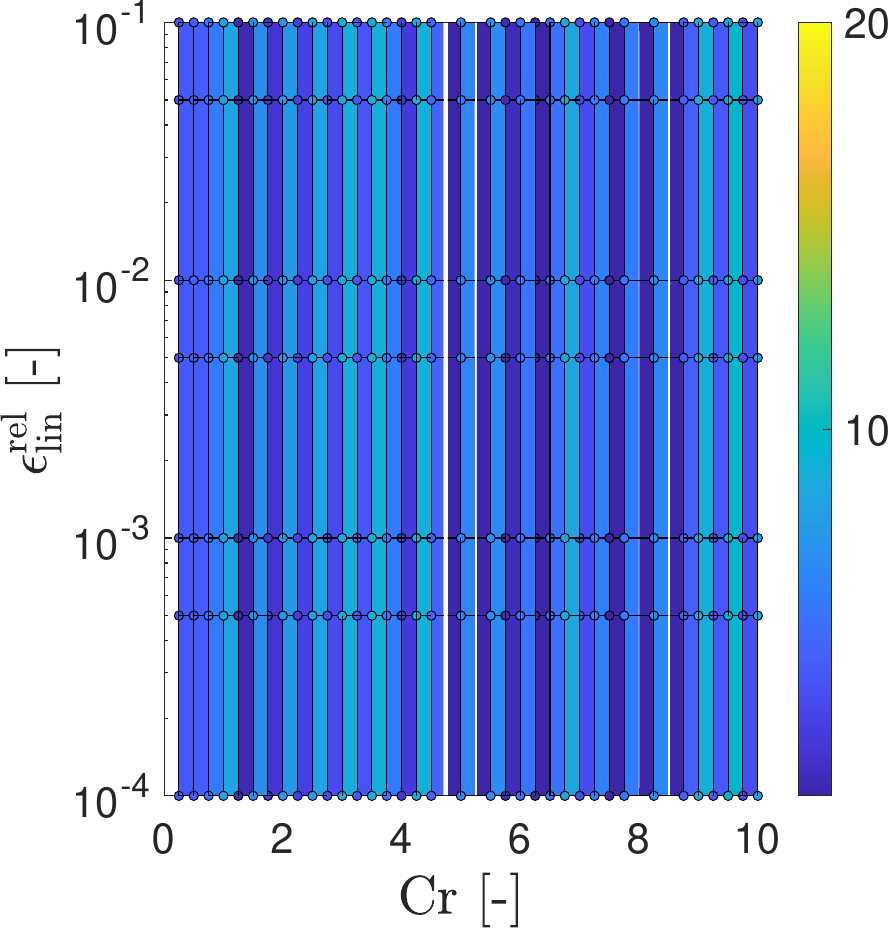}
		\caption{$\ve{u}_h^{n+1}\cdot\nabla\ve{u}_h^{n+1}$, $\nu_h = \nu_h^{n+1}$}
	\end{subfigure}
	\begin{subfigure}{.245\textwidth}
		\centering
		\includegraphics[width=\linewidth, draft=\draftSec]{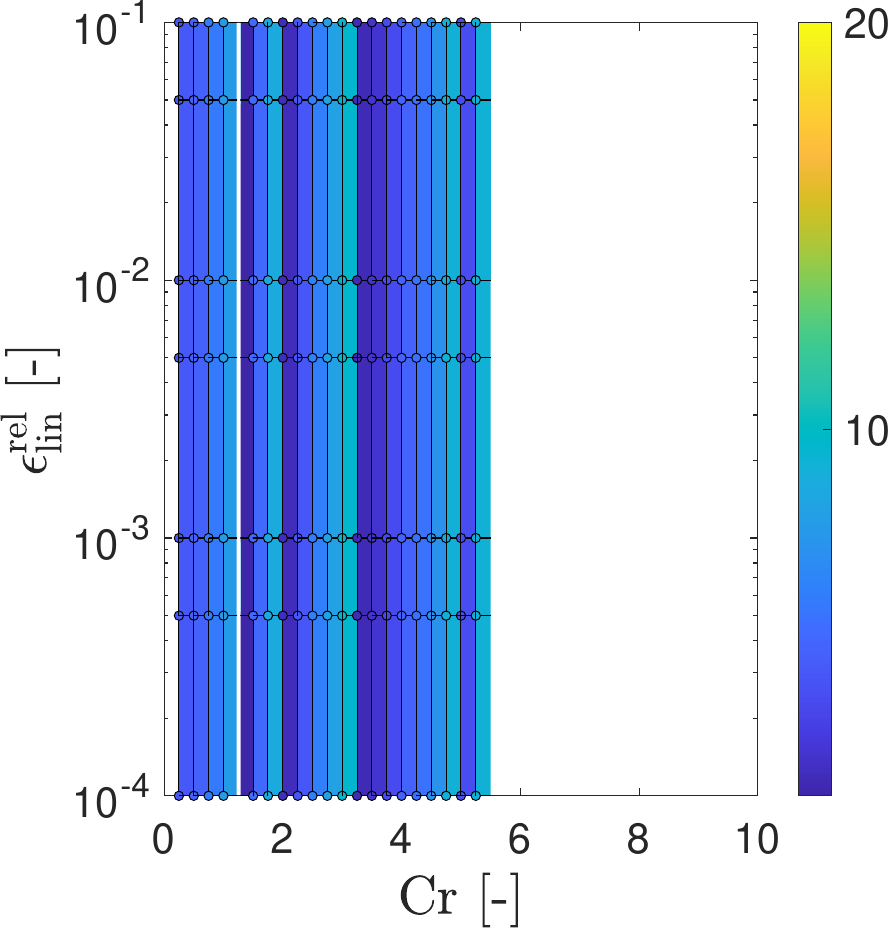}
		\caption{$\ve{u}_h^\#\cdot\nabla\ve{u}_h^{n+1}$, $\nu_h = \nu_h^{n+1}$}
	\end{subfigure}
	\begin{subfigure}{.245\textwidth}
		\centering
		\includegraphics[width=\linewidth, draft=\draftSec]{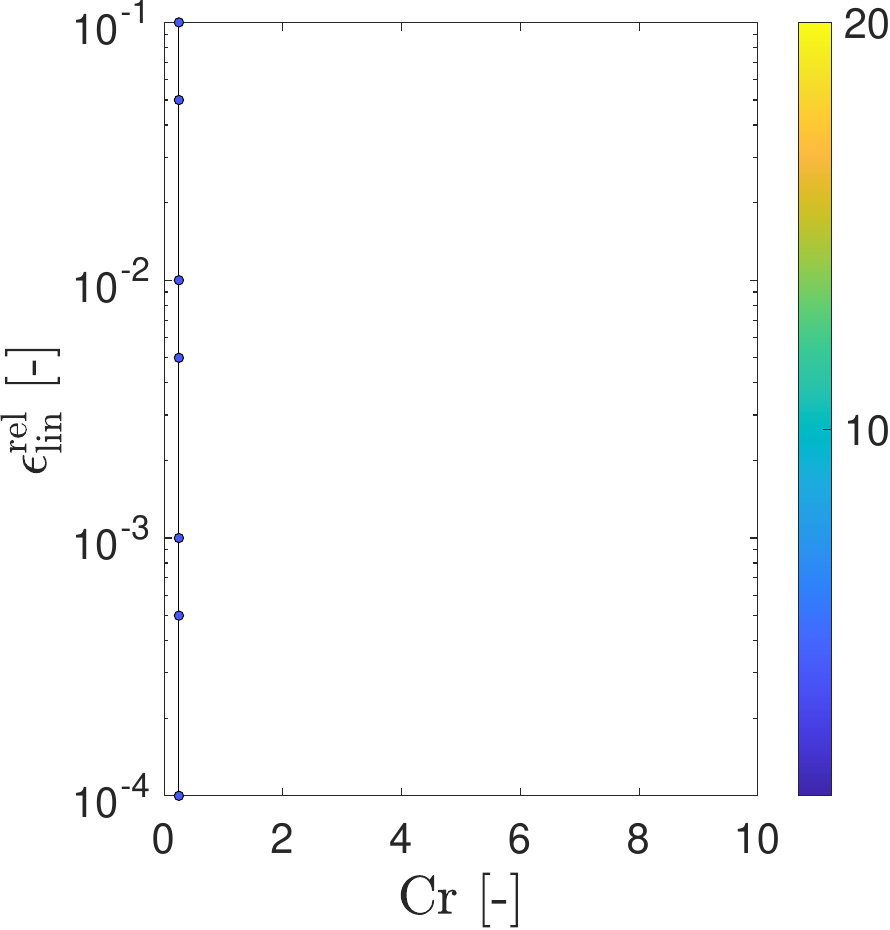}
		\caption{$\left[\convectiveterm{\ve{u}}{\ve{u}}\right]^\#$, $\nu_h = \nu_h^{n+1}$}
	\end{subfigure}
	\\
	\begin{subfigure}{.245\textwidth}
		\centering
		\includegraphics[width=\linewidth, draft=\draftSec]{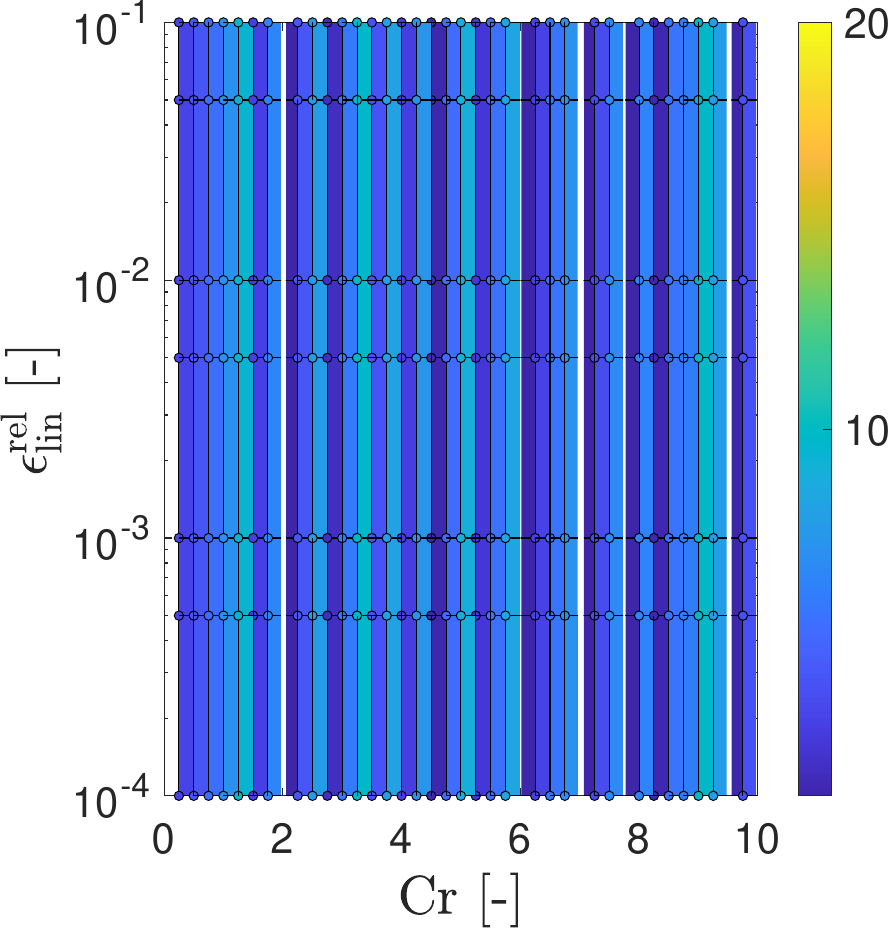}
		\caption{$\ve{u}_h^{n+1}\cdot\nabla\ve{u}_h^{n+1}$, $\nu_h = \nu_h^\#$}
	\end{subfigure}
	\begin{subfigure}{.245\textwidth}
		\centering
		\includegraphics[width=\linewidth, draft=\draftSec]{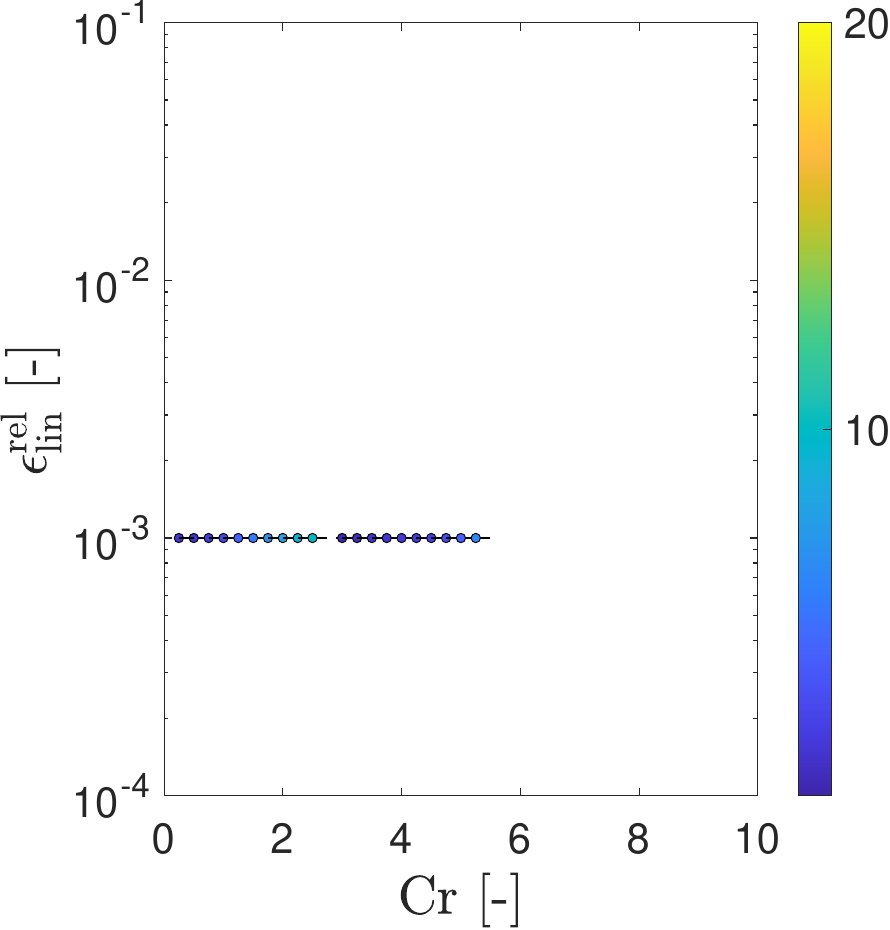}
		\caption{$\ve{u}_h^\#\cdot\nabla\ve{u}_h^{n+1}$, $\nu_h = \nu_h^\#$}
	\end{subfigure}
	\begin{subfigure}{.245\textwidth}
		\centering
		\includegraphics[width=\linewidth, draft=\draftSec]{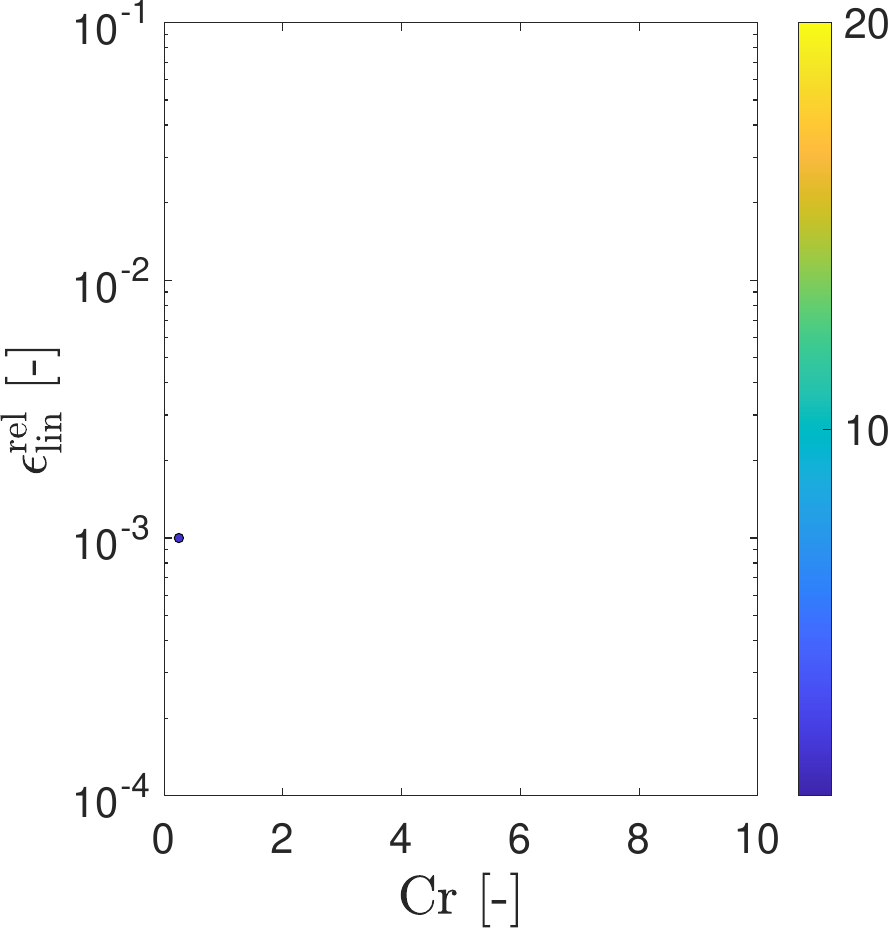}
		\caption{$\left[\convectiveterm{\ve{u}}{\ve{u}}\right]^\#$, $\nu_h = \nu_h^\#$}
	\end{subfigure}
	\caption{Linear solver iterations of the viscous step for various linearization variants
		of the splitting scheme with $k_u = 2$  over the target Courant number Cr and inner linear Krylov
		solver tolerance $\epsilon_\mathrm{lin}^\mathrm{rel}$
		(and accordingly scaled absolute inner Krylov solver tolerance).
		\textit{Linear} problems are solved with $\epsilon_\mathrm{lin}^\mathrm{rel}=10^{-3}$ only.
		\previouslyrevised{Points omitted indicate solver divergence,
			\revised{but linear schemes (e) and (f) consider only $\epsilon_\mathrm{lin}^\mathrm{rel} = 10^{-3}$.}}
	}
	\label{fig:backward_facing_step_linear_solver_behavior_splitting}
\end{figure}

Turning our attention to the results obtained for the splitting scheme,
comparing the average iteration counts per time step as
shown in Fig.~\ref{fig:backward_facing_step_linear_solver_behavior_splitting},
we observe a strict maximum Courant number independent of the linear solver
tolerance when an explicit formulation is chosen.
The fully implicit convective term allows for a wide range of
admissible target Courant numbers and linear solver tolerances.
The linear implicit convective term allows reaching a target
Courant number of more than 5 independent
of the linear solver tolerance within a nonlinear problem.
This can probably be improved by using a (linearly implicit)
skew-symmetric version of the convective term, i.e.,
replacing $(\ve{u}^{\star}\cdot\nabla)\ve{u}^{n+1}$ by
\begin{align*}
	(\ve{u}^{\star}\cdot\nabla)\ve{u}^{n+1} + \frac{\nabla\cdot\ve{u}^{\star}}{2}\ve{u}^{n+1}\, ,
\end{align*}
which---differently from other variants---does not
require $\nabla\cdot\ve{u}^{\star}=0$ for stability.
Finally, linearizing the viscous term does not result in noticeable restrictions,
which is against the observations made for the coupled solution approach.

\subsubsection{\previouslyrevised{Linear solver scalability}}

\previouslyrevised{
	Finally, we \revised{illustrate} scaling with respect to increased spatial resolution for velocity degrees $k_u=2,3,4$
	and corresponding pressure degrees $k_p=k_u-1$.
	The focus is on the Schur complement preconditioner and
	the viscous step in the projection method,
	as other steps involve simpler problems based on mass or Poisson operators,
	for which the adopted strategies are known to scale well.

	\revised{Within this section, we treat the nonlinear
	convective and viscous
	terms linearly implicit to avoid a nonlinear solver,
	while the linear solver tolerances are $\epsilon_\mathrm{lin}^\mathrm{abs}=10^{-12}$
	and $\epsilon_\mathrm{lin}^\mathrm{rel}=10^{-3}$.
	This choice includes the convective term on the left-hand side,
	while the nonlinear viscosity field
	is updated in an outer accelerated fixed-point scheme,
	see Algs.~\ref{alg:splitting_scheme} and \ref{alg:newton_coupled_all}.
	Revisiting Figs.~\ref{fig:backward_facing_step_linear_solver_behavior_coupled}
	and \ref{fig:backward_facing_step_linear_solver_behavior_splitting},
	the present choice shows the best performance for the coupled solution approach,
	which can suffer from inadequate Schur complement approximation depending on the
	problem parameters, while for the projection scheme, the scaling results are similar
	for all linearization variants using the present parameters.
	The tested configuration presents results for the fastest option identified in Sec.~\ref{sec:nonlin_lin_solver_performance}.

	The rheological parameters considered correspond to blood at
	a hematocrit value of 45\%~\cite{Kwon2008},
	see Tab.~\ref{tab:blood_parameters}.
	We hence present scaling for a particular parameter set only, noting that
	a dominant convective term challenges
	iterative solvers, and extreme viscosity gradients do so as well, see
	Tab.~\ref{tab:cavity_results_genNewtonian}.
	}

	The number of DoFs is increased by a factor of 8 due to uniform refinement between refinement levels with
	the	resulting DoF counts listed in Tab.~\ref{tab:backward_facing_step_scalability}.
	The time step is chosen adaptively targeting a Courant number of $\mathrm{Cr} = 0.8$,
	lying well below the maximal values achieved as listed in
	Tab.~\ref{tab:backward_facing_step_comparison} to rule out temporal instability issues.
}

\begin{table}[h!]
	\centering
	\caption{\previouslyrevised{Total DoF counts for uniform mesh refinement of an initial grid
		of the backward facing step, starting from a geometry involving approximately $10^4$ DoFs for various
		velocity and pressure polynomial degrees $k_u = k_p + 1 $.}}
	{
		\scriptsize
		\label{tab:backward_facing_step_scalability}
		\begin{tabular}{||c | c | c | c ||}
			\hline
			&&&\\[-1.25ex]
			level
			& $k_u=2$
			& $k_u=3$
			& $k_u=4$
			\\[1.25ex]
			\hline\hline
			&&&
			\\[-1.25ex]
			0 & $9.80 \times 10^3$ & $9.86 \times 10^3$ & $1.10 \times 10^3$ \\
			1 & $7.83 \times 10^4$ & $7.88 \times 10^4$ & $8.78 \times 10^4$ \\
			2 & $6.27 \times 10^5$ & $6.31 \times 10^5$ & $7.02 \times 10^5$ \\
			3 & $5.01 \times 10^6$ & $5.05 \times 10^6$ & $5.62 \times 10^6$ \\
			4 & $4.01 \times 10^7$ & $4.04 \times 10^7$ & $4.50 \times 10^7$ 
			\\[1.5ex]
			\hline
		\end{tabular}
	}
\end{table}

\previouslyrevised{
	Fig.~\ref{fig:backward_facing_step_scaling} shows the average GMRES iterations needed to reach convergence over all time steps
	of the coupled solver using the Cahouet--Chabard preconditioner~\eqref{eqn:schur_complement_cahouet_chabard}
	and the viscous step~\eqref{eqn:viscous_step_weak}, both employing $hp$ multigrid
	as described in Sec.~\ref{sec:matrix-free}. These steps and related iteration counts are the only
	ones in the respective schemes that are impacted by convective terms and variable viscosity.
	In Fig.~\ref{fig:backward_facing_step_scaling}, two effects overlap:
		First and foremost, the element integral containing the viscous terms scales with $h_e^{-2}$, while the other terms
		(stemming from mass and convective operators)
		grow slower. Hence, on fine grids, the viscous terms increasingly dominate,
		such that the Cahouet--Chabard preconditioner performs better as already
		observed in Sec.~\ref{sec:numerical_examples_cavity},
		see Tab.~\ref{tab:cavity_results_Newtonian}.
		%
		%
		Additionally, a constant Courant number for adaptive timestep selection~\eqref{eqn:CFL_timestep_selection}
		leads to decreased timestep size with increased order. This improves the extrapolation
		in time used as initial guess and increases the contribution of the mass operator,
		such that higher polynomial degrees might profit slightly.

	The employed multigrid strategy performs well under standard mesh quality requirements
	and assuming that spatial refinement can be employed to construct the multigrid hierarchy,
	especially when using low polynomial degrees.
}

\pgfplotstableread[comment chars={c}]{./results/backward_facing_step/scalability_tests/p2_CS.txt}\DataBFSScalePTwoCS
\pgfplotstableread[comment chars={c}]{./results/backward_facing_step/scalability_tests/p3_CS.txt}\DataBFSScalePThreeCS
\pgfplotstableread[comment chars={c}]{./results/backward_facing_step/scalability_tests/p4_CS.txt}\DataBFSScalePFourCS
\pgfplotstableread[comment chars={c}]{./results/backward_facing_step/scalability_tests/p2_DS.txt}\DataBFSScalePTwoDS
\pgfplotstableread[comment chars={c}]{./results/backward_facing_step/scalability_tests/p3_DS.txt}\DataBFSScalePThreeDS
\pgfplotstableread[comment chars={c}]{./results/backward_facing_step/scalability_tests/p4_DS.txt}\DataBFSScalePFourDS

\pgfplotsset{width=0.5\linewidth}
\begin{figure}[h!]
	\centering
	\begin{tikzpicture}
		\begin{axis}[
			xlabel={Refinement level},
			ylabel={Average iteration count},
			grid=both,
			xmin=-0.5,
			xmax=4.5,
			ymin=0,
			ymax=15,
			legend entries={
				{coupled, $k_u=2$},
				{coupled, $k_u=3$},
				{coupled, $k_u=4$},
				{viscous step, $k_u=2$},
				{viscous step, $k_u=3$},
				{viscous step, $k_u=4$}
			},
			legend style={
				nodes={scale=0.6, transform shape}},
			legend pos=north east,
			font = \normalsize,
			mark options = {scale = 0.5},
			legend cell align={left}
			]

			\addplot [ForestGreen,    mark=x,        dashed, mark options=solid, mark size=2.5pt] table [x=lvl, y=iter] {\DataBFSScalePTwoCS};
			\addplot [OrangeRed,      mark=o,        dashed, mark options=solid, mark size=2.5pt] table [x=lvl, y=iter] {\DataBFSScalePThreeCS};
			\addplot [RoyalBlue,      mark=triangle, dashed, mark options=solid, mark size=2.5pt] table [x=lvl, y=iter] {\DataBFSScalePFourCS};

			\addplot [ForestGreen,    mark=x,        mark size=5pt] table [x=lvl, y=iter] {\DataBFSScalePTwoDS};
			\addplot [OrangeRed,      mark=o,        mark size=5pt] table [x=lvl, y=iter] {\DataBFSScalePThreeDS};
			\addplot [RoyalBlue,      mark=triangle, mark size=5pt] table [x=lvl, y=iter] {\DataBFSScalePFourDS};

		\end{axis}
	\end{tikzpicture}
	\caption{\previouslyrevised{Average iteration counts under uniform refinement of the Schur complement solver for the coupled
			system and the viscous step in Eqn.~\eqref{eqn:viscous_step_weak}, respectively.
		    The multigrid preconditioners perform well under $hp$ refinement and the
		    Cahouet--Chabard preconditioner profits from dominant viscous terms as the grid is refined.}}
	\label{fig:backward_facing_step_scaling}
\end{figure}

In summary, implicit or linearly implicit formulations for the convective term
are the most robust choices with respect to target Courant number and linear solver tolerances.
\revised{The applied preconditioning strategies are shown to scale with spatial refinement for a specific, yet practically relevant parameter set corresponding to blood flow}.
Treating the convective term explicitly yields a strict time step restriction,
but simpler, and more efficient linear solvers
not considered within this work may be used in this case, see also the choices in~\cite{Fehn2018}.
The linearization of the viscous term does not lead to vast time step restrictions, especially for the
splitting scheme.
For each of the linearization variants, the splitting scheme yields a shorter time to solution,
where particularly those allowing for higher Courant numbers are interesting candidates when
focusing on implicit(-explicit) time integration schemes.
The fastest schemes presented here are the fully linearized ones with linearly implicit convective term
and linearized viscous term that nonetheless allow for Courant numbers up to 6 in the present case.
Such schemes are attractive alternatives to the fully implicit ones if the (slightly) decreased robustness and
the introduced linearization error are acceptable depending on the specific application at hand.

\section{Concluding Remarks}
\label{sec:concluding_remarks}

We presented a matrix-free higher-order discontinuous
Galerkin discretization of the Navier--Stokes equations for incompressible flows
adopting the symmetric interior penalty Galerkin method for the viscous terms with
variable viscosity.
Several (partially) linearized implicit-explicit variants of mixed and
decoupled projection-based time integration schemes are compared.
The former lead to a velocity-pressure block-system treated
via a physics-based preconditioner resting on suitable pressure Schur complement
approximations. These approximations can be delicate depending on the physical parameters
underlying the linear problem to be solved. The
splitting scheme originally proposed in~\cite{Karniadakis1991}
\previouslychanged{and extended towards non-constant viscosity by~\citet{Karamanos2000}}
has been demonstrated to handle wide parameter ranges while still leading to systems
that can be effectively preconditioned.
As well known for the constant-viscosity case, the accuracy of projection schemes
tend to decrease for highly viscous flows due to numerical boundary layers.
Optimal results are obtained for medium and high Reynolds numbers,
\previouslyrevised{approximately $\mathrm{Re}\geq100$}.
In such regimes, for variable viscosity we additionally observe a degradation of the
temporal convergence slope to linear, even when using higher-order integrators.
Nonetheless, for a suitable parameter range, the splitting scheme
can deliver accurate results and
speed-ups of 1.3--3.5 over the monolithic scheme using identical linearizations.
Comparing the fully implicit variants of the coupled and splitting schemes,
the fastest implicit-explicit variants have a relative compute time of 0.59 (coupled scheme)
and 0.32 (splitting scheme).

These findings in summary indicate that linearization---whenever the problem at hand
allows in terms of accuracy and robustness---can
significantly decrease the time to solution.
Moreover, splitting/projection schemes show better computational performance,
especially when preconditioning the arising saddle-point systems is challenging or
(quasi-) time dependent problems with high temporal resolution need to be solved.
Coupled monolithic approaches remain interesting due to the possibility of
arbitrarily high-order time integration and parameter-robustness, if efficient
preconditioning strategies exist.
Hence, conclusions from related
works~\cite{Turek1996, Turek1999, John2006_a} on the recurring questions for the
fastest general Navier--Stokes solvers
naturally still hold despite using highly tuned matrix-free algorithms.

\previouslychanged{Similar to the original scheme~\cite{Karniadakis1991}, the extension proposed in~\cite{Karamanos2000}}
to account for variable viscosity as stemming from various rheological models,
other constitutive laws or simply given as a function of space and time
suffers from a pressure boundary layer of thickness $\mathcal{O}(\sqrt{\nu \Delta t})$.
Within the present work, this problem is counteracted by an added penalty step~\eqref{eqn:penalty_step},
enforcing mass conservation and normal continuity of the velocity vector across elements.
Compared to the original scheme, this additional step helps to detect when exactly the boundary
divergence error is too large and time step sizes have to be reduced accordingly.
Therefore, the scheme can be employed in practical scenarios as demonstrated
in the numerical results section.

\previouslyrevised{
	As for the extension towards higher-order BDF schemes and extrapolation,
	we note that despite the scheme performing well in many practical scenarios,
	we cannot recommend BDF-3 and higher in general.
	To the best of our knowledge, there
	exists no proof of unconditional stability for
	BDF-3~\cite{Karniadakis1991, Guermond2003a, Guermond2006},
	while recent developments by~\citet{Huang2025stability}
	regarding related consistent
	splitting schemes are promising.
}

\section{Acknowledgements}
We gratefully acknowledge the scientific support and HPC resources
provided by
the HPC cluster Elysium of the Ruhr University Bochum,
subsidized by the DFG (INST 213/1055-1).
Moreover, this work was partially supported by the German Federal
Ministry of Research, Technology and Space (BMFTR) through project
``PDExa: Optimized software methods for solving partial differential
equations on exascale supercomputers'', grant agreement no. 16ME0637K
and the European Union -- NextGenerationEU.
D.R.Q.~Pacheco acknowledges funding by the
Federal Ministry of Research, Technology and Space (BMFTR)
and
the Ministry of Culture and Science of the German State of North Rhine-Westphalia (MKW)
under the Excellence Strategy of the Federal Government and the Länder.

\bibliographystyle{unsrtnat}
\bibliography{bibliography}

\end{document}